\documentclass[12pt]{article}
\usepackage[dvips]{graphicx}
\usepackage{epsfig}
\usepackage{latexsym,amsmath,amsfonts,amssymb}
\usepackage[latin1]{inputenc}
\usepackage[american]{babel}
\usepackage[dvips]{graphicx}
\usepackage{bbm}
\usepackage[unicode]{hyperref}
\def\im{Invent. Math.}

\def\a{\alpha}
\def\b{\beta}
\def\c{\gamma}
\def\d{\delta}
\def\eps{\epsilon}           
\def\f{\phi}               
\def\vf{\varphi}  \def\tvf{\tilde{\varphi}}
\def\vp{\varphi}
\def\g{\gamma}
\def\h{\eta}
\def\j{\psi}
\def\k{\kappa}                    
\def\l{\lambda}
\def\m{\mu}
\def\n{\nu}
\def\o{\omega}  \def\w{\omega}
\def\p{\pi}
\def\q{\theta}  \def\th{\theta}                  
\def\r{\rho}                                     
\def\s{\sigma}                                   
\def\t{\tau}
\def\u{\upsilon}
\def\x{\xi}
\def\z{\zeta}
\def\pt{\tilde{\varphi}}
\def\tt{\tilde{\theta}}
\def\lab{\label}
\def\6{\partial}
\def\wg{\wedge}
\def\bpsi{\bar{\psi}}
\def\bt{\bar{\theta}}
\def\bvf{\bar{\varphi}}

\newcommand{\td}{\mathrm{d}}

\DeclareMathOperator{\tr}{tr}

\newcommand{\be}{\begin{equation}}
\newcommand{\ee}{\end{equation}}
\newcommand{\beq}{\begin{equation}}
\newcommand{\eeq}{\end{equation}}
\newcommand{\bea}{\begin{eqnarray}}
\newcommand{\eea}{\end{eqnarray}}

\newcommand{\ba}{\begin{eqnarray}}
\newcommand{\ea}{\end{eqnarray}}

\begin{document}
\baselineskip=15.5pt
\pagestyle{plain}
\setcounter{page}{1}


\def\del{{\partial}}
\def\vev#1{\left\langle #1 \right\rangle}
\def\cn{{\cal N}}
\def\co{{\cal O}}
\def\IC{{\mathbb C}}
\def\IR{{\mathbb R}}
\def\IZ{{\mathbb Z}}
\def\RP{{\bf RP}}
\def\CP{{\bf CP}}
\def\Poincare{{Poincar\'e }}
\def\tr{{\rm tr}}
\def\tp{{\tilde \Phi}}

\def\TL{\hfil$\displaystyle{##}$}
\def\TR{$\displaystyle{{}##}$\hfil}
\def\TC{\hfil$\displaystyle{##}$\hfil}
\def\TT{\hbox{##}}
\def\HLINE{\noalign{\vskip1\jot}\hline\noalign{\vskip1\jot}}
\def\seqalign#1#2{\vcenter{\openup1\jot
   \halign{\strut #1\cr #2 \cr}}}
\def\lbldef#1#2{\expandafter\gdef\csname #1\endcsname {#2}}
\def\eqn#1#2{\lbldef{#1}{(\ref{#1})}%
\begin{equation} #2 \label{#1} \end{equation}}
\def\eqalign#1{\vcenter{\openup1\jot
     \halign{\strut\span\TL & \span\TR\cr #1 \cr
    }}}
\def\eno#1{(\ref{#1})}
\def\href#1#2{#2}
\def\half{{1 \over 2}}

\def\ads{{\it AdS}}
\def\adsp{{\it AdS}$_{p+2}$}
\def\cft{{\it CFT}}

\newcommand{\ber}{\begin{eqnarray}}
\newcommand{\eer}{\end{eqnarray}}

\newcommand{\beqar}{\begin{eqnarray}}
\newcommand{\cN}{{\cal N}}
\newcommand{\cO}{{\cal O}}
\newcommand{\cA}{{\cal A}}
\newcommand{\cT}{{\cal T}}
\newcommand{\cF}{{\cal F}}
\newcommand{\cC}{{\cal C}}
\newcommand{\cR}{{\cal R}}
\newcommand{\cW}{{\cal W}}
\newcommand{\eeqar}{\end{eqnarray}}
\newcommand{\tht}{\thteta}
\newcommand{\lm}{\lambda}\newcommand{\Lm}{\Lambda}


\newcommand{\nonu}{\nonumber}
\newcommand{\oh}{\displaystyle{\frac{1}{2}}}
\newcommand{\dsl}
   {\kern.06em\hbox{\raise.15ex\hbox{$/$}\kern-.56em\hbox{$\partial$}}}
\newcommand{\id}{i\!\!\not\!\partial}
\newcommand{\as}{\not\!\! A}
\newcommand{\ps}{\not\! p}
\newcommand{\ks}{\not\! k}
\newcommand{\D}{{\cal{D}}}
\newcommand{\dv}{d^2x}
\newcommand{\Z}{{\cal Z}}
\newcommand{\N}{{\cal N}}
\newcommand{\Dsl}{\not\!\! D}
\newcommand{\Bsl}{\not\!\! B}
\newcommand{\Psl}{\not\!\! P}
\newcommand{\eeqarr}{\end{eqnarray}}
\newcommand{\ZZ}{{\rm \kern 0.275em Z \kern -0.92em Z}\;}


\def\del{{\delta^{\hbox{\sevenrm B}}}} \def\ex{{\hbox{\rm e}}}
\def\azb{A_{\bar z}} \def\az{A_z} \def\bzb{B_{\bar z}} \def\bz{B_z}
\def\czb{C_{\bar z}} \def\cz{C_z} \def\dzb{D_{\bar z}} \def\dz{D_z}
\def\im{{\hbox{\rm Im}}} \def\mod{{\hbox{\rm mod}}} \def\tr{{\hbox{\rm Tr}}}
\def\ch{{\hbox{\rm ch}}} \def\imp{{\hbox{\sevenrm Im}}}
\def\trp{{\hbox{\sevenrm Tr}}} \def\vol{{\hbox{\rm Vol}}}
\def\rl{\Lambda_{\hbox{\sevenrm R}}} \def\wl{\Lambda_{\hbox{\sevenrm W}}}
\def\fc{{\cal F}_{k+\cox}} \def\vev{vacuum expectation value}
\def\nodiv{\mid{\hbox{\hskip-7.8pt/}}}
\def\ie{{\em i.e.}}
\def\ie{\hbox{\it i.e.}}

\def\CC{{\mathchoice
{\rm C\mkern-8mu\vrule height1.45ex depth-.05ex
width.05em\mkern9mu\kern-.05em}
{\rm C\mkern-8mu\vrule height1.45ex depth-.05ex
width.05em\mkern9mu\kern-.05em}
{\rm C\mkern-8mu\vrule height1ex depth-.07ex
width.035em\mkern9mu\kern-.035em}
{\rm C\mkern-8mu\vrule height.65ex depth-.1ex
width.025em\mkern8mu\kern-.025em}}}

\def\RR{{\rm I\kern-1.6pt {\rm R}}}
\def\NN{{\rm I\!N}}
\def\ZZ{{\rm Z}\kern-3.8pt {\rm Z} \kern2pt}
\def\IB{\relax{\rm I\kern-.18em B}}
\def\ID{\relax{\rm I\kern-.18em D}}
\def\II{\relax{\rm I\kern-.18em I}}
\def\IP{\relax{\rm I\kern-.18em P}}
\newcommand{\CS}{{\scriptstyle {\rm CS}}}
\newcommand{\CSs}{{\scriptscriptstyle {\rm CS}}}
\newcommand{\rc}{\nonumber\\}
\newcommand{\bear}{\begin{eqnarray}}
\newcommand{\eear}{\end{eqnarray}}

\newcommand{\F}{{\cal F}}
\newcommand{\LL}{{\cal L}}

\def\mani{{\cal M}}
\def\calo{{\cal O}}
\def\calb{{\cal B}}
\def\calw{{\cal W}}
\def\calz{{\cal Z}}
\def\cald{{\cal D}}
\def\calc{{\cal C}}
\def\to{\rightarrow}
\def\ele{{\hbox{\sevenrm L}}}
\def\ere{{\hbox{\sevenrm R}}}
\def\zb{{\bar z}}
\def\wb{{\bar w}}
\def\nodiv{\mid{\hbox{\hskip-7.8pt/}}}
\def\menos{\hbox{\hskip-2.9pt}}
\def\dr{\dot R_}
\def\drr{\dot r_}
\def\ds{\dot s_}
\def\da{\dot A_}
\def\dga{\dot \gamma_}
\def\ga{\gamma_}
\def\dal{\dot\alpha_}
\def\al{\alpha_}
\def\cl{{closed}}
\def\cls{{closing}}
\def\vev{vacuum expectation value}
\def\tr{{\rm Tr}}
\def\to{\rightarrow}
\def\too{\longrightarrow}


\def\a{\alpha}
\def\b{\beta}
\def\c{\gamma}
\def\d{\delta}
\def\e{\epsilon}           
\def\f{\phi}               
\def\vf{\varphi}  \def\tvf{\tilde{\varphi}}
\def\vp{\varphi}
\def\g{\gamma}
\def\h{\eta}
\def\j{\psi}
\def\k{\kappa}                    
\def\l{\lambda}
\def\m{\mu}
\def\n{\nu}
\def\o{\omega}  \def\w{\omega}
\def\q{\theta}  \def\th{\theta}                  
\def\r{\rho}                                     
\def\s{\sigma}                                   
\def\t{\tau}
\def\u{\upsilon}
\def\x{\xi}
\def\z{\zeta}
\def\pt{\tilde{\varphi}}
\def\tt{\tilde{\theta}}
\def\lab{\label}
\def\6{\partial}
\def\wg{\wedge}
\def\atanh{{\rm arctanh}}
\def\bpsi{\bar{\psi}}
\def\bt{\bar{\theta}}
\def\bvf{\bar{\varphi}}

%

\newfont{\namefont}{cmr10}
\newfont{\addfont}{cmti7 scaled 1440}
\newfont{\boldmathfont}{cmbx10}
\newfont{\headfontb}{cmbx10 scaled 1728}
\newcommand{\re}{\,\mathbb{R}\mbox{e}\,}
\newcommand{\hyph}[1]{$#1$\nobreakdash-\hspace{0pt}}
\providecommand{\abs}[1]{\lvert#1\rvert}
\newcommand{\Nugual}[1]{$\mathcal{N}= #1 $}
\newcommand{\sub}[2]{#1_\text{#2}}
\newcommand{\partfrac}[2]{\frac{\partial #1}{\partial #2}}
\newcommand{\bsp}[1]{\begin{equation} \begin{split} #1 \end{split} \end{equation}}
\newcommand{\calF}{\mathcal{F}}
\newcommand{\calO}{\mathcal{O}}
\newcommand{\calM}{\mathcal{M}}
\newcommand{\calV}{\mathcal{V}}
\newcommand{\bbZ}{\mathbb{Z}}
\newcommand{\bbC}{\mathbb{C}}
\newcommand{\cK}{{\cal K}}

\newcommand{\Thq}{\Theta\left(\r-\r_q\right)}
\newcommand{\Dq}{\d\left(\r-\r_q\right)}
\newcommand{\kten}{\kappa^2_{\left(10\right)}}
\newcommand{\pbi}[1]{\imath^*\left(#1\right)}
\newcommand{\ho}{\hat{\omega}}
\newcommand{\tth}{\tilde{\th}}
\newcommand{\tf}{\tilde{\f}}
\newcommand{\tj}{\tilde{\j}}
\newcommand{\tw}{\tilde{\omega}}
\newcommand{\tz}{\tilde{z}}
\newcommand{\prj}[2]{(\partial_r{#1})(\partial_{\j}{#2})-(\partial_r{#2})(\partial_{\j}{#1})}
\def\atanh{{\rm arctanh}}
\def\sech{{\rm sech}}
\def\csch{{\rm csch}}
\newcommand{\Pufl}{P_{\textrm{unfl}}}
\newcommand{\hc}{h_{1,\textrm{c}}}
\allowdisplaybreaks[1]

\numberwithin{equation}{section}

\newcommand{\Tr}{\mbox{Tr}}    


%
\renewcommand{\theequation}{{\rm\thesection.\arabic{equation}}}
\begin{titlepage}
\vspace{0.1in}

\begin{center}
\Large \bf On the holographic dual of ${\cal N}=1$ SQCD  with massive flavors
\end{center}
\vskip 0.2truein
\begin{center}
Eduardo Conde  ${}^{*}$\footnote{eduardo@fpaxp1.usc.es},
J\'er\^ome Gaillard ${}^{*\, \dagger}$\footnote{pyjg@swansea.ac.uk}
and
Alfonso V. Ramallo ${}^{*}$\footnote{alfonso@fpaxp1.usc.es}\\
\vspace{0.2in}
${}^{*}$ 
\it{
Departamento de  F\'\i sica de Part\'\i  culas, Universidade
de Santiago de
Compostela\\and\\Instituto Galego de F\'\i sica de Altas
Enerx\'\i as (IGFAE)\\E-15782, Santiago de Compostela, Spain
\\
\vspace{0.2in}
${}^{\dagger}$ 
{\it Department of Physics, Swansea University \\ Singleton Park, Swansea SA2 8PP, United Kingdom}
}

\vspace{0.2in}
\end{center}
\vspace{0.2in}
\centerline{{\bf Abstract}}
We construct holographic duals to ${\cal N}=1$ SQCD with a quartic superpotential and unquenched massive flavors. Our backgrounds are generated by D5-branes wrapping two-dimensional submanifolds of an internal space. The flavor degrees of freedom are introduced by means of D5-branes extended along two-dimensional calibrated surfaces, and act as sources of the different supergravity fields. The backgrounds we get include the backreaction of the flavor branes and generalize the geometries obtained so far to the case in which the  fundamental matter is massive. The supergravity solutions we find are regular everywhere and depend on a radial function which can be determined from the distribution of flavor branes used as sources. We also work out the holomorphic structure of the model and explore some of its observable consequences.

\smallskip
\end{titlepage}
\setcounter{footnote}{0}

\tableofcontents

\section{Introduction}
The AdS/CFT correspondence \cite{Maldacena:1997re} is one of the greatest conceptual developments in the study of the dynamics of gauge theories of the recent years (see \cite{Aharony:1999ti} for a review).  This correspondence, which is in agreement with early ideas of  't Hooft \cite{'tHooft:1973jz} on the large $N_c$ limit of QCD, has provided a whole set of analytical tools to study gauge theories in the strongly coupled regime. Besides, through the holographic principle of quantum gravity, the correspondence has established a fascinating  and far-reaching connection between gauge theories and the physics of black holes. 

Although the final goal would be finding a dual to QCD, a less ambitious and more feasible objective is trying to construct gravitational backgrounds dual to minimal supersymmetric models. In this paper we will concentrate on the model \cite{Malda-Nunez:N=1} dual to ${\cal N}=1$ super-Yang-Mills (SYM) theory, which is based on the geometry obtained by Chamseddine and Volkov in ref. \cite{Chamseddine:1997nm}.  This geometry can be regarded as generated by a set of $N_c$ D5-branes wrapping a compact two-cycle of a Calabi-Yau cone. If the size of the cycle is small, the low-energy description of the wrapped D5-branes is effectively (3+1)-dimensional. However, at larger energies the Kaluza-Klein modes on the compact cycle show up and mix with the four-dimensional  field theory degrees of freedom. Nevertheless, the model of \cite{Malda-Nunez:N=1} successfully encodes confinement and chiral symmetry breaking in a geometric setup (see \cite{MNreviews} for various reviews).

Another step in the process of approaching the holographic gravitational models to phenomenology is the addition of flavor, \ie\  of fields transforming in the fundamental representation of the gauge group (quarks). From the point of view of the gravitational theory, adding quarks to a given gauge theory corresponds to incorporating additional branes to the setup \cite{Karch:2002sh}. These flavor branes should extend along the gauge theory directions and  wrap a non-compact cycle in the internal manifold, in order to make their worldvolume symmetry a global flavor symmetry. If the number $N_f$ of flavor branes is much smaller than the number $N_c$ of color branes, one can reasonably  neglect the effect of the flavor branes on the geometry and treat them as probes. This is the so-called quenched approximation which, on the field theory side, amounts to considering the quarks as external non-dynamical objects that do not run in the loops. For the model dual to ${\cal N}=1$ SYM the flavor branes are also D5-branes which wrap a non-compact submanifold of the internal space in such a way that  ${\cal N}=1$ is preserved \cite{Nunez:2003cf} (see \cite{Erdmenger:2007cm} for a review of similar studies in several other models). 

In this paper we are interested in studying unquenched flavor in the holographic dual of ${\cal N}=1$ SYM. In this case one has to compute the backreaction by solving the equations of motion of a system of gravity with brane sources. Generically, these sources modify the Einstein equations and the Bianchi identities of some Ramond-Ramond field strengths. We will follow the approach  initiated in \cite{Casero:2006pt}, in which one has a large number of flavor brane sources which are delocalized and one has to deal with a continuous smeared distribution of branes (see \cite{Bigazzi:2005md} for an earlier implementation of this idea in the context of non-critical string theory). In this approach the sources do not contain Dirac $\delta$-functions, which greatly simplifies the task of solving the equations of motion. On the field theory side this setup corresponds to the so-called Veneziano limit, in which both $N_c$ and $N_f$ are large but their ratio is kept fixed. 
In refs. \cite{Casero:2007pt,HoyosBadajoz:2008fw, Caceres} different aspects of the supergravity duals of ${\cal N}=1$ SYM with smeared flavor branes were studied, whereas this approach has been also successfully applied to other  types of backgrounds (see \cite{Nunez:2010sf} for a detailed review). 

The ${\cal N}=1$  flavors added in refs. \cite{Casero:2006pt,Casero:2007pt,HoyosBadajoz:2008fw} are massless, which amounts to considering flavor branes extended along the full range of the holographic coordinate $r$.  The corresponding supergravity solutions are singular in the IR. This is, actually, a common  feature of most  massless flavored solutions found so far with the smearing technique (see, for example, refs. \cite{Benini:2006hh,Benini:2007gx,Benini:2007kg} for the D3-D7 systems on the conifold). This curvature singularity can be qualitatively understood as due to the fact that, for massless flavors, all branes pass through the origin $r=0$ and, therefore, the brane density is highly peaked at $r=0$ (an exception to this behavior is the solution recently found in \cite{Conde:2011sw} for the gravity dual of Chern-Simons-matter theories with flavors).

To remove the IR singularity one can consider massive quarks or, equivalently, a family of flavor branes which do not reach the origin (another possibility is to add temperature and to hide the singularity behind a horizon, as was done in ref. \cite{Bigazzi:2009bk}). For the D3-D7 system these regular solutions  for massive flavors were found in refs. \cite{smearingmethod,Bigazzi:2008ie,Bigazzi:2008qq}. As argued in \cite{Benini:2006hh}, passing from the massless to the massive case in these systems just amounts to substituting in the ansatz $N_f$ by $N_f S(r)$, where $S(r)$ is a profile function that interpolates between zero in the IR and one in the UV. To calculate $S(r)$ one has to perform a microscopic calculation of the flavor brane charge density, whose result is not universal since  it depends both on the characteristics of the  unflavored system and on the particular family of flavor brane embeddings. 

In this paper we find supergravity backgrounds dual to ${\cal N}=1$ SYM theories with unquenched massive quarks. The first step in our analysis will be finding the precise deformation of the background which corresponds to the backreaction induced by massive flavors. We will show that the compatibility with the  ${\cal N}=1$ supersymmetry implies  a certain type of deformation which is also parameterized by a profile function $S(r)$. When this function $S$ is identically equal to one we recover the results of \cite{Casero:2006pt} for massless quarks. However it is important to point out that, in this D5-brane case,  the massive quark ansatz cannot be recovered by performing the $N_f\to N_f S(r)$ substitution in the  massless ansatz of \cite{Casero:2006pt}. 

From our ansatz we will be able to obtain a consistent system of first-order BPS equations which can be partially integrated and reduced to a second-order master equation which is the generalization to this massive case of the equation derived in \cite{HoyosBadajoz:2008fw} for massless quarks. To solve this master equation (and the full BPS system) one needs to know first the profile function $S(r)$ which, as mentioned above, is not universal and depends on the family of embeddings of the flavor D5-branes. Such families are generated by acting with isometries on a fiducial representative embedding. It turns out that only a particular set of these families produces a backreaction which is compatible with our ansatz. For this reason we must generalize the results of \cite{Nunez:2003cf} and find new classes of supersymmetric embeddings of flavor D5-branes. In order to carry out this analysis we will introduce a convenient set of complex coordinates suitable to represent the metric and forms of the $SU(3)$-structure of our geometry. Employing these variables we will be able to find a family of compatible embeddings and to compute the corresponding profile function $S(r)$. 

For massive quarks the  function $S(r)$ vanishes when $r$ is less than a certain value $r_0$, which is related to the mass of the quarks. For $r\le r_0$ the BPS system coincides with the unflavored one, which corresponds to the fact that the quarks are effectively integrated out in this low-energy region. As shown in refs. \cite{Casero:2006pt,HoyosBadajoz:2008fw} there exists a one-parameter family of solutions of the unflavored system which are regular at $r=0$. Our flavored solutions coincide with these in this $0\le r\le r_0$ region and, although a potential threshold singularity could appear at $r=r_0$, we will show how to engineer brane distributions which give rise to geometries that are regular everywhere. 

The rest of this paper is organized as follows. In section \ref{unflavoredMN} we review the basic features of the holographic dual to unflavored ${\cal N}=1$ SYM. In section 
\ref{massive} we study the addition of massive flavor to the  ${\cal N}=1$ background and we present our ansatz for the backreaction induced by a smeared distribution of flavor branes. In this section we will also present the result of the partial integration of the BPS system, as well as the master equation for massive flavors. The holomorphic structure of the model is worked out in section \ref{holomorphic}. In section \ref{charge-distributions} we develop a technique to compute the charge distribution function $S(r)$. In this method $S(r)$ is obtained by comparing the Wess-Zumino action for the continuous set of branes and that of a single representative embedding. By applying this procedure we will discover that not all the families of embeddings produce a backreaction compatible with our ansatz. In section \ref{simple-embeddings} we find a simple class of compatible embeddings and we compute the corresponding profile function.

The problem of the threshold singularities is analyzed in section \ref{Smoothing}, where we show how to avoid them and how one can construct regular flavored backgrounds. 
In section \ref{numerics} we integrate numerically the master equation and we provide numerical solutions for the different functions of the ansatz. Some observable consequences of our model are analyzed in section \ref{observables}. In section \ref{conclusions} we summarize our main results and we discuss some further lines of research. The paper is completed with several appendices. In appendix \ref{SUSY} we study in detail the realization of ${\cal N}=1$ supersymmetry for our ansatz and we analyze the corresponding BPS system. In appendix \ref{EoM} we write in detail the equations of motion satisfied by our solutions. Appendix \ref{micro} contains a microscopic calculation of the charge density for some embeddings. Finally, in appendix \ref{KS} we reconsider the Klebanov-Strassler model with unquenched massive flavors and we apply the new techniques developed in the main text to compute the D7-brane source distribution.

\section{The holographic dual of ${\cal N}=1$ SYM}
\label{unflavoredMN}

In this section we will briefly review the supergravity dual to ${\cal N}=1$ SYM found in ref. \cite{Malda-Nunez:N=1}, which is based on the four-dimensional supergravity solution obtained in \cite{Chamseddine:1997nm}.  This supergravity background is generated by $N_c$ D5-branes that wrap a compact two-cycle inside a Calabi-Yau threefold. At low energies this supergravity solution is dual to a four-dimensional gauge theory, whereas, at sufficiently high energy, the theory becomes six-dimensional. Moreover, due to a twisting procedure in the compactification, the background preserves four supercharges. The corresponding ten-dimensional metric in Einstein 
frame is given by:
\beq
\td s^2_{10}\,=\,g_s\alpha' N_c\,e^{{\Phi\over 2}}\,\,\Big[\,
\frac{1}{g_s\alpha' N_c}\td x^2_{1,3}\,+\,
\td r^2\,+\,e^{2h}\,\big(\,\td\theta^2+\sin^2\theta\, \td\phi^2\,\big)\,+\,{1\over 4}\,(\tilde\omega^i-A^i)^2\,\Big]\,\,,
\label{metric}
\eeq
where $\Phi$ is the dilaton. In the remaining of the paper we will use units where $g_s \alpha'=1$. The angles
$\theta\in [0,\pi]$ and 
$\phi\in [0,2\pi)$ parameterize a two-sphere which is
 fibered by the one-forms 
$A^i$ $(i=1,2,3)$, which can be regarded as  the components of an $SU (2)$ non-abelian gauge vector field. Their expressions can be written in terms of a function 
$a(r)$ and the angles $(\theta,\phi)$ as follows:
\beq
A^1\,=\,-a(r) \td\theta\,,
\,\,\,\,\,\,\,\,\,
A^2\,=\,a(r) \sin\theta\, \td\phi\,,
\,\,\,\,\,\,\,\,\,
A^3\,=\,- \cos\theta\, \td\phi\,.
\label{oneform}
\eeq
The   $\tilde\omega^i\,$'s appearing in eq. \eqref{metric} are the $SU (2)$ left-invariant one-forms,
satisfying 
$\td\tilde\omega^i=-{1\over 2}\,\epsilon_{ijk}\,\tilde\omega^j\wedge \tilde\omega^k$, which parameterize a
three-sphere and can be represented in
terms of three angles $\tilde\theta$, $\tilde\phi$ and $\psi$:
\bear
\tilde\omega^1&=& \cos\psi \,\td\tilde\theta\,+\,\sin\psi\sin\tilde\theta\,\td\tilde\phi\,\,,\rc[0.1in]
\tilde\omega^2&=&-\sin\psi \,\td\tilde\theta\,+\,\cos\psi\sin\tilde\theta \,
\td\tilde\phi\,\,,\rc[0.1in]
\tilde\omega^3&=&\td\psi\,+\,\cos\tilde\theta\, \td\tilde\phi\,\,.
\eear
The three angles $\tilde\theta$, $\tilde\phi$ and $\psi$ take values in the
range $0\le\tilde\theta\le\pi$, $0\le\tilde\phi< 2\pi$ and
$0\le\psi< 4\pi$. For a metric ansatz such as the one written in 
(\ref{metric}) one  obtains a supersymmetric solution when
the functions $a(r)$, $h(r)$ and the dilaton $\Phi$ are:
\beq
\begin{aligned}
a(r)&={2r\over \sinh (2r)}\,\,,\rc[0.1in]
e^{2h}&=r\coth (2r)\,-\,{r^2\over \sinh^2 (2r)}\,-\,
{1\over 4}\,\,,\rc[0.1in]
e^{-2\Phi}&=e^{-2\Phi_0}{2e^h\over \sinh (2r)}\,\,,
\end{aligned}
\label{MNsol}
\eeq
where $\Phi_0$ is the value of the dilaton at $r=0$. Near the origin $r=0$ the function 
$e^{2h}$ in (\ref{MNsol})  behaves as $e^{2h}\sim r^2$ and the metric is non-singular. The solution of the type IIB supergravity includes a
Ramond-Ramond three-form $F_{(3)}$ given by:
\beq
F_{(3)}\,=\,-{N_c\over 4}\,\big(\,\tilde\omega^1-A^1\,\big)\wedge 
\big(\,\tilde\omega^2-A^2\,\big)\wedge \big(\,\tilde\omega^3-A^3\,\big)\,+\,{N_c\over 4}\,\,
\sum_a\,F^a\wedge \big(\,\tilde\omega^a-A^a\,\big)\,\,,
\label{RRthreeform}
\eeq
where $F^a$ is the field strength of the  $SU (2)$ gauge field $A^a$, defined as:
\beq
F^a\,=\,\td A^a\,+\,{1\over 2}\epsilon_{abc}\,A^b\wedge A^c\,\,.
\label{fieldstrenght}
\eeq
When the $A^a$'s are given by (\ref{oneform}),   the different components of $F^a$ are:
\beq
F^1\,=\,-a'\,\td r\wedge \td\theta\,\,,
\,\,\,\,\,\,\,\,\,\,
F^2\,=\,a'\sin\theta\, \td r\wedge \td\phi\,\,,
\,\,\,\,\,\,\,\,\,\,
F^3\,=\,(\,1-a^2\,)\,\sin\theta \,\td\theta\wedge \td\phi\,\,,
\label{Fa-unflavored}
\eeq
where the prime denotes derivative with respect to $r$. 

One can readily verify that, due  to the relation (\ref{fieldstrenght}), the three-form $F_{(3)}$ written in (\ref{RRthreeform}) is closed, \ie\ it satisfies the Bianchi identity 
$\td F_{(3)}\,=\,0$. Moreover,  the  field strength (\ref{RRthreeform}) satisfies the flux quantization condition corresponding to $N_c$ color D5-branes, namely:
\beq
-{1\over 2\kappa_{10}^2\,T_5}\,\,\int_{{\mathbb S}^3}\,\,F_{(3)}\,=\,N_c\,\,,
\label{flux}
\eeq
where the three-sphere is the one parameterized by the three angles $\tilde\theta$, $\tilde\phi$ and $\psi$ at a fixed value of all the other coordinates and, in order to check (\ref{flux}),  one should take into account that, in our units,  $T_5=1/(2\pi)^5$ and $2\kappa_{10}^2=(2\pi)^7$.

It was argued in \cite{Malda-Nunez:N=1} that the background written above is dual to ${\cal N}=1$ SYM in four dimensions plus some Kaluza-Klein (KK) adjoint matter. The four-dimensional theory is obtained by reducing the six-dimensional theory living on the D5-branes with the appropriate topological twist. The latter is necessary to realize the  ${\cal N}=1$ supersymmetry on the curved space \cite{Malda-Nunez:nogo} . The KK modes in the four-dimensional theory have masses of the order $1/\sqrt{g_s\,\alpha' \,N_c}$.  Since this mass is of the order of the strong coupling scale, the dynamics of the KK modes cannot be decoupled from the dynamics of confinement. A proposal for a concrete lagrangian of the vector   ${\cal N}=1$  multiplet   and the different KK modes has been written in \cite{HoyosBadajoz:2008fw} (see also \cite{AndrewsDorey}).  Schematically, this  lagrangian has the form :                                                         
\beq
L= \Tr[-\frac{1}{4}F_{\mu\nu}^2 -i \bar{\lambda}\gamma^\mu D_\mu \lambda + 
L(\Phi_k, W_k,W)]\,\,,
\eeq
where $\Phi_k$ and $W_k$  represent the infinite number of 
massive chiral and vector multiplets and $W$ denotes the curvature of the massless ${\cal N}=1$ vector  multiplet $V=(\lambda, A_{\mu})$. 

Let us finish this section by recalling that there exists another solution of type IIB supergravity  which is directly related to the one written above. In this solution the metric and the RR three-form are also given by the ansatz (\ref{metric})-(\ref{Fa-unflavored}) but, in this case, the function $a(r)$ vanishes, $e^{2h}=r$ and 
$e^{2\Phi-2\Phi_0}\,=\,{e^{2r}\over 4\sqrt{r}}$.  Actually, these functions are just the UV limit ($r\to\infty$) of the ones written in (\ref{MNsol}). On the other hand, at $r=0$ this new background has a (bad) singularity that is solved by the turning on of the function $a(r)$ in the solution (\ref{MNsol}), which makes the $A^a$ a non-abelian  one-form connection with components  along the three ${SU} (2)$ directions. This way of resolving the singularity is related,  on the field theory side, with the phenomena of confinement and R-symmetry breaking of ${\cal N}=1$ SYM.

\section{Addition of massive flavors}
\label{massive}
Let us now introduce flavors by means of pairs of chiral multiplets $Q$ and $\tilde Q$ transforming in the fundamental and antifundamental representations of both the gauge group $SU(N_c)$ and  the flavor group $SU(N_f)$. The lagrangian for the $(Q, \tilde Q)$ fields is given by the usual kinetic terms and the Yukawa interaction between  the quarks and the KK modes, which can be schematically written as:
\beq
L_{Q, \tilde Q}\,=\,\int\,\td^4\theta\,\big(\,Q^{\dagger}\,e^{-V}\, Q\,+\,
\tilde Q^{\dagger}\,e^{V}\,\tilde Q\,\big)\,+\,
\,\int \td^2\theta\,\tilde Q\,\Phi_k\,Q\,\,.
\eeq
In the effective low-energy theory obtained by integrating out the massive modes, the Yukawa coupling between $(Q, \tilde Q)$  and the $\Phi_k$ gives rise to a quartic term for the quark fields (see \cite{Casero:2006pt,Casero:2007pt,HoyosBadajoz:2008fw} for details). 

On the gravity side the addition of flavors can be performed by means of flavor branes, which add an open string sector to the unflavored closed string background. For the ${\cal N}=1$ geometry of section \ref{unflavoredMN} the flavor branes are D5-branes extended along a non-compact cycle of the Calabi-Yau threefold \cite{Nunez:2003cf}. If the branes reach the origin $r=0$ of the geometry, the corresponding flavor fields are massless. If, on the contrary, the D5's do  not reach $r=0$, the quark fields are massive (the minimal value of $r$ attained by the brane is related to the mass of the quark fields). 

In this paper we are interested in  getting a holographic dual of the ${\cal N}=1$ model with unquenched matter, in which the dynamics of fundamentals is encoded in the background. To achieve this goal we must go beyond the probe approximation and find a solution of the equations of motion derived from an action of the type:
\beq
S\,=\,S_{IIB}\,+\,S_{branes}\,\,,
\label{total-action}
\eeq
where $S_{IIB}$ is the action of ten-dimensional type IIB supergravity  and $S_{branes}$ denotes the sum of the Dirac-Born-Infeld (DBI) and Wess-Zumino (WZ) actions for the flavor branes. Generically, the branes act as sources for the different supergravity fields. In particular, the WZ term of  $S_{branes}$  is a source term for the RR fields which induces a violation of the Bianchi identity of the corresponding RR field strength. In our case, the WZ term of the action of a set of D5-branes is:
\beq
S_{WZ}=\,T_{5}\,\,\sum_{i=1}^{N_f}\,\,
\int_{{\cal M}_6^{(i)}}\,\,\imath^*\big( C_{(6)}\big)\,\,,
\label{WZ-localized}
\eeq
where $C_{(6)}$ is the RR six-form potential and $i^*\big( C_{(6)}\big)$ denotes its pullback to the D5-brane worldvolume. Let us rewrite (\ref{WZ-localized})  as a ten-dimensional integral, in terms of a charge distribution four-form $\Omega$:
\beq
S_{WZ}=\,T_5\,\,\int_{{\cal M}_{10}}\,\,
C_{(6)}\wedge \Omega\,\,.
\label{WZ-smeared}
\eeq
The term (\ref{WZ-smeared}) induces a violation of the Bianchi identity of $F_{(3)}$.
In order to determine it, let us write supergravity plus branes  action (\ref{total-action})  in terms of the RR seven-form $F_{(7)}$ and its six-form potential $C_{(6)}$. This action contains a contribution of the form:
\beq
-{1\over 2\kappa_{10}^2}\,\,{1\over 2}\,\,\int_{{\cal M}_{10}}\,e^{-\Phi}\,\,F_{(7)}\wedge *F_{(7)}\,+\,T_{5}\,\int_{{\cal M}_{10}}\,C_{(6)}\wedge \Omega\,\,.
\label{C6-action}
\eeq
The equation of motion of $C_{(6)}$ derived from (\ref{C6-action})  gives rise to the Maxwell equation for $F_{(7)}$ with $\Omega$ playing the role of a source, which is just:
\beq
\td\Big(e^{-\Phi}\,*F_{(7)}\,\Big)\,=\,-2\kappa_{10}^2\,T_{5}\,\,\Omega\,\,.
\label{Maxwell-F7}
\eeq
Taking into account that, in our units, $2\kappa_{10}^2\,T_{5}\,=\,(2\pi)^2$ and that
 $F_{(3)}=-e^{-\Phi}\,*\,F_{(7)}$, we get  that (\ref{Maxwell-F7}) is equivalent to
 the following violation of Bianchi identity of  $F_{(3)}$:
\beq
\td F_{(3)}\,=\,4\pi^2\,\Omega\,\,.
\label{Bianchi-F3}
\eeq
The four-form $\Omega$ is just the RR charge distribution due to the presence of the D5-branes. Clearly, $\Omega$ is non-zero on the location of the sources. In a localized setup, in which the $N_f$ branes are on top of each other,  $\Omega$ will contain Dirac $\delta$-functions and finding the corresponding backreacted geometry is technically a very complicated task. For this reason we will separate the $N_f$ branes and we will distribute them homogeneously along the internal manifold in such a way that, in the limit in which $N_f$ is large, they can be described by a  continuous 
charge distribution $\Omega$.

As we will detail below, the continuous set of flavor branes that we will use in our construction can be generated by acting with the isometries of the background on a representative fiducial embedding  and, therefore, all the branes of the continuous set are physically equivalent. Actually, we will  not choose an arbitrary distribution of branes. First of all, we will require that all branes are mutually supersymmetric (and thus they will not exert force on each other)  and that they preserve the same supercharges as the unflavored background.  Moreover, we will also require that the deformation induced on the metric is mild enough, in such a way that it reduces to squashing the unflavored metric (\ref{metric}) by means of squashing functions that depend only on the radial coordinate $r$. One can prove that the most general squashing of this type compatible with the ${\cal N}=1$ supersymmetry of the unflavored background is the one in which the size of one of the fibered directions  in the metric (\ref{metric}) is different from the other two. Accordingly, we will adopt  the following ansatz for the Einstein frame metric of the flavored theory:
\begin{equation} 
\label{eq:MetricAnsatz}
	\begin{aligned}
    \td s^2 &= e^{2 f(r)} \Big[ \td x_{1,3}^2 + e^{2k(r)} \td r^2 + e^{2 h(r)} ( \td\theta^2 + \sin^2 \theta\, \td\phi^2)\,\,+ \\[0.1in]
    &\quad+ \frac{e^{2 g(r)}}{4} \big((\tilde{\omega}^1 + a(r) \td\theta)^2 + (\tilde{\omega}^2 - a(r) \sin \theta\, \td\phi)^2\big) + \frac{e^{2k(r)}}{4} (\tilde{\omega}^3 + \cos \theta\, \td\phi)^2 
    \Big]\,\,.
  \end{aligned}
\end{equation}
Notice that the ansatz (\ref{eq:MetricAnsatz}) is exactly the same as the one considered in \cite{Casero:2006pt} for the case of massless flavors.

Let us next consider the deformation of the RR three-form $F_{(3)}$. Clearly,  due to the modified Bianchi identity (\ref{Bianchi-F3}) that must  be satisfied in the flavored case, $F_{(3)}$ cannot have the same form as in (\ref{RRthreeform}). Actually, we will slightly modify (\ref{RRthreeform}) and we will 
 adopt the following ansatz for the RR three-form $F_{(3)}$:
\beq
F_{(3)}\,=\,-{N_c\over 4}\,\,
(\tilde \omega^1-B^1)\wedge (\tilde \omega^2-B^2)\wedge(\tilde \omega^3-B^3)\,+\,
{N_c\over 4}\,\sum_a\,(F^a\,+\,f^a)\wedge (\tilde \omega^a\,-\,B^a)\,\,,\qquad
\label{F3ansatz-flavored}
\eeq
where $B^a$ is an  ${SU} (2)$ one-form gauge connection and $F^{a}$ is its two-form field strength, defined as in (\ref{fieldstrenght}), namely:
\beq
F^{a}\,=\,\td B^{a}\,+\,{1\over 2}\,\epsilon^{abc}\,\,B^b\wedge B^c\,\,.
\label{F=dB}
\eeq
In (\ref{F3ansatz-flavored}), the  $f^a$ are two-forms that parameterize the violation of the Bianchi identity and thus the flavor deformation of the RR three-form. Indeed, when $f^a=0$ the three-form $F_{(3)}$ is closed by construction, due to the relation (\ref{F=dB}) between $F^{a}$ and $B^{a}$.  We will take, as in \cite{Casero:2006pt}, the following ansatz for $B^{a}$:
\beq
B^1\,=\,-b(r)\,\td\theta\,\,,\qquad
B^2\,=\,b(r)\,\sin\theta\, \td\phi\,\,,\qquad
B^3\,=\,-\cos\theta\,\td\phi\,\,,
\eeq
where $b(r)$ is different from the fibering function $a(r)$ of the metric (they are equal in the background of \cite{Malda-Nunez:N=1}). By applying the definition (\ref{F=dB}), we get that the different components of the two-form field strength $F^a$ are:
\beq
F^1\,=\,-b'\,\td r\wedge \td\theta\,\,,\qquad
F^2\,=\,b'\,\sin\theta\,\td r\wedge \td\phi\,\,,\qquad
F^3\,=\,(1-b^2)\,\sin\theta\,\td\theta\wedge \td\phi\,\,.
\eeq
We will adopt for the flavor deformation two-forms $f^a$ an ansatz that parallels $F^a$, namely:
\beq
f^1\,=\,-L_1(r)\,\td r\wedge \td\theta\,\,,\qquad
f^2\,=\,L_1(r)\,\sin\theta\,\td r\wedge \td\phi\,\,,\qquad
f^3\,=\,L_2(r)\,\sin\theta\,\td\theta\wedge \td\phi\,\,,\qquad
\label{ansatz-fs}
\eeq
where $L_1$ and $L_2$ are two functions of the radial variable to be determined.  Actually, after a detailed study of the realization of supersymmetry for the metric ansatz (\ref{eq:MetricAnsatz}) one can show that (\ref{ansatz-fs}) gives rise to the most general form of $F_{(3)}$. By computing the exterior derivative of (\ref{F3ansatz-flavored}) and applying (\ref{Bianchi-F3}),  one gets the following  expression of the smearing form $\Omega$:
\begin{equation}
\begin{aligned}
\Omega\,&=\,-{N_c\over 16\pi^2}\,\,\sin\theta\,\td\theta\wedge \td\phi\wedge\,
\Big[\,L_2\,\tilde\omega^1\wedge\tilde\omega^2\,-\,L_2'\,\td r\wedge \tilde\omega^3\,\Big]
\,\,+\\[0.1in]
&\quad+{N_c\,L_1\over 16\pi^2}\,\td r\wedge \Bigg[\,\td\theta\wedge 
\tilde\omega^2\wedge\tilde\omega^3\,+\,\td\phi\wedge \Big(
\sin\theta \,\tilde\omega^1\wedge\tilde\omega^3+\cos\theta \,\td\theta\wedge 
\tilde\omega^2\Big)\Bigg]\,\,.
\end{aligned}
\label{Omega-L1-L2}
\eeq
One can now study the realization of ${\cal N}=1$ supersymmetry in type IIB supergravity for  a background with metric and RR three-form given by the ansatz written in (\ref{eq:MetricAnsatz}) and (\ref{F3ansatz-flavored}).  This analysis is performed in detail in appendix \ref{SUSY} and leads to a system of first-order BPS equations for the different functions of the ansatz. Combining these equations, a partial integration is possible. Let us summarize in this section the results of this study of the BPS equations. First of all, one can verify that the functions $L_1$ and $L_2$ parameterizing $f^a$ and $\Omega$ are not independent. Actually, 
from the BPS system one can prove that $L_1$ can be written in terms of the derivative of $L_2$ as follows:
\beq
L_1\,=\,-{L_2'\over 2\cosh (2r)}\,\,.
\label{L1-L2prime}
\eeq
Therefore, if we define the function $S(r)$ as:
\beq
N_f \,S(r)\equiv -N_c\,L_2(r)\,\,,
\label{S-definition}
\eeq
then, the two-forms $f^a$ of (\ref{ansatz-fs}) become:
\bear
&&f^1\,=\,-{N_f\over 2N_c}\,\,{S'(r)\over \cosh (2r)}\,\,\td r\wedge \td\theta\,\,,
\qquad
f^2\,=\,{N_f\over 2N_c}\,\,{S'(r)\over \cosh (2r)}\,\,
\sin\theta\,\td r\wedge \td\phi\,\,,\qquad\rc\rc
&&f^3\,=\,-{N_f\over N_c}\,\,S(r)\,\sin\theta\,\td\theta\wedge \td\phi\,\,,
\eear
whereas the smearing form $\Omega$ can be written in terms of $S$ as:
\beq
\begin{aligned}
\Omega\,&=\,{N_f\over 16\pi^2}\,\,\sin\theta\,\td\theta\wedge \td\phi\wedge\,
\Big[\,S\,\tilde\omega^1\wedge\tilde\omega^2\,-\,S'\,\td r\wedge \tilde\omega^3\,\Big]
\,\,+\\[0.1in]
&
\quad
+{N_f\over 32\pi^2}\, {S'\over \cosh (2r)}\,
\td r\wedge \Bigg[\,\td\theta\wedge 
\tilde\omega^2\wedge\tilde\omega^3\,+\,\td\phi\wedge \Big(
\sin\theta \,\tilde\omega^1\wedge\tilde\omega^3+\cos\theta \,\td\theta\wedge 
\tilde\omega^2\Big)\Bigg]\,\,.
\end{aligned}
\label{Omega-S(r)}
\eeq
Moreover,  the function $b$ parameterizing the one-forms $B^{a}$ can be written as:
\beq
b(r)\,=\,{2r+\eta(r)\over \sinh(2r)}\,\,,
\eeq
where $\eta(r)$ is defined as the following integral involving $S$:
\beq
\eta(r)\,=\,-{N_f\over 2N_c}\,\Bigg[\,
\tanh(2r)\,S(r)\,+\,2\int_{0}^{r}\,\td\rho\,
\tanh^2(2\rho)\,S(\rho)\,\Bigg]\,\,.
\label{eta-S}
\eeq
It follows from these results that the RR three-form $F_{(3)}$ in (\ref{F3ansatz-flavored}) is determined in terms of a unique function $S(r)$. Notice that the case of massless flavors studied in \cite{Casero:2006pt} is recovered by taking $S=1$ in our formulas. Indeed, in this case only the first term on the right-hand side of (\ref{Omega-S(r)}) is non-zero and the charge density distribution $\Omega$ is independent of the radial variable. Moreover, by computing the integral in (\ref{eta-S})
 one can show that our ansatz for $F_{(3)}$ is reduced to the one adopted in \cite{Casero:2006pt}. 
 
 In the case of massive flavors one expects the charge distribution to depend non-trivially on the radial coordinate and, actually, to vanish for values of $r$ smaller than a certain scale related to the mass of the quarks. In our approach this non-trivial structure is encoded in the  dependence of the function $S$ on the radial variable.  Notice also that $S$ should approach the massless value $S=1$ as $r\to\infty$ since the quarks are effectively massless in the deep UV. The way in which the profile function $S(r)$ interpolates between the IR and UV values depends on the particular set of D5-branes that constitutes our delocalized source and should be obtained by means of a microscopic calculation of the charge density (see below).

Another interesting observation is that, contrary to the backgrounds with massive flavors  studied in \cite{smearingmethod,Bigazzi:2008ie,Bigazzi:2008qq},  in this case passing from the massless to the massive case is not equivalent to substituting $N_f$ by $N_f\, S(r)$ in the massless ansatz. Indeed, it is immediate to check that making this substitution only the first line in (\ref{Omega-S(r)}) is generated, while the last two components of $\Omega$ (which are essential for the consistency of the approach) are missing. Notice that these last two terms in $\Omega$ are precisely those in (\ref{Omega-L1-L2}) which are proportional to the function $L_1$ which, according to (\ref{L1-L2prime}), always vanishes when $r\to\infty$. 
This means that, in the UV, the two-forms $f^a$ that implement the flavor deformation of $F_{(3)}$ are non-vanishing only along the third ${SU} (2)$ direction, while the other two components are excited when we move towards the IR. Interestingly, this structure is reminiscent of the way in which the singularity of the particular solution of the unflavored theory  reviewed at  the end of section \ref{unflavoredMN} is resolved in the full solution (\ref{metric})-(\ref{Fa-unflavored}), namely by turning on the function $a(r)$ and making the two-form $F^{a}$ the field strength of a non-abelian magnetic monopole.

Actually, as shown in appendix \ref{SUSY}, it turns out that one  can also integrate partially the BPS system for the functions of the metric in terms of $S(r)$. First of all, 
the function $f$ is related to the dilaton $\Phi$ as:
\beq
f\,=\,{\Phi\over 4}\,\,.
\label{f-dilaton}
\eeq
Moreover, the dilaton can be related to the other functions $h$, $g$ and $k$ as:
\beq
e^{-2\Phi}\,=\,2 e^{-2\Phi_0}\,\,
{e^{h+g+k}\over \sinh(2r)}\,\,,
\label{dilaton-integral}
\eeq
where $\Phi_0$ is a constant. In order to represent the remaining functions of the metric (\ref{eq:MetricAnsatz}) let us define, following \cite{HoyosBadajoz:2008fw}, 
the functions $P(r)$ and $Q(r)$  in terms of $a$ and $g$ as: 
\beq
Q\,=\,\big(a\cosh (2r)-1\big)\,e^{2g}\,\,,\qquad\qquad
P\,=\,a\,e^{2g}\,\sinh(2r)\,\,.
\label{P-Q-def}
\eeq
The inverse of this relation is:
\beq
e^{2g}\,=\,P\,\coth(2r)\,-\,Q\,\,,\qquad\qquad
a\,=\,{P\over P\cosh(2r)\,-\,Q\sinh(2r)}\,\,.
\eeq
It is demonstrated in appendix \ref{SUSY}  that, from the BPS system,  one can express $h$ and $k$ in terms $P$, $Q$ and $S$, namely:
\beq
e^{2h}\,=\,{1\over 4}\,\,
{P^2-Q^2\over P\coth(2r)-Q}\,\,,\qquad 
e^{2k}\,=\,{P'+N_f\,S(r)\over 2}\,\,.
\label{h-k-P-Q}
\eeq
It follows from eqs. (\ref{f-dilaton})-(\ref{h-k-P-Q}) that the dilaton and the functions of the metric are determined in terms of $P$, $Q$ and $S$. Actually,  the function $Q$ can be integrated in terms of the profile $S(r)$, namely:
\beq
Q\,=\,\coth(2r)\,\,\Bigg[\,\int_0^r\,\td\rho\,
{2N_c\,-\,N_f\,S(\rho)\over \coth^2(2\rho)}\,+\,q_0\,\Bigg]\,\,,
\label{Q-integral}
\eeq
where $q_0$ is a constant of integration. Moreover,  as in \cite{HoyosBadajoz:2008fw}, one can find a master equation:
\begin{equation}
	P'' + N_f\,S' + (\,P' + N_f\,S\,) \left( \frac{P' - Q' + 2 N_f\,S}{P + Q} + \frac{P' + Q' + 2 N_f\,S}{P - Q} - 4 \coth (2r) \right) = 0\,\,.
\label{master-eq}
\end{equation}
One can first notice that in the case $S=1$  eq. (\ref{master-eq})  reduces to the equation found in 
 \cite{HoyosBadajoz:2008fw}. Otherwise, knowing the function $S$ (from a microscopic  description of  the smearing), one can  get $Q$ from (\ref{Q-integral}) and
solve the second-order master equation (\ref{master-eq})
for $P$. As argued above, each solution of this equation will give a complete solution of the problem. Moreover, in appendix \ref{EoM} we have explicitly written the equations of motion derived from the type IIB supergravity plus sources action. One can check that any solution of the BPS system also solves the second-order equations of motion written in appendix \ref{EoM}.

Finding an analytic solution of this master equation is probably not possible, but we will be able to find numerical solutions, and their asymptotics. In order to achieve this goal we will have first to identify a family of supersymmetric embeddings whose backreaction on the background is compatible with our ansatz and then we must be able to compute the corresponding profile function $S(r)$. In the next section we will start to develop the machinery necessary to carry out this calculation.

\section{Holomorphic structure}
\label{holomorphic}
As stated at the end of section \ref{massive}, to find the profile function $S(r)$ we must analyze the families of supersymmetric embeddings of the flavor D5-branes. This problem was addressed in ref. \cite{Nunez:2003cf} by looking at the realization of kappa symmetry for probe D5-branes in the unflavored background described in
section \ref{unflavoredMN}. The analysis  of \cite{Nunez:2003cf} was performed in terms of the angular coordinates of the metric (\ref{metric}) and some particularly interesting embeddings were found. For our present purposes we clearly need a more systematic approach, which could allow us to study different families of embeddings and to determine whether or not  their backreaction is consistent with our ansatz (\ref{eq:MetricAnsatz})-(\ref{ansatz-fs}). As the internal manifold of our background is complex, it is quite natural to work in a system of holomorphic coordinates. The purpose of this section is to define these complex coordinates and to uncover the holomorphic structure of our background.

Let us begin by introducing a set of four complex variables $z_i$ ($i=1 \ldots 4$) parameterizing a deformed conifold, \ie\ satisfying the following quadratic equation:
\beq
z_1z_2-z_3z_4=1\,.
\label{conifold-def}
\eeq
We will also introduce a radial variable $r$, related to the $z_i$ as:
\beq
\sum_{i=1}^{4}\,|\,z_i\,|^2\,=\,2\cosh(2r)\,\,.
\label{conifold-radial}
\eeq
In order to find a useful parameterization of the $z_i$'s, let us arrange them as the following $2\times 2$ complex matrix $Z$:
\beq
Z=\left(
\begin{array}{cc}
z_3&z_2\\
-z_1&-z_4
\end{array}
\right)\,.
\eeq
Then, the defining equations (\ref{conifold-def}) and (\ref{conifold-radial}) can be written in matrix form as:
\beq
\det\left(Z\right)=1\,,\qquad\qquad\qquad\tr\left(Z\,Z^{\dagger}\right)=2\cosh(2r)\,.
\label{conifold-matrix}
\eeq
It is immediate to verify that the matrix
\beq
Z_0=\left(
\begin{array}{cc}
0&e^{r}\\
-e^{-r}&0
\end{array}
\right)
\label{Z-zero}
\eeq
is a particular solution of (\ref{conifold-matrix}). The general solution of this equation can  be found by realizing that the equations in (\ref{conifold-matrix}) exhibit the following ${SU} (2)_L\times {SU}(2)_R$ symmetry:
\beq
Z\to L\,Z\,R^{\dagger}\,,\qquad \quad L\in {SU}(2)_L\,,\quad R\in {SU}(2)_R\,.
\label{eqn:isometry.action}
\eeq
A generic point in the conifold  can be obtained by acting with isometries on 
the point  (\ref{Z-zero}). Actually, if we parameterize the ${SU} (2)$ matrices above in terms of Euler angles as:
\bear
&&L=\left(
\begin{array}{cc}
a&-\bar{b}\\
b&\bar{a}
\end{array}
\right)\quad\quad\quad\quad
\begin{array}{c}
a=\cos\frac{\th}{2}\,e^{i\frac{\j_1+\f}{2}}\,\,,\\\\
b=\sin\frac{\th}{2}\,e^{i\frac{\j_1-\f}{2}}\,\,,
\end{array}\rc\rc
&&R=\left(
\begin{array}{cc}
k&-\bar{l}\\
l&\bar{k}
\end{array}
\right)\quad\quad\quad\quad
\begin{array}{c}
k=\cos\frac{\tilde\th}{2}\,e^{i\frac{\j_2+\tilde\f}{2}}\,\,,\\\\
l=-\sin\frac{\tilde\th}{2}\,e^{i\frac{\j_2-\tilde\f}{2}}\,\,,
\end{array}
\label{eqn:LR}
\eear
then, the four complex variables $z_1, z_2, z_3, z_4$  that solve (\ref{conifold-matrix}) are given by:
\beq
\begin{aligned}
&z_1=-e^{-\frac{i}{2}(\f+\tf)}\left(e^{r+i\frac{\j}{2}}\sin\frac{\th}{2}\,\sin\frac{\tth}{2}-e^{-r-i\frac{\j}{2}}\cos\frac{\th}{2}\,\cos\frac{\tth}{2}\right)\,,\\[0.1in]
&z_2=e^{\frac{i}{2}(\f+\tf)}\left(e^{r+i\frac{\j}{2}}\cos\frac{\th}{2}\,\cos\frac{\tth}{2}-e^{-r-i\frac{\j}{2}}\sin\frac{\th}{2}\,\sin\frac{\tth}{2}\right)\,,\\[0.1in]
&z_3=e^{\frac{i}{2}(\f-\tf)}\left(e^{r+i\frac{\j}{2}}\cos\frac{\th}{2}\,\sin\frac{\tth}{2}+e^{-r-i\frac{\j}{2}}\sin\frac{\th}{2}\,\cos\frac{\tth}{2}\right)\,,\\[0.1in]
&z_4=-e^{-\frac{i}{2}(\f-\tf)}\left(e^{r+i\frac{\j}{2}}\sin\frac{\th}{2}\,\cos\frac{\tth}{2}+e^{-r-i\frac{\j}{2}}\cos\frac{\th}{2}\,\sin\frac{\tth}{2}\right)\,,
\end{aligned}
\label{eqn:MNz}
\eeq
where $\psi=\psi_1+\psi_2$.  We will show below that these holomorphic coordinates are very convenient to analyze the supersymmetric embeddings in our flavored backgrounds. It is also useful to introduce a new set of  complex variables $w_i$, related to the $z_i$ by means of the following linear combinations:
\beq
w_1=\frac{z_1+z_2}{2}\,,\qquad w_2=\frac{z_1-z_2}{2i}\,,\qquad w_3=\frac{z_3-z_4}{2}\,,\qquad w_4=\frac{z_3+z_4}{2i}\,.
\label{w-def}
\eeq
These variables satisfy:
\beq
(w_1)^2\,+\,(w_2)^2\,+\,(w_3)^2\,+\,(w_4)^2\,=\,1\,\,,
\eeq
and there is an obvious ${SO} (4)$  invariance that is obtained by rotating the $w_i$'s. The so-called ${SO} (4)$-invariant (1,1)-forms are defined as \cite{Herzog:2001xk}:
\beq
\h_1=\delta^{ij}\td w_i\wedge \td\wb_j\,,\qquad\h_2=\left(\delta^{ij}w_i\td\wb_j\right)\wedge\left(\delta^{kl}\wb_k\td w_l\right)\,,\qquad\h_3=\eps^{ijkl}w_i\wb_j\td \bar{w}_k\wedge \td w_l\,.
\label{eqn:etas}
\eeq
In terms of the radial and angular coordinates these forms are given by:
\beq
\begin{aligned}
\h_1&=-i\left(\cosh(2r)\,\td r\wedge\left(\tw^3+\cos\th\,\td\f\right)-\frac{1}{2}\sinh(2r)\left(\sin\th\,\td\th\wedge\td\f+\sin\tth\,\td\tth\wedge\td\tf\right)\right)\,,\\[0.1in]
\h_2&=i\sinh^2(2r)\,\td r\wedge\left(\tw^3+\cos\th\,\td\f\right)\,,\\[0.1in]
\h_3&=-i\left(\frac{1}{4}\sinh(4r)\left(\sin\th\,\td\th\wedge\td\f-\sin\tth\,\td\tth\wedge\td\tf\right)\!-\frac{1}{2}\sinh(2r)\left(\td\th\wedge\tw^2+\sin\th\,\td\f\wedge\tw^1\right)\!\right).
\end{aligned}
\label{etas-angles}
\eeq
The fundamental  two-form $J$ of the ${SU} (3)$-structure can be written in terms of the $\eta_i$ forms, which  will be very useful in what follows. In order to find the corresponding expression, let us notice that $J$ has been written in (\ref{J-Omega-ansatz}) in terms of the angle $\alpha$ that rotates the projections of the Killing spinors.  By using the value of $\alpha$ written in (\ref{alpha}) and the relations (\ref{cos-sin-alpha}), one easily proves that $J$ can be written  as:
\beq
\begin{aligned}
e^{-{\Phi\over 2}}\,\,J=&\frac{e^{2k}}{2}\td r\wedge\left(\tw^3+\cos\th\,\td\f\right)+\frac{e^{2g}}{4}\frac{a\,\cosh(2r)-1}{\sinh(2r)}\left(\td\th\wedge\tw^2+\sin\th\,\td\f\wedge\tw^1\right)-\\
&-\frac{e^{2g}}{4}\left(\frac{a\,\cosh(4r)-\cosh(2r)}{\sinh(2r)}\sin\th\,\td\th\wedge\td\f+\frac{\cosh(2r)-a}{\sinh(2r)}\sin\tth\,\td\tth\wedge\td\tf\right)\,.
\end{aligned}
\label{J}
\eeq
In terms of the $\eta_i$'s, one can rewrite $J$ as:
\beq
e^{-{\Phi\over 2}}\,\,J=\frac{1}{2i}\left[\frac{e^{2k}}{\sinh^2(2r)}\,\h_2-a\,e^{2g}\left(\h_1+\frac{\cosh(2r)}{\sinh^2(2r)}\,\h_2\right)+e^{2g}\frac{a\,\cosh(2r)-1}{\sinh^2(2r)}\,\h_3\right]\,.
\label{J-etas}
\eeq
Let us now check that the complex variables $z_i$ defined in \eqref{eqn:MNz} are good holomorphic coordinates for the internal manifold. Indeed, 
since our six-dimensional internal manifold is a complex manifold, we can write its metric in terms of the $(1,1)$-form $J$. Actually, if one writes $J$ as:
\beq \label{eq:Jcomplex}
	J = \frac{i}{2} h_{\a \bar{\b}}\, \td z^{\a} \wedge \td \bar{z}^{\bar{\b}}\,\,,
\eeq
which is allowed thanks to the fact that $J$ is a $(1,1)$-form, then one can prove that
the metric of the internal space can be written as:
\beq \label{eq:metricComplex}
	ds^2_{6} = \frac{1}{2} h_{\a \bar{\b}} \big( \td z^{\a} \otimes \td \bar{z}^{\bar{\b}} +\td \bar{z}^{\bar{\b}} \otimes \td z^{\a} \big)\,\,,
\eeq
where we have split the ten-dimensional metric (\ref{eq:MetricAnsatz}) as 
$\td s^2\,=\,e^{{\Phi/2}}\,\td x^2_{1,3}+\td s^2_6$.  The $h_{\a \bar{\b}}$ coefficients appearing in (\ref{eq:Jcomplex}) and (\ref{eq:metricComplex}) can be read from (\ref{J-etas}) by using the relation between the $\eta_i$'s and the $z_i$ coordinates (see (\ref{eqn:etas}) and (\ref{w-def})). Moreover, by using again (\ref{alpha})  and 
(\ref{cos-sin-alpha}), one can write the three-form  $\Omega_{hol}$ of (\ref{J-Omega-ansatz})
 as:
\beq
\Omega_{hol}\,=\, -\frac{1}{\sinh(2r)} e^{2\Phi+g+h+k} \frac{1}{z_3} \td z_1 \wedge \td z_2 \wedge \td z_3\,\,.
\eeq
Furthermore, taking into account   (\ref{dilaton-integral}), we can write $\Omega_{hol}$ as:
\beq
\Omega_{hol}\,=\, -{e^{2\Phi_0}\over 2}\,\,
{\td z_1 \wedge \td z_2 \wedge \td z_3\over z_3}\,\,,
\eeq
which shows that  $\Omega_{hol}$ is, indeed, a holomorphic $(3,0)$-form for the complex structure corresponding to the coordinates (\ref{eqn:MNz}). 

The RR six-form potential $C_{(6)}$, defined as
$F_{(7)}\,=\,-e^{\Phi}\,*\,F_{(3)}\,=\,\td C_{(6)}$, can also be written in terms of the $\eta_i$ one-forms. In fact, it follows from the first of the BPS conditions in (\ref{eq:gencalcondition}) that 
 $C_{(6)}$ can be written in terms of the form $J$ as:
\beq
C_{(6)}\,=\,e^{{3\Phi\over 2}}\,\,
\td^4x\wedge J\,\,,
\label{C6}
\eeq
where $\td^4x\,=\,\td x^0\wedge \td x^1\wedge \td x^2\wedge \td x^3$.  Obviously, since $J$ can be written in terms of the $z^i$ variables, the six-form $C_{(6)}$ can also be written as a $(1,1)$ form in the internal space. Notice that $C_{(6)}$  is related to the calibration form of a D5-brane, whose pullback to the worldvolume determines if the embedding is supersymmetric or not. Having $C_{(6)}$ written in complex coordinates is very convenient from the technical point of view since it will allow us to analyze the different supersymmetric embeddings by employing the full machinery of the complex variables.

Another relevant quantity that should be invariant under the ${SO} (4)$ isometry is the smearing form $\Omega$ in (\ref{Omega-S(r)}), since it is giving us the charge distribution of the system. It  is a (2,2)-form which can be cast in terms of (1,1)-forms as follows:
\beq
16\p^2\Omega=-\frac{2N_f\,S }{\sinh^2 2r}\,\,\eta_1\wedge\left(\h_1+\frac{2\cosh 2r}{\sinh^2 2r}\h_2\right)+\frac{N_f\,S'\,}{\sinh^3 2r}\,\eta_2
\wedge\left(\h_1-\frac{1}{\cosh 2r}\h_3\right)\,.
\label{eqn:Omega}
\eeq

\subsection{Supersymmetric embeddings}

It is  now straightforward to show that any embedding defined with holomorphic functions of the complex coordinates is supersymmetric. Let us study the case of an embedding extended in the Minkowski directions, and defined in the internal space in the following way:
\beq
	z_2 = F(z_1) \,, \qquad z_3 = G(z_1) \,, \qquad \bar{z}_2 = \bar{F}(\bar{z}_1) \,, \qquad \bar{z}_3 = \bar{G}(\bar{z}_1)\,\,,
\eeq
where, for definiteness, we have chosen $z_1$ and $\bar z_1$ as worldvolume coordinates in the internal space. Recall that $z_4 = z_3^{-1} (z_1\,z_2-1)$. The calibration form $\cK$ for a D5-brane in Einstein frame is given by:
\beq
	\cK = e^{\Phi} \td^4x \wedge J\,=\,e^{-{\Phi\over 2}}\,C_{(6)}\,\,.
\eeq
By using \eqref{eq:Jcomplex} one can easily get the pullback of this calibration form on the worldvolume of the embedding, namely:
\beq
	\pbi{\cK} = i\,e^{\Phi} K\, \td^4x \wedge \td z_1 \wedge \td \bar{z}_1\,\,,
\eeq
where we have  defined the function $K$ as:
\beq
K =\frac{1}{2} \big(h_{1 \bar{1}} +\bar{F}' h_{1 \bar{2}}+\bar{G}' h_{1 \bar{3}}+F' h_{2 \bar{1}}+F' \bar{F}'h_{2 \bar{2}}+ F' \bar{G}'h_{2 \bar{3}}+G' h_{3 \bar{1}}+G' \bar{F}' h_{3 \bar{2}} + G' \bar{G}' h_{3 \bar{3}}\big)\,\,.
\eeq
Now, we look at the induced metric $d\hat s^2_6$
on the worldvolume of the embedding. We get from \eqref{eq:metricComplex}:
\beq
\td\hat s^2_6\,=\,e^{{\Phi/2}}\,\td x^2_{1,3}\,+\,2K\,\td z_1\,\td \bar z_1\,\,.
\eeq
Therefore,  $\det \hat{g} = e^{2\Phi} K^2$, and one has
\beq
			\sqrt{- \det \hat{g}}\,\, \td^4 x\wedge \td z_1 \wedge \td \bar{z}_1 = i\,e^{\Phi} K \,\, \td^4 x\wedge \td z_1 \wedge \td \bar{z}_1 = \pbi{\cK}\,\,.
\eeq
This means that the embedding is supersymmetric, proving explicitly that all holomorphic embeddings are supersymmetric.

\section{Charge distributions}
\label{charge-distributions}
The supersymmetric D5-brane embeddings we are looking at are characterized by two algebraic equations of the type:
\beq
F_1(z_i)\,=\,0\,\,,\qquad\qquad
F_2(z_i)\,=\,0\,\,,
\label{general-embedding}
\eeq
which define a non-compact two-cycle ${\cal C}_2$ in the internal six-dimensional manifold.  As argued above, the preservation of supersymmetry is ensured if the two functions in (\ref{general-embedding}) are holomorphic. However, in the brane setup we are considering we will not deal with a particular embedding of the flavor D5-branes but, instead, with a family of equivalent embeddings. This family can be generated from a particular representative of the form (\ref{general-embedding}) by acting with the ${SU} (2)_L\times {SU} (2)_R$ isometries of the conifold. Let us recall how these symmetries act on the holomorphic coordinates.  Under  ${SU} (2)_L$ the  holomorphic coordinates transform as $z_i\to\tilde z_i$, where
\beq
\left(
\begin{array}{cc}
\tilde{z}_3&\tilde{z}_2\\
-\tilde{z}_1&-\tilde{z}_4
\end{array}
\right)
=
\left(
\begin{array}{cc}
\alpha z_3+\bar{\beta}z_1&\alpha z_2+\bar{\beta}z_4\\
-\bar{\alpha}z_1 +\beta z_3&-\bar{\alpha}z_4+\beta z_2
\end{array}
\right)\,,
\label{left-SU2}
\eeq
with $|\alpha|^2+|\beta|^2=1$. Similarly, the  ${SU} (2)_R$ transformation is:
\beq
\left(
\begin{array}{cc}
\tilde{z}_3&\tilde{z}_2\\
-\tilde{z}_1&-\tilde{z}_4
\end{array}
\right)
=
\left(
\begin{array}{cc}
\bar{\gamma} z_3-\delta z_2&\bar{\delta} z_3+\gamma z_2\\
-\bar{\gamma} z_1+\delta z_4&-\bar{\delta} z_1-\gamma z_4
\end{array}
\right)\,,
\label{right-SU2}
\eeq
where  the complex constants $\gamma$ and $\delta$ satisfy the condition $|\gamma|^2+|\delta|^2=1$. We want now to determine the charge distribution four-form $\Omega$ (parameterized by the profile function $S(r)$) for a given family of embeddings. Actually, we will employ a procedure which does not require performing the detailed analysis of the whole family and that allows to extract the function $S(r)$ by studying one single particular embedding belonging to the family \cite{MetodoAngel}. This method is based on the comparison between the action for the whole set of $N_f$ flavor branes and the one corresponding to a representative embedding. We can choose to compare either the DBI or the WZ part of the actions, since supersymmetry guarantees that they are the same. The WZ term of the action of the full set of D5-branes is given by the following ten-dimensional integral:
\beq
S_{WZ}^{smeared}\,=\,T_5\,\int_{{\cal M}_{10}}\,\Omega\wedge
C_{(6)}\,\,,
\eeq
whereas the action of one of the embeddings is just:
\beq
S_{WZ}^{single}\,=\,T_5\,\int_{{\cal M}_6}\,\pbi{C_{(6)}}\,\,,
\label{WZ-single}
\eeq
with ${\cal M}_6$ being the worldvolume of the representative embedding chosen and
$\pbi{C_{(6)}}$ denotes the pullback of  $C_{(6)}$ to ${\cal M}_6$. Since all the embeddings of the family are  related by isometries, they are equivalent and their actions should be the same. Thus, we should have:
\beq
S_{WZ}^{smeared}\,=\,N_f\,S_{WZ}^{single}\,\,.
\label{match-macro-micro}
\eeq
The left-hand side of (\ref{match-macro-micro}) can be obtained by plugging the expressions of $\Omega$ and $C_{(6)}$ written in  (\ref{Omega-S(r)}) and (\ref{C6}) respectively. After integrating over the angular coordinates, one gets a remarkably simple expression, namely:
\beq
S_{WZ}^{smeared}\,=\,
2\p\,N_f\,T_5\,\int\td^4x\,\td r\,e^{2\Phi}\,
\left(\,e^{2k}\,S\,+\,{1\over 2}\,e^{2g}\,\tanh(2r)\,S'\,\right)\,.
\label{smearedWZ-S}
\eeq
The non-compact two-cycle ${\cal C}_2$ that the D5-branes wrap can be parameterized by the radial coordinate $r$ and an angular variable. After integrating over  the latter, the WZ action (\ref{WZ-single}) can be represented as:
\beq
S_{WZ}^{single}\,=\,
2\pi \,T_5\,\int\td^4x\,\td r\,e^{2\Phi}\,
{\cal S}(r)\,\,,
\label{singleWZ-calS}
\eeq
where the function ${\cal S}(r)$ is related to the integral of the pullback of $J$ along the two-cycle by means of the expression:
\beq
\int_{{\cal C}_2}\,\pbi{J}\,=\,2\pi\,\int dr\,\,e^{{\Phi\over 2}}\,\,{\cal S}(r)\,\,.
\label{J-calC2}
\eeq
Notice that the dependence on the dilaton of the right-hand side of eqs. (\ref{singleWZ-calS}) and (\ref{J-calC2}) is consistent with the one displayed in (\ref{J-etas}) and (\ref{C6}). By plugging (\ref{smearedWZ-S}) and (\ref{singleWZ-calS}) into (\ref{match-macro-micro}) we arrive at the following relation between  the profile $S(r)$ and the function ${\cal S}(r)$:
\beq
e^{2k}\,S\,+\,{1\over 2}\,e^{2g}\,\tanh(2r)\,S'\,=\,{\cal S}(r)\,\,.
\label{profile-equation}
\eeq
The function ${\cal S}$ appearing on the right-hand side of (\ref{profile-equation}) depends both on the embedding and on the different functions of our ansatz. In the case in which ${\cal S}$ depends only on the functions $k$ and $g$ and this dependence is  the same as on the left-hand side of (\ref{profile-equation}), it is possible to obtain the profile function $S$ from (\ref{profile-equation}). However, this is a highly non-trivial condition which most families of embeddings do not satisfy. To illustrate this fact let us consider the families of massive embeddings obtained by acting with the ${SU} (2)_L\times {SU} (2)_R$ isometries on the two non-compact two-cycles found in \cite{Nunez:2003cf}. The first of this two-cycles is the so-called unit-winding embedding (see section 6.1 in \cite{Nunez:2003cf}), which has the following representation in terms of the real coordinates of the metric (\ref{eq:MetricAnsatz}):
\beq
\sinh r\,=\,{\sinh r_q\over \sin\theta}\,\,,\qquad
\tilde\theta\,=\,\theta\,\,,\qquad
\tilde\phi\,=\,\phi\,\,,\qquad
\psi\,=\,\pi\,\,,
\label{unit-winding-real}
\eeq
where $r_q$ is a constant. In terms of the complex  coordinates (\ref{eqn:MNz}), one can easily show that (\ref{unit-winding-real}) is a particular solution of the following two holomorphic equations:
\beq
z_1\,z_2\,=\,\cosh^2{r_q}\,\,,\qquad\qquad
z_3\,+\,z_4\,=\,0\,\,.
\label{unit-winding-complex}
\eeq
By using (\ref{unit-winding-real}) and (\ref{J}) it is straightforward to compute the pullback of $J$ and to obtain ${\cal S}(r)$. One gets:
\beq
{\cal S}(r)\,=\,\sqrt{
1\,-\,{\sinh^2 r_q\over \sinh^2r}}\,\,\Bigg(\,e^{2k}\,+\,
e^{2g}\,\,a\,{\sinh^2r\over \sinh^2r\,-\,\sinh^2 r_q}\,\,\cosh^2r\,\Bigg)\,\,.
\label{calS-unit-winding}
\eeq
Notice that the right-hand side of (\ref{calS-unit-winding}) contains the function $a(r)$, which is not present on the left-hand side of (\ref{profile-equation}). Therefore, the determination of the profile $S$ is not possible in this case. Similarly, one can consider the so-called zero-winding embeddings of section of 6.2 of  \cite{Nunez:2003cf}. In this case the cycle is characterized by the equations:
\beq
\sinh(2r)\,=\,{\sinh(2r_q)\over \sin \theta}\,\,,\qquad
\sin\tilde\theta\,=\,-{\cos\theta\over \cosh (2r_q)}\,\,,\qquad
\tilde\phi\,=\,\tilde\phi_0\,\,,\qquad
\psi\,=\,\pi\,\,,
\label{zero-winding-real}
\eeq
which solve the following system of two complex equations:
\beq
z_1\,z_2\,=\,{1\over 2}\,\,,\qquad\qquad
z_1\,-\,e^{-2r_q}\,e^{-i\tilde\phi_0}\,z_4\,=\,0
\,\,.
\label{zero-winding-complex}
\eeq
In (\ref{zero-winding-real}) and (\ref{zero-winding-complex})  $\tilde\phi_0$ is a constant. By computing  ${\cal S}(r)$ in this case one gets:
\beq
{\cal S}(r)\,=\,\sqrt{
1\,-\,{\sinh^2 (2r_q)\over \sinh^2(2r)}}\,\,\Bigg(\,e^{2k}\,+\,
e^{2g}\,\,{\sinh^2(2r)\over \sinh^2(2r)\,-\,\sinh^2 (2r_q)}\,\,
\Big(\,2a\,\cosh(2r)\,-\,1\,\Big)\,\,\Bigg)\,\,,
\label{calS-zero-winding}
\eeq
which, as in (\ref{calS-unit-winding}), contains the function $a(r)$ and, as a consequence, it is not of the form displayed in (\ref{profile-equation}).

Our interpretation of the fact that ${\cal S}$ is not of the form (\ref{profile-equation}) for the embeddings (\ref{unit-winding-real}) and (\ref{zero-winding-real}) is that their backreaction is not compatible with our ansatz. Notice that our $F_{(3)}$ in (\ref{F3ansatz-flavored}), as well as the charge-density four-form $\Omega$ in (\ref{Omega-S(r)}), are dictated by the ${SU}(3)$-structure of the ${\cal N}=1$ supersymmetry and they are highly asymmetric with respect to the exchange $(\theta,\phi)\leftrightarrow (\tilde\theta,\tilde\phi)$.  This interpretation is supported by an independent microscopic calculation of $\Omega$, which we present in appendix \ref{micro}. Indeed,  we show in this appendix that a simple embedding whose 
${\cal S}$ is not of the form (\ref{profile-equation}) gives rise to a charge-density $\Omega$ which does not fit into our ansatz. Thus, in order to proceed further with our formalism we have to find a concrete example of compatible embeddings and we have  to determine the corresponding charge profile.

Fortunately, we have been able to find a simple  family of embeddings  for which ${\cal S}(r)$  depends on the functions $k$ and $g$ in the same way as the left-hand side of (\ref{profile-equation})  and, as a consequence, one can directly read the  profile function $S(r)$  for this configuration. The profile $S(r)$ obtained in this way can be used to get $Q(r)$ from (\ref{Q-integral}), and as an input for solving the master equation (\ref{master-eq}). In the next section we will present these embeddings and we will determine the corresponding profile.

\section{A simple class of embeddings}
\label{simple-embeddings}
As shown above, the massive embeddings found in \cite{Nunez:2003cf} do not seem to produce a backreaction compatible with our ansatz and with its underlying $SU(3)$-structure. In principle, we should consider the logical possibility that such a compatible set of embeddings does not exist.  In this section we will discard this possibility by finding a family of embeddings for which (\ref{profile-equation}) can be solved and a simple expression for the profile function $S$ can be found. We confirm this fact in appendix 
\ref{micro} by means of an explicit microscopic calculation in the UV region of large $r$ of the charge density four-form $\Omega$. By considering the full distribution of flavor branes we will indeed show that, for the embeddings discussed in this section, the density $\Omega$ is of the form displayed in (\ref{Omega-S(r)}) and we will find an expression of $S(r)$ which is just the limit of the one found by solving (\ref{profile-equation}) for large $r$. 

In terms of the holomorphic coordinates (\ref{eqn:MNz}) the simplest embeddings one can think of are those characterized by two linear relations of the $z_i$'s. Many of these embeddings are related by the action of the $SU(2)_L\times SU(2)_R$ symmetry and they belong to the same set. Instead of considering the full set we will only deal with a particular representative. By using the machinery developed above it is rather easy to consider systematically the different linear embeddings, to compute the pullback of the fundamental two-form $J$ and to verify if the function ${\cal S}(r)$ depends on the functions of the ansatz as the left-hand side of (\ref{profile-equation}). Most of these linear embeddings do not give rise to a compatible charge density.  Let us now work out an example for which everything works fine. 
The representative embedding we want to focus on can be written in terms of the holomorphic coordinates \eqref{eqn:MNz} as the following two linear equations:
\beq
z_3=A\,z_1\,\,,\qquad\qquad\qquad
z_4=B\,z_2\,\,,
\label{eqn:the.emb}
\eeq
where $A$ and $B$ are two complex constants. 
We can parameterize the two-surface defined by \eqref{eqn:the.emb} in terms of, for example, $z_1$:
\beq
z_2=\frac{1}{1-A\,B}\frac{1}{z_1}\,,\qquad z_3=A\,z_1\,\qquad z_4=\frac{B}{1-A\,B}\frac{1}{z_1}\,.
\label{complex-2-surface}
\eeq
This allows us to get the relation between $r$ and $z_1$:
\beq
2\cosh(2r)=(1+|A|^2)|z_1|^2+\frac{1+|B|^2}{|1-A\,B|^2}\frac{1}{|z_1|^2}\,,
\label{cosh(r)-zs}
\eeq
where we have used the relation between $r$ and the holomorphic coordinates written in (\ref{conifold-radial}). From (\ref{cosh(r)-zs}) we can compute the minimum distance $r_q$ that this embedding reaches, namely:

\beq
 \cosh(2r_q)=\frac{\sqrt{1+|A|^2}\sqrt{1+|B|^2}}{|1-A\,B|}\,.
\eeq
Notice that this minimum distance depends on the modulus of the constants $A$ and $B$ , as well as on the phase of $A\, B$. In order to compute the function ${\cal S}(r)$ for these embeddings, let us compute the pullback of $J$. It is quite useful to work with the complex coordinates $z_i$ and to obtain first the pullback of the $\eta_i$ forms. Actually, the pullbacks of the ${SO} (4)$ invariant (1,1)-forms can be cast nicely as:
\begin{align}
\pbi{\h_1}&=\frac{1}{2}\left((1+|A|^2)|z_1|^2+\frac{1+|B|^2}{|1-A\,B|^2}\frac{1}{|z_1|^2}\right)\frac{\td z_1\wedge\td \bar{z}_1}{|z_1|^2}=\cosh (2r)\,\frac{\td z_1\wedge\td \bar{z}_1}{|z_1|^2}\,,\rc\rc
\pbi{\h_2}&=-\frac{1}{4}\frac{\left((1+|B|^2)-(1+|A|^2)|1-A\,B|^2|z_1|^2\right)}{|1-A\,B|^4|z_1|^4}\frac{\td z_1\wedge\td \bar{z}_1}{|z_1|^2}=\rc\rc
&=\left(\cosh^2(2r_{q})-\cosh^2(2r)\right)\frac{\td z_1\wedge\td \bar{z}_1}{|z_1|^2}\,,\rc\rc
\pbi{\h_3}&=\frac{|A+\bar{B}|^2}{|1-A\, B|^2}\frac{\td z_1\wedge\td \bar{z}_1}{|z_1|^2}=\,\cosh^2(2r_{q})\,\frac{\td z_1\wedge\td \bar{z}_1}{|z_1|^2}\,.
\end{align}
From these pullbacks we can readily compute the pullback of $J$, namely:
\beq
e^{-{\Phi\over 2}}\,\,\pbi{J}=\left(e^{2k}\,\frac{\cosh(4r)+1-2\cosh^2 (2r_{q})}{\sinh^2(2r)}+e^{2g}\,\frac{2\cosh^2(2r_{q})-2}{\sinh^2(2r)}\right)\frac{i}{4}\frac{\td z_1\wedge\td \bar{z}_1}{|z_1|^2}\,.
\eeq
Magically, the pullback of $J$ does not contain the function  $a$, and it is ready for comparison with the smeared action. In order to obtain the actual value of ${\cal S}(r)$
we need to express $\td z_1\wedge\td \bar{z}_1=\td r\wedge\td (\textrm{angular})$. With this purpose in mind we will parameterize $z_1$ as:
\beq
z_1=u\,e^{i\,\th}\,\,.
\eeq
Then, one has:
\beq
{\td z_1\wedge\td \bar{z}_1\over |z_1|^2}= -2i\,\frac{\td u}{u}\wedge\td\th\,\,,
\eeq
and since  from (\ref{cosh(r)-zs}) it follows that:
\beq
\frac{\td u}{u}=\pm\frac{\sinh(2r)}{\sqrt{\cosh^2(2r)-\cosh^2(2r_{q})}}\,,
\eeq
we can write:
\beq
	\int_{{\cal C}_2}\imath^*(J)=2\p\!\int\!\td r\,e^{{\Phi\over 2}}\left(\!e^{2k}\frac{\sqrt{\cosh(4r)-\cosh (4r_{q})}}{\sqrt{2}\sinh(2r)} +e^{2g}\frac{\sqrt{2}\tanh(2r)\cosh(2r)\,\sinh^2(2r_{q})}{\sinh^2(2r)\,\sqrt{\cosh(4r)-\cosh (4r_{q})}}\right).
\eeq
Thus, the function ${\cal S}(r)$ in this case is given by:
\beq
{\cal S}\,=\,
e^{2k}\,\frac{\sqrt{\cosh(4r)-\cosh (4r_{q})}}{\sqrt{2}\sinh(2r)} +e^{2g}\,\tanh(2r)\,\frac{\sqrt{2}\cosh(2r)\,\sinh^2(2r_{q})}{\sinh^2(2r)\,\sqrt{\cosh(4r)-\cosh (4r_{q})}}\,\,.
\eeq
Plugging this result in the right-hand side of (\ref{profile-equation}) and taking into account that the coefficients of $e^{2k}$ and $e^{2g}\,\tanh2r$ are related by a derivative:
\beq
\frac{\td}{\td r}\,\Bigg[\,
\frac{\sqrt{\cosh(4r)-\cosh (4r_{q})}}{\sqrt{2}\sinh(2r)}\,\Bigg]
=2\,
\,\frac{\sqrt{2}\cosh(2r)\,\sinh^2(2r_{q})}{\sinh^2(2r)\,\sqrt{\cosh(4r)-\cosh (4r_{q})}}\,\,,
\eeq
one immediately gets that the profile function $S(r)$  for this family is given by:
\beq 
S(r)=\,\frac{\sqrt{\cosh(4r)-\cosh (4r_{q})}}{\sqrt{2}\sinh(2r)}\,\,
\Theta(r-r_q)= \sqrt{1-\frac{\sinh^2(2r_q)}{\sinh^2(2r)}}\,\Theta(r-r_q)\,\,,
\label{S-wall}
\eeq
where we have taken into account that $r\ge r_q$ on the cycle.  Notice that $S(r)\to 1$ as $r\to\infty$ and the massive solution becomes the solution of \cite{Casero:2006pt} in the far UV, as it should (see figure \ref{fig:Width}). Notice also that $S(r)=1$ in (\ref{S-wall})  for the massless case $r_q=0$ and, therefore, we recover the results of \cite{Casero:2006pt}  in this case. As mentioned above, in appendix \ref{micro} we have checked the form of $\Omega$ and the expression of $S$ for these embeddings by means of a microscopic calculation in the UV, where the unflavored model reduces to the ``abelian" model discussed at the end of section \ref{unflavoredMN}.

Near $r=r_q$ the profile $S(r)$ in (\ref{S-wall})  vanishes as:
\beq
S(r)\sim 2\sqrt{\cosh (2r_q)}\,\,
\sqrt{r-r_q}\,\,,
\eeq
which means that $S(r)$ is continuous at $r=r_q$. However $S'(r)$ diverges as
$1/\sqrt{r-r_q}$ when $r\to r_q$.  Since $S'(r)$  enters into the energy momentum tensor of the branes (see  eq. (\ref{emtensor-components}) in appendix \ref{EoM}), it follows from Einstein's equations that this divergence will induce  the divergence of the Ricci tensor at $r=r_q$. This divergence is due to the hard-wall effect that we are introducing in our configuration when the flavor branes are added and it should be thought as the gravitational analogue of the threshold effects of field theory. In the next section we will propose a way to resolve this singularity in our string duals.

\section{Removing the threshold singularity}
\label{Smoothing}

Let us consider the class of embeddings studied in section \ref{simple-embeddings}. We will show how one can engineer a brane setup such that the unwanted singularity of $S'(r)$ at $r=r_q$ disappears. The idea is to consider branes whose tips reach different radial positions and perform an average over the value $r_q$ of the radial coordinate of the tip of the flavor branes. Actually, this is the way in which the threshold singularity is removed in the Klebanov-Strassler model with massive flavors studied in \cite{Bigazzi:2008qq}. Indeed, in appendix \ref{KS} we reconsider this last model with the tools developed here and we explicitly show how  averaging over a certain phase is equivalent to  a particular superposition of flavor branes ending on different radial positions.  Moreover, in appendix \ref{KS} we will also compute the function $S(r)$ for a set of branes ending on a fixed $r_q$ (\ie\ the analogue of (\ref{S-wall}) for the Klebanov-Strassler model) and we will also explore some other possibilities to perform the superposition of branes with different positions of their tips.

Inspired by the resolution of the threshold singularity in the Klebanov-Strassler model, we will consider a flavor brane distribution containing branes with different $r_q$'s. 
Furthermore, we will allow $r_q$ to vary in a certain  finite interval and  we will weight the different values of $r_q$ with a non-negative measure function $\rho(r_q)$, which should be conveniently normalized. In this way the hard wall at $r=r_q$ will be substituted by a shell of non-vanishing width. If the resulting profile function $S$ and its first radial derivative are continuous the geometry will be free of threshold singularities. 
As we will see explicitly below, if the measure function is smooth enough the resulting profile will fulfill  the conditions to have a regular supergravity solution.

For convenience let us redefine the radial coordinate as:
\beq
x = \cosh(4r)\,\,,\qquad\qquad x\ge 1\,\,.
\eeq
We will also denote $x_q=\cosh(4r_q)$. We will  consider distributions of branes  having $x_q$'s  in the interval $x_0\le x_q\le x_0+\d$, which correspond to having quarks of different masses. The resulting charge density distribution is additive and can be obtained by integrating over $x_q$  the profile functions (\ref{S-wall})  multiplied by the measure $\rho(x_q)$.  Since the branes with a given $x_q$ only contribute to the charge density distribution $S(x)$ for $x\ge x_q$, one has:
\beq \label{eq:NfIntegral2}
	S(x)=\int_{x_0}^{x}\td x_{q}\,\,\r(x_q)\,\,\frac{\sqrt{x-x_{q}}}{\sqrt{x-1}}\,.
\eeq
The measure function $\rho(x_q)$ must obey the  normalization condition:
\beq
\int_{1}^{\infty}\,\td x_{q}\,\,\r(x_q)\,=\,1\,\,.
\label{rho-normalization}
\eeq
When the measure $\rho$ is a $\delta$-function of the type $\rho(x_q)=\delta(x_q-x_{\bar q})$ the profile (\ref{eq:NfIntegral2}) reduces to (\ref{S-wall}) which, as we have seen, leads to background with a threshold singularity. To resolve this singularity we just consider measures with a finite width $\delta$ and we will regard $\delta$ as a regularization parameter of the threshold effect. As $\delta\to 0$ we will recover
(\ref{S-wall}). Below we work out two simple prescriptions for the functional form of $\rho$. In both cases $\rho(x_q)$ is non-vanishing only in a finite interval 
$x_0\le x_q \le x_0 +\d$ and the resulting  $S(x)$ and $S'(x)$ are continuous   and thus they source a regular geometry. Moreover, the profile functions for both measures are actually very similar if one compares distributions with the same width, as one can appreciate in the right plot of figure \ref{fig:Width}.

\subsection{Flat measure}

As first example of weighting measure we consider the situation in which all the embeddings  with different tips in the interval $x_0\le x_q \le x_0 +\d$  weight the same. This election corresponds to choosing a rectangular step function in the interval 
$x_0\le x_q \le x_0 +\d$ which, conveniently normalized, reads:
\beq
\r(x_q)\,=\,{\Theta (x_q-x_0)\,-\,\Theta (x_q-x_0-\d)\over \d}\,\,.
\label{step-measure}
\eeq
Performing the integral \eqref{eq:NfIntegral2} for this measure, we get:
\beq
	\begin{aligned}
		S(x) &={2\over 3}\,\, \frac{(x-x_0)^{3/2}}{\d\,\sqrt{x-1}} \quad \text{when } x_0 \leq x \leq x_0 +\d\,\,, \\\\
		S(x) &= {2\over 3}\,\, \frac{(x-x_0)^{3/2}-(x-x_0-\d)^{3/2}}{\d\,\sqrt{x-1}} \qquad \text{when } x \ge x_0+\d\,\,,
	\end{aligned}
\label{S-step}
\eeq
and  it is understood that $S(x)=0$ for $x\le x_0$.  In figure \ref{fig:Width} (left) we have plotted the function $S(x)$ for different values of the width $\delta$. As shown in this figure, when $\delta$ is increased $S(x)$ grows slower in the transition region and, thus, $S(x)$ is a milder function of $x$. 

Let us now consider the issue of the regularity of $S(x)$. The potentially  dangerous points  are $x=x_0, x_0+\delta$, where the measure $\rho$ is discontinuous. It can be straightforwardly checked that $S$ and its first derivative are continuous at these two points. Actually, one has:
\beq
	\begin{aligned}
	&S(x_0) = 0\,, \quad\quad\quad\quad
	S(x_0+\d) ={2\over 3}\, \frac{\sqrt{\d}}{\sqrt{x_0+\d-1}} \,, \\\\
	&S'(x_0) = 0\,, \quad\quad \quad\quad 
	S'(x_0+\d)=\frac{2\d+3x_0-3}{3\sqrt{\d}(x_0+\d-1)^{3/2}} \,.
	\end{aligned}
\eeq
Moreover, it follows from (\ref{S-step}) that $S(x)$ vanishes as $(x-x_0)^{3/2}$ as we approach the endpoint of the charge distribution at $x=x_0$. Thus, this profile function gives rise to a solution without threshold singularities, as claimed. 

\begin{figure}
	\begin{tabular}{c c}
	\includegraphics[width=0.45\textwidth]{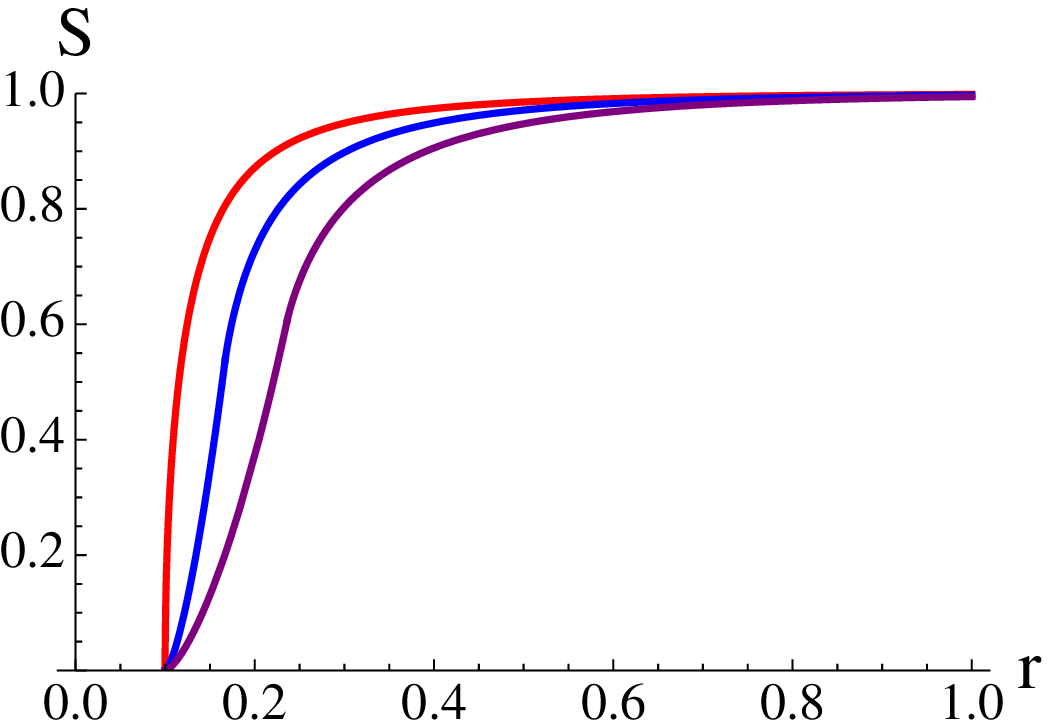} & \includegraphics[width=0.45\textwidth]{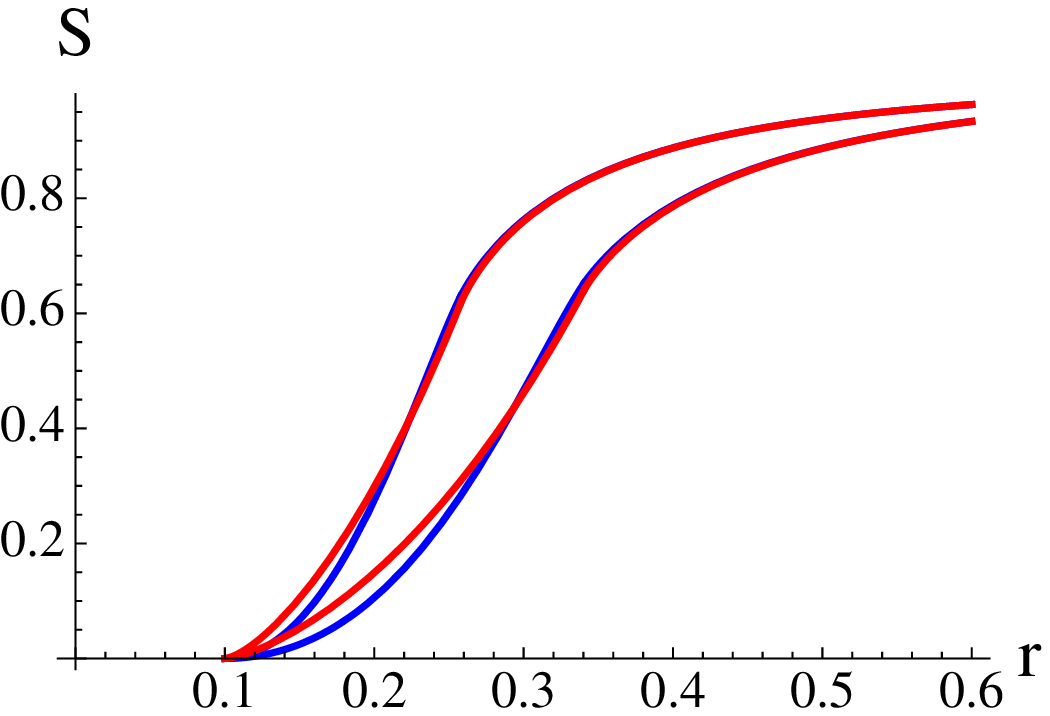}
	\end{tabular}
\caption{We show plots of the function $S$ for the flat measure on the left. The red curve is the singular profile, the blue one is for $\delta = 0.15$ and the purple one is for $\delta = 0.4$. On the right, we plot, for $\delta = 0.5$ and $\delta = 1$, $S$ for the flat measure (red) and the peaked one (blue) and we see that there is almost no difference.}
\label{fig:Width}
\end{figure}

\subsection{Peaked measure}

In our previous example we have considered a weighting function $\rho$ which is discontinuous at $x_0$ and $x_0+\delta$. We now want to explore the possibility of having a measure which vanishes continuously at these endpoints. To choose this new measure, we think of taking a distribution that reproduces a mass peak with finite width for the quark. It means that we choose a distribution that looks like a peak of finite width, a bit like a Gaussian function, but we want something simpler to be able to perform the integration. For that reason, we choose\footnote{We thank  \'Angel Paredes for discussions on this point.}:
\beq
	\r(x_q) = \frac{8}{\p \,\d^2}\sqrt{(x_q - x_0)(x_0+\d-x_q)}\,\,
	\Big[\,\,\Theta (x_q-x_0)\,-\,\Theta (x_q-x_0-\d)\,\Big]
\eeq
The integral \eqref{eq:NfIntegral2} now gives:
\beq
	\begin{aligned}
		S(x) &= \frac{16}{15\p \,\d^{3/2}} \frac{1}{\sqrt{x-1}} \left[ 2(x^2 + x_0^2 + \d^2 +x_0 \d -2 x x_0-x \d) E \left(\frac{x-x_0}{\d} \right) \right. \\
		&\qquad \qquad \qquad \qquad \left. + (x_0+\d-x)(x-x_0-2\d) K  \left(\frac{x-x_0}{\d} \right) \right] \quad \text{when } x_0 \leq x \leq x_0+\d \,\,,\\
		S(x) &= \frac{16}{15\p \,\d^{2}} \frac{\sqrt{x-x_0}}{\sqrt{x-1}} \left[ 2(x^2 + x_0^2 + \d^2 + x_0 \d-2x x_0 - x \d) E \left(\frac{\d}{x-x_0} \right) \right. \\
		&\qquad \qquad \qquad \qquad \qquad \left. + (x-x_0-\d)(2x_0+\d-2x) K  \left(\frac{\d}{x-x_0} \right) \right]  \quad \text{when }  x \ge x_0+\d\,\,,
	\end{aligned}
	\label{profile-peaked}
\eeq
where, again,  it is understood that $S(x)=0$ for $x\le x_0$. The functions 
$K$ and $E$  in (\ref{profile-peaked}) are complete elliptic integrals of the first and second kind respectively. They are defined as:
\beq
E(k) = \int_0^{\pi/2}\td t\, \sqrt{1-k\,\sin^2(t)}\,\,, 
\qquad\qquad
K(k) = \int_0^{\pi/2}\td t\, \frac{1}{\sqrt{1-k\,\sin^2(t)}}\,\,.
	\eeq
If we now look at the properties of our solution at $x=x_0, x_0+\d$, we get:
\beq
	\begin{aligned}
	&S(x_0) = 0\,, \quad\quad  \quad\quad 
	S(x_0+\d) = \frac{32\sqrt{\d}}{15\p \sqrt{x_0+\d-1}} \,, \\\\
	&S'(x_0) = 0\,, \quad\quad \quad\quad 
	S'(x_0+\d)=\frac{8(3\d+5x_0-5)}{15\p\sqrt{(x_0+\d-1)^3\d}} \,, 
	\end{aligned}
\eeq
which shows that $S$ and $S'$ are regular and, therefore, 
this measure leads to another solution free of threshold singularities. In order to compare the profile (\ref{profile-peaked}) with the one  obtained with the flat measure (eq.  (\ref{S-step})) we have plotted in figure \ref{fig:Width} (right) the two functions $S(x)$ for two values of $\delta$. It is rather clear from this figure that the election of $\rho$ does not influence much $S$ (and even less the functions of the ansatz) and, therefore, the physically relevant parameter is the width. For this reason, in the numerical calculations of this paper we will use the simpler result (\ref{S-step}).

\section{Solutions  of the master equation}
\label{numerics}

After discussing the details of our setup in the previous sections, we now move on to the task of finding explicit type IIB supergravity solutions. As we argued in section \ref{massive}, it is enough to solve the master equation \eqref{master-eq}, since all the other functions of our ansatz will follow. For each $P$, we will have a background preserving four supersymmetries that solves the equations of motion of type IIB supergravity (see appendix \ref{EoM}).

The master equation involves the profile $S(r)$ and the function $Q(r)$ (see \eqref{Q-integral}). Notice that in the cases $S=0$ and $S=1$, this master equation has been extensively studied in the literature \cite{HoyosBadajoz:2008fw}. These cases are precisely the IR and UV limits of our profiles $S(r)$, so then the asymptotics of our solutions are already known. What we have to find is a smooth matching between them.

We cannot provide an exact analytical solution of the master equation, but we can give analytical expansions in the relevant regions (around $r=0$, $r=r_0$, and $r=\infty$), and solve numerically in between them.

\subsection{Analytical matching}

On general grounds we expect $S(r)$  to be null up to a certain point $r=r_0$, where we have enough energy to start seeing the effects of virtual quarks running in the loops.  Then it starts growing because as the energy increases it is easier to produce the quarks. It should eventually stabilize around $S(r)=1$ since we will have enough energy so that the flavors appear to be massless. Although we know the specific functional form of $S(r)$ in some cases, let us keep the discussion more general and assume that $S(r)$ can be expanded in a kind of power series around $r_0$ as the one below:
\beq
S(r)=\Theta(r-r_0)\left[S_{1}(r-r_0)^{1/2}+S_{2}(r-r_0)+S_{3}(r-r_0)^{3/2}+\cO\left((r-r_0)^{2}\right)\right]\,.
\label{eqn:S3exp}
\eeq
 It is important to notice that, according to this expansion, although $S(r)$ is continuous, the $\left[\frac{n+1}{2}\right]$-th derivative will not be,  when  $S_{n}$ is the first non-zero coefficient of the expansion. Note that $S$ calculated with both the flat measure and the peaked measure are included in this expansion (we only have the odd coefficients for the former, and the even ones for the latter).

Of course up to $r=r_0$ the solution of the master equation will be the unflavored one. This solution was written close to $r=0$ in \cite{HoyosBadajoz:2008fw}:
\beq
\Pufl(r)=2N_c\,\b\left[ r+\frac{4}{15}\left(1-\frac{1}{\b^2}\right)r^3+\frac{16}{525}\left(1-\frac{1}{3\b^2}-\frac{2}{3\b^4}\right)r^5+\cO\left(r^7\right)\right]\,,
\label{eqn:genPufl}
\eeq
where\footnote{The parameter $h_1$ found in \cite{HoyosBadajoz:2008fw} is related to $\b$ by $h_1 = 2N_c\,\b.$} $\b\geq1$. Unfortunately, far from this point, it is only known numerically (except for the case $\b=1$, where the previous expansion truncates to the exact solution $P=2N_c\, r$ of \cite{Malda-Nunez:N=1})\,\,. The different functions of this solution behave near $r=0$ as:
\begin{equation}
\begin{aligned}
&e^{2h}=N_c\left[\b \,r^2+ \frac{4}{45\b}\left(-12\b^2+15\b-8\right)r^4+\cO\left(r^6\right)\right]\,,\\[0.1in]
&e^{2g}=N_c\left[\b+\frac{4}{15\b}\left(6\b^2-5\b-1\right)r^2\! +\!\frac{16}{1575\b^3}\!\left(3\b^4+35\b^3-36\b^2-2\right)r^4 \!+\! \cO\left(r^6\right)\right]\,,\\[0.1in]
&e^{2k}=N_c\left[\b+\frac{4}{5\b}\left(\b^2-1\right)r^2+\frac{16}{315\b^3}\left(3\b^4-\b^2-2\right)r^4+\cO\left(r^6\right)\right]\,,\\[0.1in]
&e^{4(\Phi-\Phi_0)} = \frac{4}{N_c^3\b^3}\left[1+\frac{16}{9\b^2}r^2+\frac{32}{405\b^5}\left(-15\b^2+31\right)r^4+\cO\left(r^6\right)\right]\,,\\[0.1in]
&a=1+\left(-2+\frac{4}{3\b}\right)r^2+\frac{2}{45\b^3}\left(75\b^3-116 \b^2+40 \b+8\right)r^4+\cO\left(r^6\right)\,.
\end{aligned}
\label{IR-behavior}
\eeq
This solution is regular in the IR. Near $r=0$ the different curvature invariants are:
\beq
\begin{aligned}
		&R = \frac{e^{-\Phi_0/2}}{3} \frac{2^{7/4}}{N_c^{5/8}\b^{21/8}} + \cO(r)\,, \\[0.1in]
		&R_{\mu \nu} R^{\mu \nu} = \frac{31e^{-\Phi_0}}{27} \frac{2^{5/2}}{N_c^{5/4}\b^{21/4}} + \cO(r)\,, \\[0.1in]
		&R_{\mu \nu \rho \sigma} R^{\mu \nu \rho \sigma} = \frac{e^{-\Phi_0}}{45} \frac{976 -3072 \b^2 +3456 \b^4}{2^{1/2} N_c^{5/4} \b^{21/4}} +\cO(r)\,.
\end{aligned}
\eeq

From $r_0$ on, $S\neq0$, and we will have to solve the master equation with initial conditions given by the unflavored solution: $P(r_0)=\Pufl(r_0)$, $P'(r_0)=\Pufl'(r_0)$. The form of the solution will depend on the form of $S(r)$ around $r=r_0$.

To solve the master equation in power series close to the matching point $r=r_0$, we need to know the expression for $Q(r)$, which can be obtained from (\ref{Q-integral}):
\begin{equation}
\begin{aligned}
	&Q(r)=N_c\bigg[2r_0\coth(2r_0)-1+\frac{\sinh(4r_0)-4r_0}{\sinh^2(2r_0)}(r-r_0)-\frac{2N_f}{3N_c}S_1 \tanh(2r_0) (r-r_0)^{3/2}+\\
	&\qquad\qquad\quad+\left(\frac{4\left(2r_0 \coth(2r_0) -1\right)}{\sinh^2 (2r_0)}- \frac{N_f}{2N_c} \tanh(2r_0) S_2\right)(r-r_0)^2+\cO\left((r-r_0)^{5/2}\right)\bigg]\,.
\end{aligned}
\label{eqn:Qexp}
\end{equation}
To arrive at (\ref{eqn:Qexp}) we have fixed the integration constant $q_0$ in (\ref{Q-integral}) to match the unflavored solution at $r=r_0$. This matching is achieved if one takes $q_0=0$. Notice that, in $Q$, $N_f$ appears only through the combination $N_f/N_c$ and $N_c$ is just an overall factor. Actually, the master equation (\ref{master-eq}) can be written in terms of $P/N_c$, $Q/N_c$, $N_f/N_c$ and $S$ and no other term depends on $N_c$.

As no term will be singular in the master equation at $r=r_0$, the uniqueness and existence theorem for ordinary differential equations guarantees the existence of a unique smooth solution (actually as smooth as $\int\td r S$) for this second order differential equation. Therefore, let us propose an expansion for $P(r)$ as:
\beq
\begin{aligned}
N_c^{-1}P(r)=&N_c^{-1}\Pufl(r_0)+N_c^{-1}\Pufl'(r_0)(r-r_0)+P_3(r-r_0)^{3/2}+P_4(r-r_0)^2+\\[0.1in]
&+P_5(r-r_0)^{5/2} +\cO\left((r-r_0)^3\right)\,.
\end{aligned}
\label{eqn:Pexp}
\eeq
Plugging the expansions \eqref{eqn:S3exp}, \eqref{eqn:Qexp} and \eqref{eqn:Pexp} in the master equation \eqref{master-eq}, we obtain the following solution:
\beq
	\begin{aligned}
		P_3 &= -\frac{2N_f}{3N_c} S_1\,,\\
		P_4 &= \frac{1}{2}\left(N_c^{-1}\Pufl''(r_0)-\frac{N_{f}}{N_c}\,S_2\right)\,,\\
		P_5 &= -\frac{2N_f}{5N_c} S_3 + \frac{8N_f}{15N_c}S_1\,N_c^{-1} \Pufl'(r_0) \frac{N_c^{-1}\Pufl(r_0) -2 r_0 + \tanh(2r_0)}{\big(2r_0 \coth(2r_0) -1 \big)^2 - N_c^{-2}\Pufl^2(r_0)}\,.
	\end{aligned}
\eeq

An important lesson to extract from here is the following: our background will present no curvature discontinuity as long as $P''$ is continuous (if only $P'$ is continuous, then the Ricci scalar will have a finite jump at $r=r_0$). So in this case, no curvature singularity will amount to having $S_1=S_2=0$.

In the UV we have $S\to1$. The asymptotic value $S=1$ will be reached  exponentially, in a fashion that depends on the particular details of the measure used to compute $S$, although the first subleading term is universal (given by the abelian limit):
\beq
 S=1-\left(\frac{\d+2\cosh4r_0}{2}-1\right)e^{-4r}+\cO\left(e^{-8r}\right)\,.
 \eeq
 As we mentioned, the case $S=1$ has been studied already in \cite{HoyosBadajoz:2008fw}, where two possible UV analytical expansions were found, dubbed Class I (linearly growing $P$) and Class II (exponentially growing $P$). We will also have two possible UV behaviors, and the analytical expansions will have the same coefficients as those in \cite{HoyosBadajoz:2008fw} for the solutions with linearly growing $P$, and the same leading coefficients for the solutions with exponentially growing $P$. As argued in the next subsection, we are interested in the Class I behavior.

\subsection{Numerical matching}
\label{matching}

If we solve the master equation \eqref{master-eq} numerically, we find, regardless of the specific profile $S(r)$ we use, two qualitatively different behaviors as we go to $r\to\infty$, which are in correspondence with the two classes of UV described in section 4 of \cite{HoyosBadajoz:2008fw}. Indeed, in the deep UV, the massive flavors we are introducing can be considered massless. We have checked that our numerical solutions comply with the UV asymptotic behaviors described in \cite{HoyosBadajoz:2008fw}.

\begin{figure}
\centering
\includegraphics[width=0.65\textwidth]{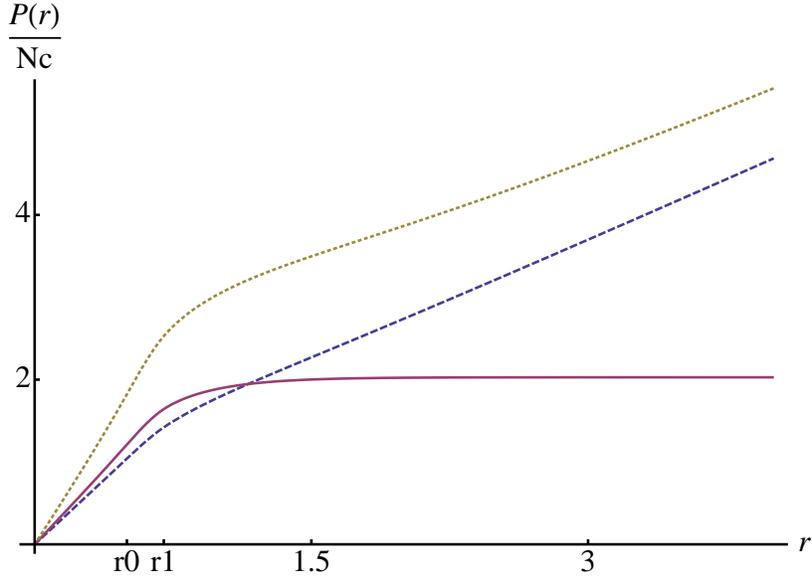}
\caption{Numerical solutions for $N_c^{-1}P$  for different values of $N_f/N_c$, keeping fixed the profile (flat measure) and $r_0$ and $\d$ (in the plot, $\cosh (4r_1) = \cosh(4r_0)+\d$). The blue dotted line corresponds to $N_f/N_c=1$. The purple line corresponds to the conformal case $N_f/N_c=2$. And the olive dotted line corresponds to $N_f/N_c=3$. Notice the expected UV asymptotic behaviors.}
\label{fig:Ps}
\end{figure}

We find that in general the flavored solution only matches nicely (meaning that the solution will reach infinity) with the unflavored solution \eqref{eqn:genPufl} if we choose $\b$ to be bigger than some critical value $\b_c$, which is only known numerically (see figure \ref{fig:h1c}) and bigger than 1. This means in particular that the unflavored solution cannot be that of \cite{Malda-Nunez:N=1}. We observe the following:

Assume the unflavored $P$ up to $r_0$ is given by the numerical solution characterized in the IR by \eqref{eqn:genPufl}. Then there exists a $\b_c$ such that:
\begin{itemize}
\item For $\b<\b_c$, $P$ will eventually start decreasing, crossing $Q$ at some finite value of the radial coordinate and making $e^{2h}=0$ at that point. This solution is then singular.
\item For $\b=\b_c$, $P$ will reach infinity linearly. This solution has precisely the same asymptotics as those described as Class I in \cite{HoyosBadajoz:2008fw}, characterized by a linearly growing $P$ and a linearly growing dilaton. 
\item For $\b>\b_c$, $P$ will reach infinity exponentially. This solution possesses the asymptotics dubbed as Class II in the previous reference, characterized by an exponentially growing $P$, and an asymptotically constant dilaton.
\end{itemize}

\begin{figure}
\centering
\includegraphics[width=0.6\textwidth]{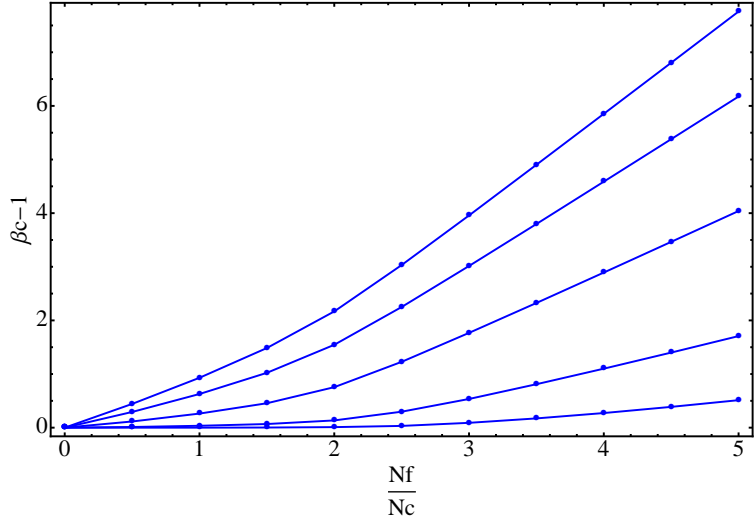}
\caption{We plot the different values of $\beta_c-1$ as one varies the ratio $N_f/N_c$. The different curves are for different quark masses: moving from the upper curves to the lower ones, the values used are $r_0=0,0.15,0.3,0.7,1.2 $; and fixed width $\delta=0.2$. Notice that as the $r_0$ increases (the mass increases), the growth of $\beta_c$ with $N_f/N_c$ is less and less noticeable, and the solution in the unflavored region is almost that of \cite{Malda-Nunez:N=1} ($\beta_c\simeq1$). This was to be expected since the more massive the flavors, the less they affect the IR dynamics.}
\label{fig:h1c}
\end{figure}

So the IR expansion \eqref{eqn:genPufl} can be connected with any of the two known UV behaviors as long as we choose the parameter $\b$ appropriately. For an interpretation of our solutions as gravity duals of  $\cN=1$ SQCD we are interested in the ones with asymptotically linear dilaton \cite{Casero:2006pt}, \ie~the ones which have $\b=\b_c$. Notice that the IR effects of the flavors will be codified in the dependence of $\b_c$ with $N_f/N_c$. We can then regard $\beta_c$  as a measure of the deformation induced by the flavors in the IR. In figure \ref{fig:h1c} we explore the dependence of $\beta_c$ on the number of flavors and their mass.

Even if we fix $\b=\b_c$, and for a given ratio $N_f/N_c$, we can still play with several parameters in the profile $S(r)$, like $r_0$, $\d$ or even with the functional form of $S$ itself. The reader may wonder what would be the effect of that. We find that the qualitative behavior of the metric functions does not change. For instance, varying the width of the mass distribution of the quarks $\d$, just makes more or less sharp the transition from the unflavored region to the flavored one. We gathered in figure \ref{fig:metricfs} the plots of the various metric functions for some particular values of the parameters, just to exhibit explicitly this transition from unflavored to flavored background that happens around $r_0$.

\begin{figure}
\centering
\includegraphics[width=0.65\textwidth]{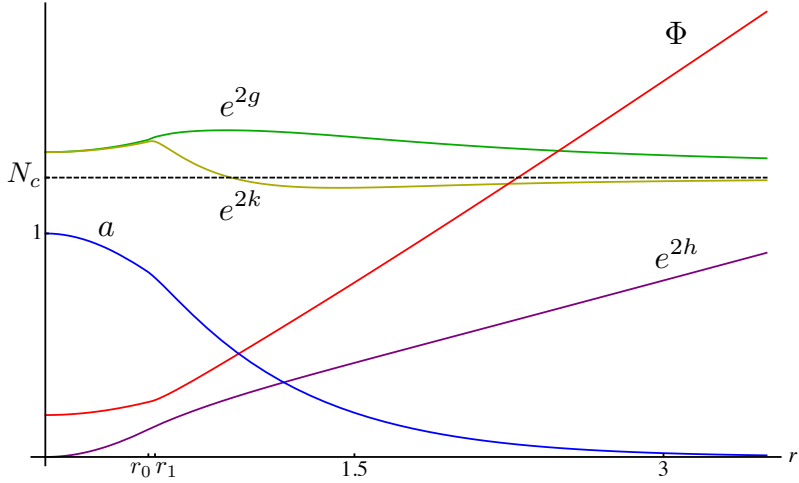}
\caption{Metric functions for a case with $N_f\neq2N_c$. We have used the flat measure profile with $r_0=0.5$, $\delta=0.5$. All the functions have the expected asymptotics. Notice in particular the linearly growing dilaton, in red.}
\label{fig:metricfs}
\end{figure}

\subsection{The solution for massless flavors}

Let us take $r_0\to 0$ in our expressions, keeping a finite width $\d$ for the measure (recall that also taking $\d\to0$ gives back the singular solution of \cite{Casero:2006pt}). This makes the lightest quark we are introducing massless. Nonetheless, due to the non-zero width, some of the quarks are massive; notice however that their mass can be chosen to be as small as one wants. In that respect, this solution is not a typical massless-flavor solution, as in \cite{Casero:2006pt}.

Let us consider the following expansion for the profile function $S(r)$:
\beq
S(r)=S_{1}\,r+S_{2}\,r^2+S_{3}\,r^3+\cO(r^4)\,.
\eeq
We set the first coefficient to zero because we are imposing $S(0)=0$. This expansion encompasses the results coming from the two measures we chose in section \ref{Smoothing}.

For this $S(r)$, we have to integrate the differential equation for $Q$ in a series expansion. We get:
\beq
Q(r)=N_c\left[\frac{4}{3}r^2-\frac{N_{f}\,S_1}{2N_c}r^3-\left(\frac{16}{45}+\frac{2N_{f}\,S_2}{5N_c}\right)r^4+\left(\frac{2N_{f}\,S_1}{9N_c}-\frac{N_{f}\,S_3}{3N_c}\right)r^5+\cO(r^6)\right]\,.
\eeq
The expansion we get for $P$ now is:
\beq
\begin{aligned}
P&=N_c\Bigg[2\b\, r-\frac{5N_f}{4N_c}S_1r^2+\frac{8\,\b}{15}\left(1-\frac{1}{\b^2}+\frac{9}{256}\frac{N_{f}^2\,S_1^2}{N_c^2\b^2}-\frac{9}{8}\frac{N_{f}\,S_2}{N_c\,\b}\right)r^3\\
&\;\;+\left(\!-\frac{7N_f}{18N_c}S_3+\frac{N_{f}}{N_c}S_1\left(-\frac{34}{135}+\frac{7}{27\b}-\frac{7}{45\b^2}+\frac{7}{360}\frac{N_{f}\,S_2}{N_c\,\b}+\frac{7}{1280}\frac{N_{f}^2\,S_1^2}{N_c^2\b^2}\right)\!\right)r^4 +\cO(r^5)\Bigg],
\end{aligned}
\eeq
where $\b$ is a free parameter. We find the following IR asymptotics for the metric functions and the dilaton:
\beq
\begin{aligned}
&e^{2h}=N_c\left[\b\,r^2-\frac{5N_f}{8N_c}S_1\,r^3-\frac{16\b}{15}\left(1+\frac{2}{3\b^2}-\frac{5}{4\b}-\frac{9}{1024}\frac{N_{f}^2\,S_1^2}{N_c^2\b^2}+\frac{9}{32}\frac{N_{f}\,S_2}{N_c\,\b}\right)r^4+\cO\left(r^5\right)\right],\\[0.1in]
&e^{2g}=N_c\left[1-\frac{5N_f}{8N_c}S_1\,r+\frac{8\b}{5}\!\left(\!1-\frac{1}{6\b^2}-\frac{5}{6\b}+\frac{3}{512}\frac{N_{f}^2\,S_1^2}{N_c^2\b^2}-\frac{3}{16}\frac{N_{f}\,S_2}{N_c\,\b}\!\right)\!r^2\!+\!\cO\left(r^3\right)\right]\,,\\[0.1in]
&e^{2k}=N_c\left[1-\frac{3N_f}{4N_c}S_1\,r+\frac{4\b}{5}\left(1-\frac{1}{\b^2}+\frac{9}{256}\frac{N_{f}^2\,S_1^2}{N_c^2\b^2}-\frac{1}{2}\frac{N_{f}\,S_2}{N_c\,\b}\right)r^2+\cO\left(r^3\right)\right]\,,\\[0.1in]
&e^{4(\Phi-\Phi_0)}=\frac{4}{N_c^3\b^3}\left[1+2\frac{N_{f}\,S_1}{N_c\,\b}r+\left(\frac{16}{9\b^2}+\frac{21}{8}\frac{N_{f}^2\,S_1^2}{N_c^2\b^2}+\frac{N_{f}\,S_2}{N_c\,\b}\right)r^2+\cO\left(r^3\right)\right]\,, \\[0.1in]
&a=1-\left(2-\frac{4}{3\b}\right)r^2-\frac{N_{f}\,S_1}{6N_c\,\b}\left(3-\frac{5}{\b}\right)r^3+\cO\left(r^4\right)\,.
\end{aligned}
\eeq
Solving the master equation numerically, we find the same UV behaviors as in the previous subsection, that is, $P$ grows either linearly or exponentially as $r\to\infty$. Again, the linear behavior can only be reached by choosing $\b$ equal to a critical value $\b_c$ (see figure \ref{fig:h1c}). 

We have checked that the solution above presents no curvature singularity in the IR if we choose $S_{1}=0$. For instance, the Ricci scalar near $r=0$ is given by:
\begin{equation}
	R = 3 e^{-\Phi_0/2} \frac{N_f\, S_1}{2^{5/4}N_c^{13/8}\b^{13/8}} \frac{1}{r} + \cO(r^0)\,\,,
\end{equation}
and the metric is clearly singular at $r=0$  if $S_1\not=0$.

Note that what is done in this subsection might be thought as a regular way to introduce massless flavors, as opposite to what happens in \cite{Casero:2006pt}, where the geometry is singular in the far IR. This statement should be taken with a grain of salt: if we interpret $N_f\,S(r)$ as giving the number of flavors that are effectively massless at a given scale $r$ (see next section), we are clearly reading off $S(r=0)=0$ that there are no massless flavors in the far IR. But given that the tips of some branes reach the origin of the space, there are certainly massless quark states in the dual theory. One could conjecture about the existence of some field-theoretical counterpart of the fact that the tips of the branes should be spread in order not to generate a curvature singularity. Unfortunately, we cannot assert any strong claim in this regard.

\section{On the dual QFT interpretation}
\label{observables}

Up to now we have only dealt with the problem of finding a regular supersymmetric solution for unquenched massive quarks. This solution should capture, in a holographic
setup, those flavor effects for which the fact that the fundamentals are massive is important. Notice that, in the UV, our solution reduces to the one in \cite{Casero:2006pt}. Thus, we expect our formalism to be relevant in the description of the IR physics of the model. In this section we work out explicitly some of the effects of massive flavors. In section \ref{Seiberg} we analyze the realization of Seiberg duality in our massive solutions. In section \ref{Wilson} we study we study the Wilson loops in our background. Finally, in section \ref{k-strings} we perform the calculation of k-string tensions within our formalism.

\subsection{Seiberg duality}
\label{Seiberg}

Seiberg duality is an interesting feature of $\cN=1$ four-dimensional gauge theories with flavors. In this section, we briefly comment on the particularities of Seiberg duality in the presence of massive flavors, and explain how these features are realized in our holographic setup.

In his original paper \cite{Seiberg:1994pq}, Seiberg argued that the IR dynamics of SQCD could be understood with the usual ``electric'' description, that of an $SU(N_c)$  gauge theory with $N_f$ flavors; or alternatively via a ``magnetic'' description, consisting of an $SU(N_f-N_c)$ gauge theory with $N_f$ flavors interacting with some gauge singlets. The global anomalies of both the electric and magnetic theory match, a precise dictionary between gauge-invariant primary operators can be found (in particular the gauge singlets of the magnetic theory are related to the mesons of the electric theory) so that the 't Hooft anomaly matching conditions are satisfied, and the deformations of the two moduli spaces can be put in correspondence.

As discussed in Seiberg's original work, to understand the effect of giving a mass to the fundamentals, we can just give a mass to, say, the $N_f$-th quark flavor. We can now think what happens to the IR theory both in the electric and in the magnetic picture. In the electric picture, the massive flavor will be integrated out in the IR, so that the effective electric theory will have $N_c$ colors and $N_f-1$ flavors. From the magnetic perspective, the mass term becomes, after working out the F-term equation of the gauge singlets, a VEV for the magnetic $N_f$-th quark. The gauge group $SU(N_f-N_c)$ is then broken down through the Higgs mechanism to $SU(N_f-N_c-1)$, and thus the IR theory will have $N_f-N_c-1$ colors and $N_f-1$ flavors. This magnetic effective description is precisely the dual of the electric one just described.

The lesson we should extract is that Seiberg duality in the presence of massive flavors works very much like in the case of massless flavors, but instead of the usual duality relation $\left(N_c,\,N_f\right)\leftrightharpoons \left(N_f-N_c,\,N_f\right)$, one should have $\left(N_c,\,N_f^{\textrm{eff}}\right)\leftrightharpoons \left(N_f^{\textrm{eff}}-N_c,\,N_f^{\textrm{eff}}\right)$, where $N_f^{\textrm{eff}}$ is the number of massless flavors. Let us see how this feature is codified in our holographic dual.

The field theory whose dynamics our supergravity background is capturing is not exactly SQCD, but rather SQCD plus a quartic superpotential ($W\sim\k\,\tilde{Q}Q\tilde{Q}Q$). This field theory exhibits ``exact'' Seiberg duality \cite{Strassler:2005qs}, meaning that the duality holds along the RG flow\footnote{The idea is that the $\tilde{Q}Q\tilde{Q}Q$ term of the electric theory becomes a mass term for the gauge singlets of the magnetic theory. They can be therefore integrated out, leaving a SQCD theory with a quartic superpotential as the electric one.} and not just at the IR fixed point, and the theory is Seiberg self-dual. The way this was seen in the solution with massless flavors \cite{Casero:2006pt} (see \cite{Casero:2007pt} for a deeper and more subtle analysis) was by realizing that the BPS system is invariant under the change $\left(N_c,\,N_f\right)\leftrightharpoons \left(N_f-N_c,\,N_f\right)$. It corresponds to $Q(r)\leftrightharpoons-Q(r)$ which leaves the master equation invariant. That is, the same supergravity solution (\emph{i.e:} the same physics) can have two different interpretations, one electric and another magnetic.

In the solution for massive flavors of the present paper, it can be easily seen that the master equation \eqref{master-eq} is invariant under $\left(N_c,\,N_f S(r)\right)\leftrightharpoons \left(N_f S(r)-N_c,\,N_f S(r)\right)$ (recall equation \eqref{Q-integral}). Taking into account that $N_f S(r)$ is precisely counting how many flavors are effectively massless at a given energy scale, this is exactly what we were expecting to find from the discussion above. Note that the change $\left(N_c,\,N_f\right)\leftrightharpoons \left(N_f-N_c,\,N_f\right)$ is NOT a symmetry of the master equation with massive flavors.

\subsection{Wilson loops}
\label{Wilson}

We would like to look now at the behavior of the quark-antiquark potential in the field theory dual to our supergravity solution, that can be studied within the gauge/gravity correspondence. This topic has been treated already in \cite{Casero:2006pt} for the case of massless flavors, and extended in \cite{Bigazzi:2008gd} to the case of massive flavors. The interest of revisiting this calculation is the following:

In the original case of \cite{Casero:2006pt}, the metric presented a curvature singularity in the IR; as shown in \cite{Nunez:2009da}, this invalidates the gravity calculation at low energy. One could think of giving a mass to the quarks, that will remove the aforementioned IR singularity, but a proper gravity model for the theory with massive flavors was not available. The authors in \cite{Bigazzi:2008gd} proposed to use a gravity solution built out of a flavorless solution and a solution with massless flavors, glued at some finite $r=r_q$, modeling a mass $m_q\sim r_q$ for the quarks. This solution would correspond to taking in our formalism $S=\Theta\left(r-r_q\right)$, that introduces a very ugly curvature singularity at $r=r_q$.

With our gravity solution at hand, we can address the study of the quark-antiquark potential in a singularity-free context. Before going on, let us state that the results we obtain are in qualitative agreement with those of \cite{Bigazzi:2008gd}, where they found that the ``connected part'' of the static potential between two non-dynamical quarks (i.e: without taking into account the decay into mesons) went from a Coulomb-like law at short separation distances to a confining behavior in the IR. Moreover, depending on the mass of the quarks $m_q$, there was a first-order phase transition between these two different behaviors for masses below a certain critical mass $m_c$.

The quark-antiquark potential can be extracted from the expectation value of a Wilson loop, and the procedure for computing the latter within the gauge/gravity correspondence is well known. The idea is to introduce a probe flavor brane at $r=\infty$ (so that the probe quarks have infinite mass and are non-dynamical) extended along the Minkowski directions as well as wrapping a certain two-cycle in the internal manifold. We attach then a string to this brane, that will hang into the ten-dimensional geometry, reaching a minimum radial distance $r_0$. We have to compute the energy $E$ of the string and the separation $L$ of the quarks at the end-points of the string for different $r_0$'s. We briefly summarize the relevant formulae. For details one can have a look at \cite{Nunez:2009da}. Defining:
\beq
f^2=g_{tt}g_{x^ix^i}=e^{2\Phi}\,,\qquad g^2=g_{tt}g_{rr}=e^{2\Phi+2k}\,,\qquad V=\frac{f}{C g}\sqrt{f^2-C^2}\,,
\eeq
where $C=f(r_0)$ and we are using string frame, we have that
\beq
L=2\int_{r_0}^\infty\frac{\td r}{V}\,,\qquad E=2\int_{r_0}^\infty\td r\frac{g\,f}{\sqrt{f^2-C^2}}-2\int_0^{\infty}\td r\,g\,.
\eeq
Attaching a string to the probe flavor brane we are introducing can be done whenever it is possible to impose Dirichlet conditions on the string end-points. For our geometry, as discussed in \cite{Nunez:2009da}, this is possible when $\lim\limits_{r\to\infty}V(r)=\infty$. Since for large $r$, $V\sim e^{\Phi-k}$, this conditions holds only for the solutions with an asymptotic linear dilaton. For these solutions we plot the results in figure \ref{fig:Wilson}.
\begin{figure}
\centering
\includegraphics[width=\textwidth]{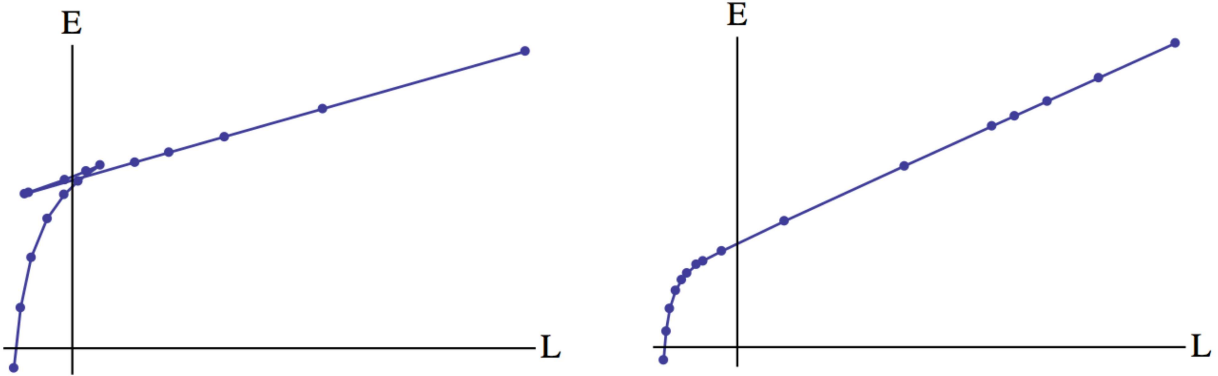}
\caption{We plot the energy of the Wilson loop $E$ vs.~the quark separation $L$. We have fixed $N_f/N_c=1$ and $r_0=0.05$. For small widths of the brane distribution we have the plot on the left, where we observe a first order phase transition. As we increase the width, the mass of the heaviest quark becomes of the order of $\Lambda_{\textrm{SQCD}}$, and the phase transition disappears, as shown in the plot on the right. Notice the similarity between these curves and the G-P (Gibbs free energy vs.~pressure) curves of the Van der Waals gas.}
\label{fig:Wilson}
\end{figure}

As mentioned above, the quark-antiquark potential exhibits two different behaviors: an inverse-power law in the UV, and confining in the IR, where the massive quarks have been integrated out, and the dynamics of the unflavored theory is recovered. The transition between these two behaviors can be smooth as in the plot on the right, or a first-order phase transition (the derivative of the energy has a finite jump), as in the plot on the left. As explained in \cite{Bigazzi:2008gd}, this behavior could be expected whenever we have two scales in the theory. In the present case, these scales are the gaugino condensate and the mass of the quarks. More precisely, our background does not have a sharp value for the mass of the quarks, but rather a distribution of masses with a certain width. The phase transition shows up when the mass of the heaviest quark is smaller than a certain critical mass $m_c$, set by the gaugino condensate.

We could have pursued a more detailed study of these phenomena, like analyzing the dependence of $m_c$ with the number of flavors, their masses, and their distribution, or exploring the decay into mesons characterized by the string breaking length. A fast analysis revealed that the way we distribute the quarks is not very relevant for these observables, and that other features follow qualitatively the behavior described in \cite{Bigazzi:2008gd}. We would like to stress though, that the present calculations are performed in a background without any pathology, giving a more solid foundation to them.

\subsection{k-string tensions}
\label{k-strings}

One of the most interesting features of the IR physics of the confining ${\cal N}=1$ theories is the existence of the so-called k-string states, \ie\ of flux tubes induced by sources with k fundamental indices. It was argued in \cite{Herzog:2001fq} that such a state can be described by D3-branes extended along one of the Minkowski spatial directions, time and wrapping a two-sphere in the IR geometry. For the unflavored geometry of section \ref{unflavoredMN}, the tensions of these k-strings obey a sine law.  It is important to notice that, in order to get the results of \cite{Herzog:2001fq}, it is crucial to find the RR two-form potential $C_{(2)}$. In the approach of \cite{Herzog:2001fq} the potential $C_{(2)}$ is converted into a NSNS two-form $B_{(2)}$ by means of an S-duality transformation and, then, the flux stabilization mechanism of \cite{Bachas:2000ik} is applied to determine the configurations that minimize the energy and to obtain the corresponding tensions. In our flavored background the Bianchi identity of $F_{(3)}$ is violated due  to the presence of  D5-brane sources (eq. (\ref{Bianchi-F3})) and, accordingly, one cannot define the RR potential in regions in which $\Omega$ is different from zero. However, in our massive flavored case, the probe D3-brane will only explore the deep IR region near $r=0$, where there are no flavor brane sources since the profile function $S(r)$  vanishes there. For this reason we will be able to define the potential 
$C_{(2)}$ in this region and we will  proceed with the analysis of the k-string states. Notice also that, in our low-energy analysis, one would not expect to find k-string breaking due to quark-antiquark pair production. However, we will clearly find screening effects due to quark loops which will modify the tensions. 

Let us begin our analysis by studying the IR geometry near $r=0$. Following \cite{Casero:2006pt}, we will  consider the submanifold \cite{Bertolini&Merlatti} defined by the conditions $\tilde\theta=\theta$, 
$\tilde\phi=2\pi-\phi$ at $r=0$. From the IR behavior (\ref{IR-behavior}) of our solutions it is straightforward to verify that the metric (\ref{eq:MetricAnsatz}) along this submanifold, in the string frame,  takes the form:
\beq
\td s^2\,=\,e^{\Phi(0)}\,\,\Bigg[\,\td x^2_{1,3}\,+\,N_c\,\b\,
\Big(\,\td\chi^2\,+\,\sin^2\chi\, 
\big(\,\td\theta^2\,+\,\sin^2\theta\, \td\phi^2\,\big)\,\Big)\,
\Bigg]\,\,,
\label{zero-r-metric}
\eeq
where the angle $\chi$ is related to the coordinate $\psi$ by means of the relation:
$\chi=(\psi-\pi)/ 2$. Clearly, the angles $\chi$, $\theta$ and $\phi$ parameterize 
a  non-collapsing three-sphere at the origin $r=0$ and we should take $\chi$ to vary in the range $0\le\chi\le \pi$. Notice also that
the constant $\b$ characterizes the size of this three-sphere. At  $r=0$ the charge density of the flavor branes vanishes and, as a consequence, there is no violation of the Bianchi identity of $F_{(3)}$. Therefore, it will be possible to represent at this point 
$F_{(3)}$ in terms of a two-form potential $C_{(2)}$ ($F_{(3)}=\td C_{(2)}$). Actually, it is straightforward to check that  $C_{(2)}$ at $r=0$ in these coordinates takes the form:
\beq
C_{(2)}\,=\,-N_c\,C(\chi)\,\sin\theta\, \td\theta\,\wedge \td\phi\,\,,
\label{C2-r=0}
\eeq
with $C(\chi)$ being the function:
\beq
C(\chi)=\,-\,\chi\,+\,{\sin (2\chi)\over 2}\,\,.
\eeq
Contrary to the approach followed in \cite{Herzog:2001fq}, we will perform our analysis directly in the D5-brane background, without performing the S-duality transformation (see also \cite{Ridgway:2007vh}). Accordingly,  let us now consider a probe D3-brane moving in our background. Its dynamics would be governed by the action:
\beq
S_{D3}\,=\,-T_3\,\int \td^4\xi\,e^{-\Phi}\sqrt{-\det\big(\hat g+F)}\,+\,
T_3\int F\wedge C_{(2)}\,\,,
\eeq
with $\hat g$ being the induced metric on the worldvolume of the D3-brane and $F$ the worldvolume gauge field.  We now consider that the D3-brane is extended in $(t,x,\theta,\phi)$ in the metric (\ref{zero-r-metric}) at $r=0$ and at fixed values of $\chi$ and of the other two Minkowski coordinates.  We will also assume that there exists an electric worldvolume gauge field $F_{0x}$ along the Minkowski direction.  In this case, the D3-brane action can be written as:
\beq
S_{D3}\,=\,\int \td t \,\td x\,\, {\cal L}\,\,,
\eeq
where we have integrated  over the angles $(\theta,\phi)$ and  ${\cal L}$ is the effective lagrangian density, given by:
\beq
{\cal L}\,=\,-4\pi T_3\,N_c\,\Big[
\beta\,\sin^2\chi\,
\sqrt{e^{2\Phi(0)}\,-\,F_{0x}^2}\,+\,
\,F_{0x}\,C(\chi)\,\Big]\,\,.
\label{D3-lagrangian}
\eeq
The equation of motion for the electric worldvolume field is:
\beq
{\partial {\cal L}\over \partial F_{0x}}\,=\,{\rm constant}\,\,,
\label{GaussLaw}
\eeq
which is nothing but Gauss' law. 
Following \cite{Camino:2001at}  the constant on the right-hand side of (\ref{GaussLaw}) is fixed by imposing the quantization condition corresponding to having $k$ fundamental strings along the $x$ direction:
\beq
{\partial {\cal L}\over \partial F_{0x}}\,=\,k\,T_{f}\,\,,
\qquad\qquad k\in {\mathbb Z}\,\,,
\eeq
where $T_f=1/(2\pi\alpha')$ is the tension of the fundamental string. This condition determines the electric field in terms of the angle $\chi$. Indeed, let us define
a new function ${\cal C}(\chi)$ as:
\beq
{\cal C}(\chi)\,\equiv\,C(\chi)\,+\,{\pi k\over N_c}\,\,.
\eeq
Then, one has:
\beq
F_{0x}\,=\,{e^{\Phi(0)}\,{\cal C}(\chi)\over
\sqrt{\beta^2\,\sin^4\chi\,+\,{\cal C}(\chi)^2}}\,\,.
\eeq
Notice that $F_{0x}$ is the momentum of a cyclic coordinate that can be eliminated from the lagrangian. The correct way to do this is by performing the Legendre transformation and computing the hamiltonian as:
\beq
H\,=\,\int dx\,\Big[\,F_{0x}\,{\partial {\cal L}\over \partial F_{0x}}\,-\,
{\cal L}\,\Big]\,\,.
\eeq
By calculating explicitly the right-hand side of this equation and writing the result in terms 
of $\chi$, we get:
\beq
H\,=\,4\pi T_3 e^{\Phi(0)}\,N_c\,\,\int dx\,
\sqrt{\beta^2\,\sin^4\chi\,+\,{\cal C}^2(\chi)}\,\,.
\label{eq:Energy}
\eeq
Let us minimize the energy with respect to $\chi$. For this purpose it is interesting to notice that  the function ${\cal C}(\chi)$ satisfies  $d {\cal C}/d\chi=-2\sin^2\chi$. 
Using this property of ${\cal C}$ it is straightforward to prove that, for a given integer $k$,  the energy is minimized for the $\chi_k$ which satisfies:
\beq
{\cal C}(\chi_k)\,=\,{\beta^2\over 2}\,\sin(2\chi_k)\,\,,
\eeq
or equivalently:
\beq
\chi_k\,-\,{\pi k\over N_c}\,+\,{\beta^2-1\over 2}\,\,\sin(2\chi_k)\,=\,0\,\,,
\label{eq:ChiEq}
\eeq
which is the equation written in \cite{Herzog:2001fq} with $\beta$ instead of the $b$ of \cite{Herzog:2001fq}. It is also immediate to find the tension of the $k$-string object, namely:
\beq
T_k\,=\,{e^{\Phi(0)}\,N_c\over 2\pi^2\,\alpha'}\,\,\beta\,\sin\chi_k\,\,
\sqrt{1\,+\,(\beta^2-1)\,\cos^2\,\chi_k}\,\,.
\label{eq:Tension}
\eeq
The worldvolume electric field corresponding to this solution is:
\beq
F_{0x}\,=\,{\beta e^{\Phi(0)}\,\cos\chi_k\over
\sqrt{1\,+\,(\beta^2-1)\,\cos^2\,\chi_k}}\,\,.
\label{Fox}
\eeq
It is interesting to point out that (\ref{eq:ChiEq}) does not change under the transformation $k\to N_c-k$ and $\chi_k\to \pi-\chi_k$. One can also check that the tension in (\ref{eq:Tension}) does not change under this transformation, while the electric field (\ref{Fox}) changes its sign.  Notice also that in the unflavored case reviewed in section \ref{unflavoredMN} one has $\beta=1$ and we recover the results in  \cite{Herzog:2001fq,Camino:2001at,Ridgway:2007vh}. The case with $\beta\not=1$ for the  generalized unflavored models with the IR behavior (\ref{IR-behavior}) was considered in \cite{Casero:2006pt}. Notice that, in our case, the parameter $\beta$ is related to the mass of the quarks and to the number of flavors by means of the matching conditions  discussed in section \ref{matching}.

Let us look at the tension of the k-string as a function of $k/N_c$, for different values of $\beta$. First, we need to solve \eqref{eq:ChiEq}. Depending on the value of $\beta$ and $k/N_c$, we find that there can be up to three different solutions. We then have to check which one corresponds to the true minimum of the energy \eqref{eq:Energy}. We notice that for $k/N_c < 1/2$, the minimum of the energy is given by the solution for $\chi_k$ closest to $0$ as we can see in figure \ref{fig:EnergyMin}, while for $k/N_c > 1/2$, it is the solution closest to $\pi$.

\begin{figure}
	\begin{tabular}{c c}
	\includegraphics[width=0.45\textwidth]{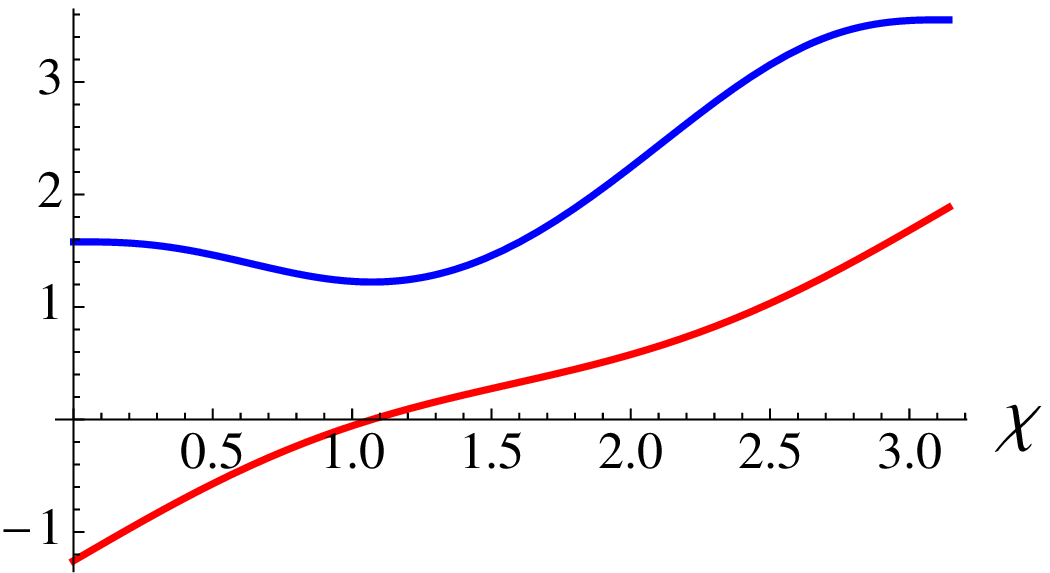} & \includegraphics[width=0.45\textwidth]{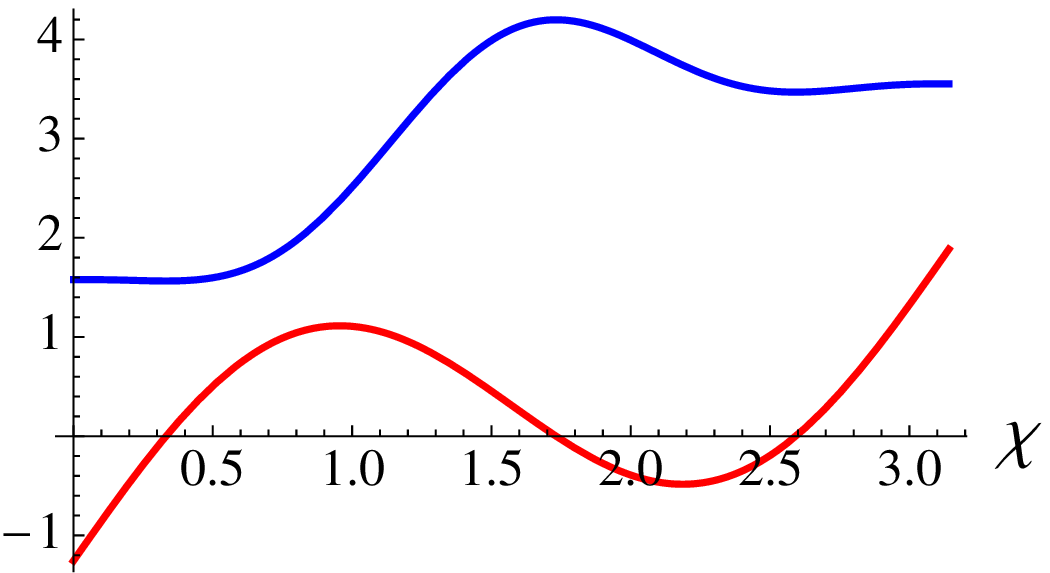}
	\end{tabular}
\caption{The (lower) red curves correspond to the function on the left-hand side of (\ref{eq:ChiEq}), whose zeros are the angles that extremize the energy (\ref{eq:Energy}). The (upper) blue curves are plots of the energy. We are taking  in all cases
 $k/N_c = 0.4$. On the left, we plot these curves for $\beta = 1.2$ and we see only one zero, which corresponds to the minimum of the energy. On the right, for $\beta =2$, we notice the appearance of two new zeros, one corresponding to a maximum of the energy, and the other one to a metastable configuration. The true minimum, however, moves towards $\chi = 0$, that is towards the north pole of the sphere, as $\beta$ is increased.}
\label{fig:EnergyMin}
\end{figure}

Knowing the correct value of $\chi_k$ for each $\beta$ and $k/N_c$, we can go on and plot the tension \eqref{eq:Tension} as a function of $k/N_c$ for various values of $\beta$, as shown in figure \ref{fig:Tension}. When $\beta$ is close to $1$ (which is the smallest value it can reach, corresponding to the unflavored case), the tension of the k-string can be approximated by:
\beq \label{eq:TensionSmallBeta}
	T_k \sim \frac{e^{\Phi(0)} N_c}{2 \pi^2 \alpha'} \beta \sin \left( \frac{\pi k}{N_c} \right)\,.
\eeq
Thus, in this low $\beta$ case, the screening effect due to the flavor is manifested in the tensions by just multiplying the sine formula by the deformation parameter $\beta$. 
In turn, $\beta$ can be related to the number of flavors and their masses by means of the matching condition studied in section \ref{numerics}. Notice that, for a given number of strings $k$, the tension of the flavored k-string is higher than the one corresponding to the unflavored theory. Actually, this is what is expected on general grounds since the screening reduces the (negative) binding energy and, therefore, it increases the total energy (\ie\ the tension).

As $\beta$ goes to infinity, the binding energy becomes smaller and smaller and the tension of a k-string is a linear function of $k$. In this case one can analytically obtain the approximate solution of (\ref{eq:ChiEq}) which corresponds to the minimum of the energy. Indeed, if $\beta$ is large the only possibility to solve (\ref{eq:ChiEq}) is by having 
$\sin (2\chi_k)$ small. One can show that when $k/N_c<1/2$ this equation is solved for 
$\chi_k\approx \pi k/\beta^2 N_c$, while for $k/N_c>1/2$ the energy is minimized for
$\chi_k\approx \pi\,-\,\pi (N_c-k)/\beta^2 N_c$. The corresponding tensions in these two cases are:
\beq
	e^{-\Phi(0)}\,T_k \sim
	\begin{cases} k T_f \,\,,&\qquad{\rm for\,}\qquad  0\le k\le N_c/2
	\,\,,\cr\cr
(N_c-k)\, T_f\,\,,&\qquad{\rm for}\, \qquad N_c/2\le k\le N_c
\,\,,
\end{cases}
\eeq
which shows that, when $\beta$ is large, the screening effects are so large that the binding energy is very small  and one can regard the flux tube as composed by non-interacting strings with vanishing binding energy. One can visualize this behavior as $\beta$ is increased in figure \ref{fig:Tension}.

\begin{figure}
	\begin{center}
	\includegraphics[width=0.5\textwidth]{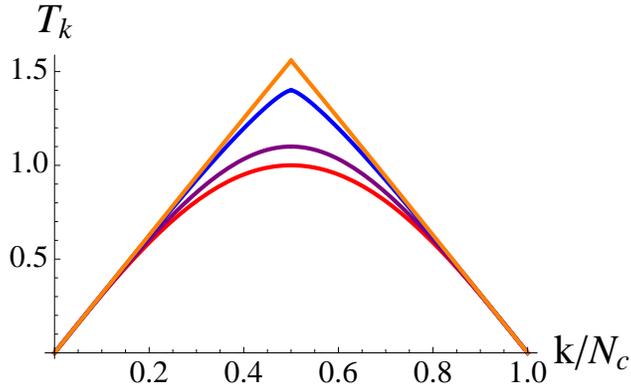}
	\end{center}
\caption{This plot corresponds to the tension of the k-string as a function of $k/N_c$. The values for $\beta$ are 1 for red, 1.1 for purple, 1.4 for blue and 2 for orange.}
\label{fig:Tension}
\end{figure}

\section{Conclusions}
\label{conclusions}

Let us summarize our main results. We have considered the addition of unquenched massive fundamental matter to the gravity dual of the ${\cal N}=1$ SQCD-like theory obtained when D5-branes wrap a two-cycle inside a Calabi-Yau threefold. The matter fields are added by means of D5-branes that are extended along a non-compact two-dimensional submanifold of the internal space. These flavor branes do not reach, in general, the origin of the holographic coordinate $r$ and their charge density depends on $r$. In order to incorporate consistently  in the backreacted background   the effects of this dependence we have modified in a non-trivial way the ansatz of \cite{Casero:2006pt}, by including new terms in the RR three-form $F_{(3)}$ which depend on a profile function $S(r)$ and on its derivative. We have shown that the BPS system can be reduced to a master equation, containing $S$ and $S'$, which is a generalization of the one found in \cite{HoyosBadajoz:2008fw}. This equation can be integrated numerically and, by matching the unflavored solution at the scale at which $S=0$, one can find a supergravity solution in the whole range of the radial coordinate. 

Our solutions satisfy Einstein equations with sources and one of the non-trivial points we have addressed is the determination of the distributions of branes whose charge density and backreaction have precisely the form that we have adopted in our ansatz. We have verified this fact by means of a macroscopic calculation (comparing the action of the full set of branes with the one corresponding to a representative), as well as by a direct microscopic calculation of the charge density in the UV. We have also shown how to get rid of the curvature singularities which appear at the position of the tip of the branes.

After all these developments we have been able to find regular supergravity backgrounds dual to ${\cal N}=1$ SQCD-like theories with massive unquenched flavors. Our results generalize those in \cite{Casero:2006pt} in the sense that our solutions incorporate the effects of the mass scale introduced by the quark mass and, at the same time, they resolve the IR curvature singularity that limits the applicability of the geometry of \cite{Casero:2006pt} to explore holographically the ${\cal N}=1$ $SU(N_c)$ gauge theory with flavors.

We have studied some observables of the field theory for which the  IR structure and the mass scale introduced by the quarks are relevant (see \cite{Barranco} for further analysis along these lines). We think that our formalism provides a framework to explore holographically the ${\cal N}=1$ SQCD-like theories in a firmer basis.

For the standard field theory interpretation, we focused mainly on solutions with linear dilaton at
infinity. However, it is also possible to get backgrounds where the
dilaton is bounded. On such solutions, we can apply a transformation found
in \cite{Gaillard:2010qg}, called rotation. It would give us a completely
regular version of the background of \cite{Gaillard:2010qg}, and for that
reason it could allow for a better understanding of the dual field theory.

We could also think of introducing new scales in our supergravity
solutions, by choosing a profile function for the flavors created by
several separated shells of branes. The function $S(r)$ would then look
like a series of steps, whose lengths would be related to the different
positions of the flavor branes. If, by engineering the profile in a smart
way, we can have the $\beta$-function of the theory almost zero for one of
the steps, the this could lead to a dual field theory exhibiting a walking
behavior \cite{Nunez:2008wi,Nunez:2009da,Elander:2009pk,Piai:2010ma}.

\section*{Acknowledgments}
We are grateful to Alejandro Barranco, Francesco Bigazzi, Aldo Cotrone, Carlos N\'u\~nez, \'Angel Paredes and Jorge Russo for many useful discussions. We are specially grateful to Carlos N\'u\~nez for  encouragement and collaboration in the initial stages of this (long) project. This  work  was funded in part by MICINN   under grant
FPA2008-01838,  by the Spanish Consolider-Ingenio 2010 Programme CPAN (CSD2007-00042), by Xunta de Galicia (Conseller\'\i a de Educaci\'on and grant INCITE09 206 121 PR) and by FEDER. E. C is supported by a Spanish FPU fellowship, and would like to thank the FRont Of Galician-speaking Scientists for unconditional support.

\vskip 1cm
\renewcommand{\theequation}{\rm{A}.\arabic{equation}}
\setcounter{equation}{0}
\appendix

\section{Supersymmetry analysis}
\label{SUSY}
In this appendix we will study the realization of supersymmetry for a background of type IIB supergravity with metric and RR three-form $F_{(3)}$ as given by the ansatz written in eqs. (\ref{eq:MetricAnsatz})-(\ref{ansatz-fs}). To perform this analysis we will choose the following basis of  vielbein one-forms for the metric (\ref{eq:MetricAnsatz}):
\begin{equation}
  \begin{aligned}
    e^{x^i} &= e^f \td x^i\,\,, \qquad i =0,1,2,3 \,\,,\\
    e^r &= e^{f+k} \td r\,\,, &
    e^\theta &= e^{f+h} \td\theta\,\,, \\
    e^\phi &= e^{f+h} \sin\theta \,\td\phi \,\,,&
    e^1 &= \frac{e^{f+g}}{2} (\tilde{\omega}^1 + a(r) \td\theta)\,\,, \\
    e^2 &= \frac{e^{f+g}}{2} (\tilde{\omega}^2 - a(r) \sin \theta\, \td\phi)\,\,, &
    e^3 &= \frac{e^{f+k}}{2} (\tilde{\omega}^3 + \cos \theta\, \td\phi)\,\,.
  \end{aligned}
  \label{vielbein}
\end{equation}
In order to find the supercharges preserved by a given background of type IIB supergravity one should find the Killing spinors $\epsilon$ which leave invariant the dilatino and gravitino field under a supersymmetry transformation. These spinors are characterized by a set of commuting projections. In the case of the unflavored background reviewed in section \ref{unflavoredMN} these projections were found in \cite{Nunez:2003cf}. We will impose that the spinors $\epsilon$ of the flavored background satisfy the same type of conditions as in \cite{Nunez:2003cf}, which can be written as:
\begin{equation} \label{eq:oldprojections}
	\begin{aligned}
		i \,\epsilon^* &= \epsilon, & \Gamma_{\theta \phi} \epsilon &= \Gamma_{12} \epsilon, & \Gamma_{r123} \epsilon &= (\cos \alpha + \sin \alpha \,\Gamma_{\phi 2}) \epsilon\,\,,
	\end{aligned}
\end{equation}
where $\Gamma_{a_1\a_2\cdots}$ denotes the antisymmetrized product of constant Dirac matrices and the $a_i$'s are  flat indices  corresponding to the vielbein (\ref{vielbein}). In (\ref{eq:oldprojections}) $\alpha = \alpha (r)$ is an angle to be determined. The result of the analysis of the supersymmetry variations can be nicely recast in terms of the two fundamental forms of the underlying geometric  ${SU} (3)$-structure of the internal complex manifold. These forms are the $(1,1)$ two-form $J$ and the holomorphic $(3,0)$ three-form $\Omega_{hol}$, which are defined in terms of fermionic bilinears as:
\beq
J\,\equiv\,{i\over 2!}\,\epsilon^{\dagger}\,\Gamma_{a_1\,a_2}\,\epsilon\,\,e^{a_1\,a_2}\,\,,
\qquad\qquad
\Omega_{hol}\,\equiv\,{e^{3f+{\Phi\over 2}}\over 3!}\,\,
\epsilon^{T}\,\Gamma_{a_1\,a_2\,a_3}\,\epsilon\,\,e^{a_1\,a_2\,a_3}\,\,,
\label{J-Omegahol-def}
\eeq
with $\epsilon$ being a Killing spinor normalized as 
$\epsilon^{\dagger}\,\epsilon\,=\,1$ and 
$e^{a_1\a_2\cdots}\,\equiv\,e^{a_1}\wedge e^{a_2}\cdots$. Actually, by using the projections (\ref{eq:oldprojections}) satisfied by $\epsilon$  one can express the
 ${SU} (3)$-structure  forms as:
\begin{equation}
	\begin{aligned}
		J &= e^{r3} + (\cos \alpha\, e^{\phi} + \sin \alpha \,e^2 ) \wedge e^{\theta} + (-\sin \alpha\, e^{\phi} + \cos \alpha \,e^2 ) \wedge e^1\,\,, \\[0.1in]
		\Omega_{hol} &=e^{3f + \Phi/2} \, \big( e^r + i\, e^3 \big) \wedge \big( (\cos \alpha \,e^{\phi} + \sin \alpha\, e^2 ) + i\, e^{\theta} \big) \wedge \big( (-\sin \alpha\, e^{\phi} + \cos \alpha\, e^2 ) + i\, e^1 \big)\,\,.
	\end{aligned}
	\label{J-Omega-ansatz}
\end{equation}
The conditions imposed by the preservation of ${\cal N}=1$ supersymmetry can be written as the following set of equations to be satisfied by the structure forms:
\begin{equation} \label{eq:gencalcondition}
	\begin{aligned}
		e^{-2f - \Phi/2} \td \big( e^{2f + \Phi} J \big) &= -e^{\Phi} *_6 F_{(3)}\,\,, \\
		\td   \Omega_{hol} &= 0 \,\,, \\ 
		\td \big( e^{4f} J \wedge J \big) &= 0\,\,, \\
		\td \big( e^{2f - \Phi / 2} \big) &= 0\,\,,
	\end{aligned}
\end{equation}
where $*_6$ denotes the Hodge dual with respect to the internal part of the metric
(\ref{eq:MetricAnsatz}). Plugging in (\ref{eq:gencalcondition})  the explicit  expressions 
of $J$ and $ \Omega_{hol}$  written in eqs. (\ref{J-Omega-ansatz}), as well as our ansatz for the metric and RR three-form, one gets  the following system of first-oder BPS equations:
\begin{align}
&f'\,=\,{\Phi'\over 4}\,=\,{e^{-g-h}\over 8\sin\alpha}\,
\Big[\,2N_c(b-a)\,+\,2a\,e^{2g}\,\cos^2\alpha\,+\,2\,e^{g+h}\,\sin(2\alpha)\,\Big]\,\,,\rc[0.1in]
&h'=-{e^{-2g-h}\over 4\sin\alpha}\,\Big[
2\,e^{h}\big(2\,e^{2g}+e^{2k}-N_c\big)\,\sin(2\alpha)+2e^{g}\,
\Big((b-a)N_c\,+\,(e^{2g}+e^{2k})a\Big)(1+\cos^2\alpha)
\Big],\rc[0.1in]
&g'\,=\,e^{-2g}\,\Big[\,(e^{2k}-N_c)\,\cos\alpha\,+\,{e^{g-h}\over 2}\,
\big[\,(a-b)N_c\,+\,(e^{2g}-e^{2k})a\,\big]\sin\alpha\,\Big]\,\,,\rc[0.1in]
&k'\,=\,{e^{-g-h}\over \sin\alpha}\,\,\big[\,
(a-b)\,N_c\,+\, (e^{2k}-e^{2g})\,a\,\big]\,\,,\rc[0.1in]
&a'\,=\,2e^{-2g}\,\Big[\,\big(bN_c\,+\,(e^{2k}-N_c)\,a\big)\cos\alpha\,+\,
2\,e^{h-g}\,\big(e^{2g}+e^{2k}-N_c\big)\sin\alpha\,\Big]\,\,,\rc[0.1in]
&b'\,=\,-L_1\,+\,{2e^{2g}\over N_c}\,\,\big[\,
a \cos\alpha\,+\,2 e^{h-g}\,\sin\alpha\,\big]\,\,,
\label{BPS-new}
\end{align}
together with the following two algebraic relations:
\beq
\begin{aligned}
\cot\alpha\,&=\,{e^{-g-h}\over 4a}\,\Big(\,e^{2g}\,a^2\,-\,4\,e^{2h}\,-\,e^{2g}\,\Big)\,\,,\\[0.1in]
b\,&=\,-{e^{g-h}\over 2\big[2\cos\alpha\,-\,e^{g-h}a\sin\alpha\big]}\,\,
\Big[\,\big(L_2\,+\,2\,\big)\,\sin\alpha\,+\,4{e^{g+h}\over N_c}\,\,\big(a\cos\alpha\,+\,2 e^{h-g}\,\sin\alpha\,\big)\,
\Big]\,\,.
\end{aligned}
\label{constraints}
\eeq
By computing the derivative of the second of the constraints in (\ref{constraints}), one gets the remarkably simple relation between the derivative of $L_2$ and $L_1$:
\beq
L_2'\,=\,\Big(4\,e^{h-g}\,\cot\alpha\,-\,2a\,\Big)\,L_1\,=\,-
{1+a^2\,+\, 4\,e^{2h-2g}\,\over a}\,\,L_1\,\,,
\label{L2-L1}
\eeq
where in the last step we have used the first equation in (\ref{constraints}) to eliminate $\alpha$. 

\subsection{Partial integration}
Although we have not been able to get an analytic expression of the general solution  of the system (\ref{BPS-new}), we have been able to simplify it and to perform a partial integration. Indeed, let us proceed as in \cite{Casero:2006pt} and begin by defining two new functions $\tilde C(r)$ and $\tilde S(r)$ as:
\bear
&&\tilde C\,\equiv\,{1+a^2+4e^{2h-2g}\over 2a}\,\,,\rc[0.1in]
&&\tilde S\,\equiv\,{\sqrt{a^4+2a^2\big(4e^{2h-2g}-1\big)\,+\,\big(1+e^{2h-2g}\big)^2}\over
2a}\,\,.
\label{C-S-definition}
\eear
From the BPS equations (\ref{BPS-new}) and (\ref{constraints})  one can show that 
$\tilde C$ and $\tilde S$ satisfy the following second-order differential equations:
\beq
\tilde C''\,-\,4\,\tilde C\,=\,0\,\,,\qquad\qquad
\tilde S''\,-\,4\tilde S\,=\,0\,\,,\qquad\qquad
\tilde C^2\,-\,\tilde S^2\,=\,1\,\,.
\eeq
These equations can be immediately  integrated as:
\beq
\tilde C\,=\,\cosh(2r)\,\,,\qquad\qquad
\tilde S\,=\,\sinh(2r)\,\,,
\eeq
where we have fixed the integration constants by imposing the $r\to 0$ behaviors $\tilde C\sim 1+2r^2$, $\tilde S\sim 2r$. Using the definitions of $\tilde C$ and $\tilde S$ in (\ref{C-S-definition}), we get the following relations:
\bear
&&{a^2+1\over 4}\,e^{2g}\,+\,e^{2h}\,=\,{a\over 2}\,\,e^{2g}\,\cosh(2r)\,\,,\rc\rc
&&a^4\,+\,2a^2\,\big(4e^{2h-2g}-1\big)\,+\,\big(1+e^{2h-2g}\big)^2\,=\,4a^2\,
\sinh^2(2r)\,\,.
\label{CS}
\eear
The relations (\ref{CS})  can be combined with the first constraint in (\ref{constraints})to give:
\beq
{a^2-1\over 4}\,e^{2g}\,-\,e^{2h}\,=\,e^{h+g}\,a\,\cot\alpha\,\,.
\eeq
Indeed, by summing this last equation and the first one in (\ref{CS}), we get:
\beq
\cot\alpha\,=\,{e^{g-h}\over 2}\,\,(a-\,\cosh(2r))\,\,.
\eeq
Using this result it is easy to find the following expressions of $\sin\alpha$ and $\cos\alpha$ in terms of the other functions:
\beq
\sin\alpha\,=\,-{2 e^{h-g}\over \sinh(2r)}\,\,,\qquad
\cos\alpha\,=\,{\cosh(2r)\,-\,a\over \sinh(2r)}\,\,.
\label{alpha}
\eeq
The following useful combinations of $\cos\alpha$ and $\sin\alpha$ can be found from the previous equations:
\beq
\begin{aligned}
&a \cos\alpha\,+\,2 e^{h-g}\,\sin\alpha\,=\,{1-a\cosh (2r)\over \sinh(2r)}\,\,,\\[0.1in]
&2\cos\alpha\,-\,e^{g-h}\,a\,\sin\alpha\,=\,2\,\coth(2r)\,\,.
\end{aligned}
\label{cos-sin-alpha}
\eeq
Using these results, the equation giving $b'$ in the system (\ref{BPS-new}) reduces to:
\beq
b'\,=\,{2e^{2g}\over N_c}\,\,{1-a\cosh(2r)\over \sinh(2r)}\,-\,L_1\,\,,
\label{bprime}
\eeq
while the second equation in (\ref{constraints}) becomes:
\beq
b\,=\,{1\over 2\cosh(2r)}\,\,\Big[L_2\,+\,2\,+\, {2e^{2g}\over N_c}\,
\big(a\cosh(2r)-1\big)\,\Big]\,\,.
\label{b-constraint}
\eeq
Moreover, the relation (\ref{L2-L1}) takes the form:
\beq
L_2'\,=\,-2\cosh(2r)\,L_1\,\,,
\label{L2-L1-partial}
\eeq
which is nothing but (\ref{L1-L2prime}). 
By combining (\ref{bprime}) and (\ref{b-constraint}) we can obtain the following differential equation for $b$:
\beq
b'\,+\,{2\cosh (2r)\over \sinh(2r)}\,b\,=\,{L_2+2\over \sinh(2r)}\,-\,L_1\,\,,
\eeq
Using (\ref{L2-L1-partial}) this equation reduces to:
\beq
b'\,+\,{2\cosh (2r)\over \sinh(2r)}\,b\,=\,{L_2+2\over \sinh(2r)}\,+\,
{L_2'\over 2\cosh(2r)}\,\,.
\label{b-ODE}
\eeq
Applying the method of variation of constants, we can integrate (\ref{b-ODE}) and get
$b$ as a function of $L_2$. The result is:
\beq
b(r)\,=\,{2r+\eta(r)\over \sinh(2r)}\,\,,
\eeq
where $\eta(r)$ is defined as the following integral involving $L_2$:
\beq
\eta(r)\,=\,\int_{0}^{r}\,\td\rho\,
\Big[\,L_2(\rho)\,+\,{\tanh(2\rho)\over 2}\,\,L_2'(\rho)\,\Big]\,\,.
\label{eta-1}
\eeq
Performing a partial integration, one can rewrite $\eta(r)$ as follows:
\beq
\eta(r)\,=\,{1\over 2}\,\tanh(2r)\,L_2(r)\,+\,\int_{0}^{r}\,\td\rho\,
\tanh^2(2\rho)\,L_2(\rho)\,\,.
\label{eta-2}
\eeq
Plugging in  (\ref{eta-2}) the definition (\ref{S-definition})  of the profile function $S$ we just get \eqref{eta-S}. 
We can check that the above formulas give the right results in the known cases. For example, in the unflavored case $L_2$ is zero and it is obvious from (\ref{eta-2}) that $\eta$ vanishes and one recovers the correct result of section \ref{unflavoredMN}. Moreover,  in the massless case studied in \cite{Casero:2006pt},  the function $L_2$ is constant. Actually, in this last case $L_2=-x$ with $x=N_f/N_c$. From (\ref{eta-1}) or (\ref{eta-2}) we get $\eta=-x r$ and the result of eq. (4.14) in \cite{Casero:2006pt} is recovered. 

There are combinations of the BPS equations in (\ref{BPS-new}) that become particularly simple. One of these combinations is:
\beq
2\Phi'\,+\,h'\,+\,g'\,+\,k'\,=\,2\coth(2r)\,\,.
\eeq
Integrating  this relation, one gets eq. (\ref{dilaton-integral}).  Other interesting combinations are:
\beq
\begin{aligned}
{d\over dr}\,\big[\,e^{2g}\,\big]\,&=\,{2\over \sinh (2r)}\,\,
\Big[\,\Big(\,e^{2k}\,-\,N_c\,\Big)\,\cosh (2r)\,-\,ae^{2g}\,+\,N_c\,b\,\Big]\,\,,\\[0.1in]
{d\over dr}\,\big[\,a\,e^{2g}\,\big]\,&=\,{2\over \sinh (2r)}\,\,
\Big[\,e^{2k}\,-\,N_c\,+\,e^{2g}\,\big(1-a\cosh(2r)\big)\,+\,N_c\,b\,\cosh(2r)\,\Big]\,\,.
\end{aligned}
\label{a-exp(g)}
\eeq
Combining  the second of these equations with (\ref{b-constraint}), one can immediately compute the derivative of the function $P$ defined in (\ref{P-Q-def}), with the result:
\beq
P'\,=\,2\,e^{2k}\,-\,N_f\,S\,\,,
\eeq
which is just  the second equation in (\ref{h-k-P-Q}). Moreover one can also compute  from (\ref{a-exp(g)}) and (\ref{b-constraint}) the derivative of $Q$. The result can be written as:
\beq
{d\over dr}\,\Big[\,{Q\over \coth(2r)}\,\Big]\,=\,{2N_c-N_f\,S\over \coth^2(2r)}\,\,,
\eeq
whose integration yields (\ref{Q-integral}).  Another combination of derivatives that becomes particularly simple when one uses the BPS equations is the following:
\beq
h'\,+\,g'\,-\,k'\,=\,a\,e^{2k-2h}\,-\,2\coth (2r)\,\,.
\eeq
From this equation we easily get:
\beq
P\,=\,{1\over 4 \sinh^2 (2r)}\,\,
\partial_r\,\Big[\,\sinh^2 (2r)\,{P^2-Q^2\over P'+N_f\,S}\,\Big]\,\,,
\eeq
which is nothing but the master equation (\ref{master-eq}).

\renewcommand{\theequation}{\rm{B}.\arabic{equation}}
\setcounter{equation}{0}

\section{Equations of motion}
\label{EoM}

In this appendix, we state for completeness the equations of motion. Let us first rewrite the action, constituted of the type IIB action and the source action:
\beq
	S = S_{IIB} + S_{\text{sources}}\,\,,
\eeq
where:
\beq
		S_{IIB} = \frac{1}{2\kappa_{10}^2} \left[ \int \sqrt{-g} \left( R - \frac{1}{2} \partial_{\mu} \Phi \partial^{\mu} \Phi \right) - \frac{1}{2} \int \left( e^{\Phi} F_{(3)} \wedge *F_{(3)} \right) \right]\,\,,
\eeq
and:
\beq
	S_{\text{sources}} = - T_5 \int \Big(e^{\Phi/2} \cK - C_{(6)}  \Big) \wedge \Omega\,\,,
\eeq
with $\cK$ being the calibration form for the D5-branes which is related to the $SU(3)$-structure as:
\beq
	\cK = e^{\Phi} \td^4 x \wedge J\,\,.
\eeq
First we give the modified Bianchi identities for the flux:
\beq
		\td F_{(3)} = 4 \pi^2 \Omega\,\,,
\eeq
where we used that $2\kappa_{10}^2 T_5 = 4\pi^2$.
The equation of motion for the flux reads:
\beq
		\td \left( e^{\Phi} * F_{(3)} \right) = 0\,\,.
\eeq
For the dilaton and the Einstein equations, we define first the following notation:
\beq
	\o_{(p)} \lrcorner \l_{(p)} = \frac{1}{p!} \o^{\mu_1 ... \mu_p} \l_{\mu_1 ... \mu_p}\,\,,
\eeq
for any two $p$-forms $\omega_{(p)}$ and $\lambda_{(p)}$. One can easily prove that, in a ten-dimensional manifold, one has:
\beq
	\int \o_{(p)} \wedge \l_{(10-p)} = - \int \sqrt{-g} \l \lrcorner (*\o)\,\,.
\eeq
Using these results, we can write the equation of motion  of the dilaton as:
\beq
	\frac{1}{\sqrt{-g}} \partial_{\mu} \left( \sqrt{-g} g^{\mu \nu} \partial_{\nu} \Phi \right) = \frac{1}{12} e^{\Phi} F_{(3)}^2 - 2 \pi^2 e^{\Phi/2} \Omega \lrcorner (* \cK)\,\,.
\eeq
Finally, the Einstein equation is:
\beq
		R_{\m \n} - \frac{1}{2} g_{\m \n} R = \frac{1}{2} \partial_{\m} \Phi \partial_{\n} \Phi -\frac{1}{4} g_{\m\n} \partial_{\r} \Phi \partial_{\r} \Phi +\frac{1}{24} e^{\Phi} \left( 6 F_{\m\r\s} F_{\n}^{\phantom{\n}\r\s} - g_{\m\n} F_{(3)}^2 \right) + T_{\m\n}^{\text{sources}}
		\,\,,
		\label{Einstein-eq}
\eeq
where $T_{\m\n}^{\text{sources}}$ is the energy-momentum tensor coming from the source action. It is given by:
\beq
	T_{\m\n}^{\text{sources}} = \frac{\pi^2}{3} e^{\Phi/2} \Big( 6 g_{\m\n}\Omega \lrcorner (*\cK) - \Omega_{\m \r_1 \r_2 \r_3} \left(* \cK \right)_{\n}^{\phantom{\n} \r_1 \r_2 \r_3} \Big)\,\,.
	\label{Tmu-nu}
\eeq
From (\ref{Einstein-eq}) and (\ref{Tmu-nu}), one can get an equation for the Ricci tensor as:
\beq
	\begin{aligned}
		R_{\m \n} = &\frac{1}{2} \partial_{\m} \Phi \partial_{\n} \Phi + \frac{1}{48} e^{\Phi} \left( 12 F_{\m \r \s} F_{\n}^{\phantom{\n} \r \s} - g_{\m \n} F_{(3)}^2 \right) \\
		&- \frac{\pi^2}{6} e^{\Phi/2} \big( 2 \Omega_{\m \r_1 \r_2 \r_3} \left(* \cK \right)_{\n}^{\phantom{\n} \r_1 \r_2 \r_3} -3 g_{\m \n} \Omega \lrcorner (*\cK) \big)\,\,.
	\end{aligned}
	\label{Ricci}
\eeq
To show more clearly the dependence of $T_{\m\n}^{\text{sources}}$ on the smearing profile of the flavor branes, we can write $T_{\m\n}^{\text{sources}}$ in flat components in the vielbein basis (\ref{vielbein}):
\beq
	\begin{aligned}
		T_{{x^i\,x^j}}^{\text{sources}} &= -\frac{N_f}{2} e^{-2g-2h-2k-\Phi/2}  \left( e^{2k} S + \frac{4e^{2h} +e^{2g} \left(a-\cosh(2r)\right)^2}{\sinh(4r)}S' \right)\eta_{ij}\,\,,
		\\[0.1in]
		T_{{rr}}^{\text{sources}} &= -\frac{N_f}{2} e^{-2g-2h-\Phi/2} S = T_{{33}}^{\text{sources}}\,\,, \\[0.1in]
		T_{{\theta\theta}}^{\text{sources}} &= -\frac{N_f}{\sinh(4r)} e^{-2g-2k-\Phi/2} S' = T_{{\phi\phi}}^{\text{sources}}\,\,, \\[0.1in]
		T_{{11}}^{\text{sources}} &= -\frac{N_f}{2} e^{-2g-2h-2k-\Phi/2}  \frac{2e^{2h} +e^{2g} \left(a-\cosh(2r)\right)^2}{\sinh(4r)}S'  = T_{{22}}^{\text{sources}}\,\,, \\[0.1in]
		T_{{\theta1}}^{\text{sources}} &= \frac{N_f}{2} e^{-g-h-2k-\Phi/2} \frac{a-\cosh(2r)}{\sinh(4r)} S' = T_{{\phi2}}^{\text{sources}}\,\,.
	\end{aligned}
	\label{emtensor-components}
\eeq
It is clear from (\ref{Ricci}) and (\ref{emtensor-components}) that to have a geometry with regular Ricci tensor one must require that, as stated in the main text, both $S$ and $S'$ are continuous functions of the radial variable. Moreover, it is also possible to verify that any solution of the first-order BPS system written in (\ref{BPS-new}) and (\ref{constraints}) is also a solution of the second-order equations of motion for supergravity plus sources written above.

\vskip 1cm
\renewcommand{\theequation}{\rm{C}.\arabic{equation}}
\setcounter{equation}{0}

\section{Microscopic computation of $\Omega$}
\label{micro}

Let us show here how the results obtained for the smearing form in section \ref{simple-embeddings}, using just the knowledge of one embedding of the family plus an ansatz for the functional form of $\Omega$ (eq. (\ref{Omega-S(r)})), could be derived from a purely microscopic computation: {\it i.e.}, by summing the contributions to the smearing form of all the embeddings in a given family.

Notice that this microscopic approach does not assume any specific ansatz for the smearing form. Obviously, one can expect it to be much harder to be carried out. Indeed, except for some very simple cases (\cite{smearingmethod,Bigazzi:2008ie,Bigazzi:2008qq,Bigazzi:2009bk}), a full reconstruction of the functional form of $\Omega$ for massive quarks from the microscopic family of embeddings giving rise to it is not available in the literature. The use of the holomorphic structure of our internal manifold developed in section \ref{holomorphic} will be instrumental  to carry out this microscopic computation.

\subsection{Holomorphic structure in the abelian limit}

For simplicity, we focus in this appendix on the UV limit ($r\to\infty$) of our backgrounds. This limit corresponds to the so-called abelian solution (see the last paragraph of section \ref{unflavoredMN}). The holomorphic structure simplifies a little bit in this limit, and one can define a new set of four complex variables $\z_i$ ($i=1,\ldots, 4$) that parameterize now a singular conifold:
\beq
\z_1\,\z_2-\z_3\,\z_4=0\,.
\label{conifold-sing}
\eeq
The radial variable $r$ is related to the $\z_i$ in this case as:
\beq
\sum_{i=1}^{4}\,|\,\z_i\,|^2\,=\,e^{2r}\,\,.
\label{sing.conifold-radial}
\eeq
The expression of these complex variables in terms of the coordinates of the internal manifold can be read from \eqref{eqn:MNz}. One just needs to take the $r\to\infty$ limit there, to obtain:
\beq
\begin{aligned}
\z_1&=-e^r\sin\frac{\th}{2}\,\sin\frac{\tth}{2}\,e^{i\frac{\j-\f-\tf}{2}}\,,&\z_2=e^r\cos\frac{\th}{2}\,\cos\frac{\tth}{2}\,e^{i\frac{\j+\f+\tf}{2}}\,,\\\\
\z_3&=e^r\cos\frac{\th}{2}\,\sin\frac{\tth}{2}\,e^{i\frac{\j+\f-\tf}{2}}\,,&\z_4=-e^r\sin\frac{\th}{2}\,\cos\frac{\tth}{2}\,e^{i\frac{\j-\f+\tf}{2}}\,.
\end{aligned}
\label{eqn:abelian.zs}
\eeq
This abelian geometry inherits the  $SU(2)_L\times SU(2)_R$ symmetry of the non-abelian one (actually the isometry group is enlarged to $SU(2)_L\times SU(2)_R\times U(1)$). Again, taking carefully\footnote{In the abelian limit: $a\to0$, $\cosh 2r\to\sinh 2r\to\frac{e^{2r}}{2}$, and $a\,e^{2r}\to1+4e^{2h-2g}$.} the limit $r\to\infty$ in the non-abelian expressions (\ref{J}) and (\ref{Omega-S(r)}) for the fundamental two-form $J$ and the smearing form $\Omega$ we get:
\begin{align}
&e^{-{\Phi\over 2}}\,\,J=\frac{e^{2k}}{2}\td r\wedge\left(\tw_3+\cos\th\,\td\f\right)-\frac{e^{2g}}{4}\sin\tth\,\td\tth\wedge\td\tf-e^{2h}\sin\th\,\td\th\wedge\td\f\,,\rc\rc
&\frac{16\p^2}{N_f}\Omega=\,\sin\theta\,\td\theta\wedge\td\phi\wedge\left(S\sin\tilde{\theta}\,\td\tilde{\theta}\wedge\td\tilde{\phi}-S'\td r\wedge\left(\td\psi + \cos\tilde{\theta}\,\td\tilde{\phi}\right)\right)\,\,.
\label{eqn:ab.Omega}
\end{align}
Then one can define $SO(4)$-invariant (1,1)-forms $\eta_i$ ($i=1,\ldots,4$) as in (\ref{eqn:etas}), and express both $J$ and $\Omega$ in this abelian setup as:
\begin{align}
&e^{-{\Phi\over 2}}\,J=2i\,e^{-2r}\left[e^{2h}\left(\eta_1+2\,e^{-2r}\eta_2-2\,e^{-2r}\eta_3\right)+\frac{e^{2g}}{4}\left(\eta_1+2\,e^{-2r}\eta_2+2\,e^{-2r}\eta_3\right)-e^{2k}e^{-2r}\eta_2\right],\rc
&\frac{16\p^2}{N_f}\,\Omega=-8e^{-4r}S\,\h_1\wedge\left(\h_1+4e^{-2r}\h_2\right)+8e^{-6r}S'\,\h_2\wedge\left(\h_1-2e^{-2r}\h_3\right)\,\,,
\end{align}
where the $\eta$'s are the abelian $(1,1)$ two-forms, which can be obtained from (\ref{etas-angles}) by keeping the leading term when $r\to\infty$.

\subsection{Abelian limit of the simple class of embeddings}
\label{ssec:micro.good}

Let us now calculate $S(r)$ for the abelian version of the class of embeddings discussed in section \ref{simple-embeddings}. The first thing to notice is that the parameterization (\ref{eqn:the.emb}) is not good in the UV limit.  Indeed, as $z_4=z_1 z_2/z_3$  when  $r\to\infty$, the two equations in (\ref{eqn:the.emb}) become the same. For this reason, to study the cycle in the UV, it is better to use instead the first two equations in (\ref{complex-2-surface})  and write the equation of the embedding as $z_1=C z_3$ and $z_1 z_2=\tilde \mu$, with $C$ and $\tilde{\m}$ being arbitrary complex constants. By taking the UV limit in which $z_i\to\z_i$, one concludes that 
the abelian limit of the particular representative of the embedding studied in section \ref{simple-embeddings} is:
\beq
\z_1=C\,\z_3\,,\qquad\qquad\z_1\,\z_2=\tilde{\m}\,.
\label{eqn:ab.nice.emb}
\eeq
 One nice thing of the abelian limit is that \eqref{eqn:ab.nice.emb} can be easily solved in terms of coordinates:
\beq
\th=\th_0\,,\quad\f=\f_0\,,\qquad\textrm{and}\qquad \frac{1}{2}\sin\tth\,e^{2r}e^{i\j}=\m\equiv\frac{1}{2}e^{2r_q}e^{i\g}\,,
\label{abelian-emb-coordinates}
\eeq
where we have parameterized the constants above as $C=\tan\frac{\th_0}{2}\,e^{-i\f_0}$, $\tilde{\m}=\m\left(\sin\frac{\th_0}{2}\,\cos\frac{\th_0}{2}\right)^{-1}$, and $r_q$ is the minimum radial distance this embedding reaches ($e^{2r_q}=|2\mu|$). If we now rotate this embedding with the $SU(2)_L\times SU(2)_R$ isometry group (see \eqref{eqn:isometry.action}), we obtain the expression of a generic embedding of the family as $f_1=0$ and $f_2=0$, with:
\begin{align}
f_1&=\z_1-\frac{b+a\,C}{\bar{a}-\bar{b}\,C}\,\z_3\,,\rc\rc
f_2&=\left((|k|^2-|l|^2)\z_1\,\z_2-k\,\bar{l}\,C^{-1}\,\z_1^2+\bar{k}\,l\,C\,\z_2^2\right)-\left(|a|^2-|b|^2-a\,\bar{b}\,C+\bar{a}\,b\,\bar{C}\right)^{-1}\tilde{\m}\,.\label{eqn:ab.emb.f12}
\end{align}
The smearing form should be computed as an appropriately weighted sum of the transverse volume forms of each embedding. The formula for real constraints was first written down in \cite{smearingmethod}, and the generalization to complex constraints like the ones we have now is immediate\footnote{The complex Dirac delta should be understood as $\delta^{(2)}(f)=\delta({\rm Re} (f))\,\delta({\rm Im} (f))$. The $\frac{1}{-2i}$ prefactor is included because $\td f\wedge\td\bar{f}=-2i\,\td{\rm Re}(f)\wedge\td{\rm Im}(f)$.}:
\beq
\Omega=\frac{1}{(-2i)^2}\int_{\mathbb{C}^4}\td\r\,\delta^{(2)}\left(f_1\right)\delta^{(2)}\left(f_2\right)\td f_1\wedge\td\bar{f_1}\wedge\td f_2\wedge\td\bar{f_2}\,,
\label{eqn:full.micro.ab.O}
\eeq
where $\r$ is the (normalized to the unity) measure of $SU(2)_L\times SU(2)_R$, multiplied by $N_f$, and is given by:
\beq
\td\r=\td a\,\td \bar{a}\,\td b\,\td \bar{b}\,\td k\,\td \bar{k}\,\td l\,\td \bar{l}\,\delta\left(|a|^2+|b|^2-1\right)\delta\left(|k|^2+|l|^2-1\right)\,\frac{N_f}{16\p^4}\,,
\eeq
A shortcut for computing \eqref{eqn:full.micro.ab.O} is to notice that all the embeddings of the present family, in virtue of the first equation in \eqref{eqn:ab.emb.f12}, sit at constant values of $\th$ and $\f$. Since it turns out that the action of $SU(2)_L$ corresponds precisely to varying these constant values over a two-sphere, the smearing form $\Omega$ necessarily exhibits a $\frac{1}{4\p}\sin\th\,\td\th\wedge\td\f$ factor. We are not interested in getting this trivial part from \eqref{eqn:full.micro.ab.O}, so we factor it out by defining an effective (complex) two-dimensional problem. We can define a new pair of effective complex variables:
\beq
\x_1=e^r\cos\frac{\tth}{2}\,e^{i\frac{\j+\tf}{2}}\,,\qquad\qquad\x_2=e^r\sin\frac{\tth}{2}\,e^{i\frac{\j-\tf}{2}}\,,
\label{eqn:effective.zs}
\eeq
and the family of embeddings over which we want to smear recasts as
\beq
	f\equiv (|A|^2 - |B|^2)\,\x_1\,\x_2 +A\bar{B}\,\x_2^2-\bar{A}B\,\x_1^2-\m=0\,,\qquad |A|^2+|B|^2=1\,.
\eeq
(Recall that $\m = \frac{1}{2} e^{2r_q} e^{i \g}$, see (\ref{abelian-emb-coordinates})). Forgetting for the moment about the correct normalization factors, the integral we want to compute is:
\beq
	W\equiv \int_{\mathbb{C}^2} \td A\, \td \bar{A}\, \td B\, \td \bar{B} \, \d( |A|^2 +|B|^2 -1) \d^{(2)} (f)\, \td f \wedge \td \bar{f}\,.
\eeq
Performing this integral requires a little bit of care with the delta functions, but other than that, it can be considered straightforward. Let us sketch how one could proceed.
To simplify the calculation, we reparameterize the integration variables as follows:
\beq
	A = \sqrt{\frac{u_1+u_2}{2}}\, e^{i \a_1} \,, \,\,\,\,\, B = \sqrt{\frac{u_1-u_2}{2}}\, e^{i \a_2}\,.
\eeq
Clearly, one has $|A|^2 + |B|^2 = u_1$ and $|A|^2 - |B|^2 = u_2$ and:
\beq
	\int_{\mathbb{C}^2} \td A \td \bar{A} \td B \td \bar{B}\, \d( |A|^2 +|B|^2 -1) =\frac{1}{2} \int_{-\infty}^\infty \td u_2 \int_{|u_2|}^\infty \td u_1 \int_0^{2\p} \td \a_1 \int_0^{2\p} \td \a_2\, \d(u_1 -1)\,.
\eeq
The integral in $u_1$ is then immediate. Rewriting $e^{i\a_2} = x_2 + i y_2$, and using that:
\beq
	\int_0^{2\p} \td \a_2 = 2 \int_{\mathbb{R}^2} \td x_2 \td y_2 \,\d(x_2^2 + y_2^2 -1)\,,
\eeq
we can write:
\beq
	W =\int_0^{2\p} \td \a_1 \int_{-1}^1 \td u_2 \int_{\mathbb{R}^2} \td x_2 \td y_2 \,\d(x_2^2 + y_2^2 -1) \d(R) \d(I) \, \td f \wedge \td \bar{f}\,,
\eeq
where $R\equiv{\rm Re}\left(f|_{u_1=1}\right)$, $I\equiv{\rm Im}\left(f|_{u_1=1}\right)$. In the new variables one has:
\beq
	f|_{u_1=1}= -\frac{1}{2} e^{i (\a_1-\a_2)} \sqrt{1 - u_2^2}\; \x_1^2 +u_2\,\x_1\,\x_2 + \frac{1}{2} e^{i(-\a_1+\a_2)} \sqrt{1-u_2^2}\; \x_2^2 - \m\,.
\eeq
Solving $R=I=0$ for $x_2$ and $y_2$, the integral of the corresponding deltas produces a factor:
\beq
	\Bigg|\frac{\td R}{\td x_2} \frac{\td I}{\td y_2}\Bigg|^{-1} = \frac{4}{1-u_2^2}\,\, \frac{1}{\big||\x_1|^4 - |\x_2|^4 \big|}\,,
\eeq
and leaves the argument of the remaining delta function as:
\beq
	\d(x_2^2+y_2^2 -1) = \d\left(\frac{(|\x_1|^2 + |\x_2|^2)^2}{(1-u_2^2)(|\x_1|^2 -|\x_2|^2)^2} (u_2 - u_{2+})(u_2 - u_{2-}) \right)\,,
\eeq
where $u_{2\pm} $ are given by:
\beq
	u_{2\pm} = \frac{4 |\m\, \x_1\, \x_2| \cos(\j-\g) \pm \big||\x_1|^2 - |\x_2|^2\big| \sqrt{(|\x_1|^2 + |\x_2|^2)^2 - 4 |\m|^2}}{(|\x_1|^2 + |\x_2|^2)^2}\,.
	\label{eqn:u2pm}
\eeq
We have at this point:
\beq
	W =\int_0^{2\p} \td \a_1 \int_{-1}^1 \td u_2  \frac{\frac{4}{1-u_2^2}}{\big||\x_1|^4 - |\x_2|^4 \big|}  \d\left(\frac{(|\x_1|^2 + |\x_2|^2)^2}{(1-u_2^2)(|\x_1|^2 -|\x_2|^2)^2} (u_2 - u_{2+})(u_2 - u_{2-}) \right) \td f \wedge \td \bar{f}\,.
\eeq
In this expression, nothing depends on $\a_1$, so one can integrate it easily. Also, both $u_{2+}$ and $u_{2-}$ are between $-1$ and $1$, so they both contribute to the integral. Using \eqref{sing.conifold-radial}, and replacing $u_{2+}$ and $u_{2-}$ by their values \eqref{eqn:u2pm}, we finally get:
\beq
	\begin{aligned}
		W &= 4\p\frac{1-\cos \tth +2 e^{4r_q-4r} \cos \tth}{\sqrt{e^{4r}- e^{4r_q}}} \td \x_1 \wedge \td \bar{\x}_1 +4\p \frac{1+\cos \tth -2 e^{4r_q-4r} \cos \tth}{\sqrt{e^{4r}- e^{4r_q}}} \td \x_2 \wedge \td \bar{\x}_2 \,-\,\\\\
		&\quad- 4\p e^{-i \tf}\, \frac{(1-2e^{4r_q-4r})\sin \tth}{\sqrt{e^{4r}- e^{4r_q}}} (\td \x_1 \wedge \td \bar{\x}_2 + \td \x_2 \wedge \td \bar{\x}_1)\,.
	\end{aligned}
\eeq
Plugging the values of $\x_1$ and $\x_2$ in \eqref{eqn:effective.zs}, and taking into account the proper normalization factors, we find exactly:
\beq
	\frac{N_f}{4\p i}\sin\th\,\td\th\wedge\td\f \wedge W = 16 \p^2\,\Omega\,,
\eeq
where $\Omega$ is the one written in \eqref{eqn:ab.Omega} with the following function 
$S(r)$:
\beq
S(r)\,=\,\sqrt{1\,-\,e^{4r-4r_q}}\,\,\Theta(r-r_q)
\,\,.
\label{abelian-profile}
\eeq
Notice that (\ref{abelian-profile}) is the limit of the function $S(r)$ written in (\ref{S-wall}) when $r$ and $r_q$ are large. This confirms our results of section \ref{simple-embeddings}.

\subsection{An example of a non-compatible embedding}

As we saw in the previous subsection, one has to work quite hard in order to obtain the smearing form $\Omega$ from the microscopic average over a family of embeddings. Certainly, the trick described in section \ref{charge-distributions} gives a much faster and simpler way to get $\Omega$. One can wonder nevertheless about the reliability of the trick, since it assumes a given functional form of $\Omega$, and the only unknown is the radial profile of the brane distribution $S(r)$.

In principle the trick can be run for any representative embedding. However, it is hard to think that any given family of embeddings, even if supersymmetric, generates (when we place flavor branes along the embeddings of the family) a backreaction of the metric that is compatible with the initial ansatz we assumed for this metric, if this is not the most general possible. It seems nonetheless that the trick is able to detect this ``compatibility property'', and we present in what follows some arguments in favor of that.

Recalling the discussion in section \ref{charge-distributions}, the trick was to compute the effective radial action of the smeared brane distribution, and to compare it with $N_f$ times the WZ effective radial action of a single brane sitting at one of the embeddings of the family over which we smear. Both actions should be equal. The smeared action always contains two terms, one proportional to $S(r)$, and another proportional to $S'(r)$:
\beq
{\cal L}^{\textrm{smeared}}_{WZ}=F_1(r)\,S(r)+F_2(r)\,S'(r)\,,
\label{smeared-compatibility}
\eeq
where $F_1$ and $F_2$ depend on the functions of the ansatz.
We conjecture that the way to detect if a family of embeddings generates a backreaction compatible with the ansatz  is to take any representative embedding of this family and to compute its WZ effective radial action. We must then check whether or not the result depends functionally on the functions of the ansatz as in (\ref{smeared-compatibility}). Let us assume that this is the case and that the effective WZ radial lagrangian density for the representative embedding is of the form:
\beq
{\cal L}^{\textrm{single}}_{WZ}=F_1(r)\,G(r)+F_2(r)\,H(r)\,,
\label{single-compatibility}
\eeq
where $F_1$ and $F_2$ are the same as in (\ref{smeared-compatibility}) and $G(r)$ and $H(r)$ are functions of $r$ which do not depend on the functions of the ansatz. In order to verify that (\ref{single-compatibility}) is of the form 
(\ref{smeared-compatibility}) one must check that:
\beq
\frac{\td G(r)}{\td r}=H(r)\,.
\eeq
If this were the case, we conjecture that the backreaction is compatible with the ansatz and, furthermore, that the profile function $S$ is proportional to $G$.

In section \ref{simple-embeddings}, we have worked out one example in which the compatibility condition was satisfied (see also appendix \ref{KS}). Moreover, in subsection \ref{ssec:micro.good} above, we have checked explicitly with an independent calculation of $\Omega$ that the trick gives the right result. In what follows, let us illustrate with an example the case in which the compatibility condition is not satisfied, and show with a microscopic calculation that indeed the resulting $\Omega$ is incompatible with the ansatz for it.

We choose to work again in the abelian background, since it is simpler and therefore the explanation will be cleaner. Let us focus on the following embedding:
\beq
\z_1=C\,\z_4\,,\qquad\qquad\z_2=\m\,,
\label{eqn:ab.bad.emb}
\eeq
where the $\z_i$'s are the complex coordinates (\ref{eqn:abelian.zs}) and
$C$ and $\m$ are constants that we parameterize as $C=\tan\frac{\tth_0}{2}\,e^{-i\tf_0}$ and $\m=\cos\frac{\tth_0}{2}\,e^{i\tf_0/2}\,e^{i\b}$. We can solve the embedding equations in (\ref{eqn:ab.bad.emb})  in terms of coordinates as:
\beq
\tth=\tth_0\,,\quad\tf=\tf_0\,,\qquad\textrm{and}\qquad e^{r}\,\cos\frac{\th}{2}=e^{r_q}\,,\quad\j+\f=2\b\,.
\eeq
It is easy then to compute the effective radial lagrangians of the smeared distribution and of a single brane extended along the embedding \eqref{eqn:ab.bad.emb}, with the result:
\begin{align}
{\cal L}^{\textrm{smeared}}_{WZ}&=2\p\, N_f\,T_{D5}\,e^{2\Phi}\left(e^{2k}\,S+\frac{e^{2g}}{2}\,S'\right)\,,\rc\rc
{\cal L}^{\textrm{single}}_{WZ}&=2\p\, T_{D5}\,e^{2\Phi}\left(e^{2k}\left(1-e^{2r_q-2r}\right)+4e^{2h}\,e^{2r_q-2r}\right)\,,\label{eqn:L.2}
\end{align}
where we have assumed that $\Omega$ should be as in \eqref{eqn:ab.Omega}. As we see, the $e^{2g}$ term in the smeared lagrangian is not present in ${\cal L}^{\textrm{single}}_{WZ}$ (we have instead an $e^{2h}$ term), so the compatibility condition is not satisfied.

Let us now check with a microscopic computation that, indeed, the family of embeddings generated by rotating \eqref{eqn:ab.bad.emb} with the $SU(2)_L\times SU(2)_R$ symmetry, generates an $\Omega$ that is not of the form of \eqref{eqn:ab.Omega}. After using the relation \eqref{conifold-sing}, the family can be characterized by $f_1=0$, $f_2=0$ with:
\begin{align}
f_1&=\bar{a}\z_1+\bar{b}\,\z_4\,,\rc\rc
f_2&=\bar{k}\z_2+\bar{l}\z_4+\bar{b}\m\,,\label{eqn:bad.emb.f1}
\end{align}
with $|a|^2+|b|^2=1=|l|^2+|k|^2$. In this case, it is easy to perform the integral \eqref{eqn:full.micro.ab.O} by making use of the following two results:
\begin{align}
&\int\td z\,\td\bar{z}\,\delta^{(2)}\left(w_1\,z-w_2\right)=-\frac{2i}{|w_1|^2}\,,\label{eqn:int1}\\
&\int\td x\,\td y\,\left(x^2+y^2+\a_1\,x+\b_1\,y+\g_1\right)\,\delta\left(x^2+y^2+\a_2\,x+\b_2\,y+\g_2\right)=\nonumber\\
&=\p\left(\g_1-\g_2+\frac{\a_2^2+\b_2^2-\a_1\a_2-\b_1\b_2}{2}\right)\,.
\label{eqn:int2}
\end{align}
The final result we get for the smearing form is:
\beq
\Omega=\frac{N_f}{16\p^2}\,\sin\tth\,\td\tth\wedge\td\tf\wedge\left(\left(1-e^{2r_q-2r}\right)\sin\th\,\td\th\wedge\td\f-2e^{2r_q-2r}\td r\wedge\left(\td\psi + \cos\th\,\td\f\right)\right)\,,
\eeq
and we see that this is clearly incompatible with \eqref{eqn:ab.Omega} (the roles of 
$(\theta,\phi)$ and $(\tilde\theta,\tilde\phi)$ are exchanged in these two expressions of
$\Omega$).

\renewcommand{\theequation}{\rm{D}.\arabic{equation}}
\setcounter{equation}{0}

\section{Revisiting the massive KS background}
\label{KS}

The flavored version of the Klebanov-Strassler (KS) background \cite{ks}, corresponding to a system of D3- and D7- branes on the deformed conifold, 
was obtained in ref. \cite{Benini:2007gx} for the case of massless flavors. This geometry was generalized in \cite{Bigazzi:2008qq} to the case in which the flavors are massive. In reference \cite{Bigazzi:2008qq}, in particular, a microscopic calculation of the D7-brane charge density was presented. In this appendix we will revisit this problem by applying  the techniques developed in the present paper and we will reproduce and generalize the results found in \cite{Bigazzi:2008qq}.

The ten-dimensional metric in Einstein frame of flavored KS mode is given by:
\beq
\td s^2_{10}\,=\,h^{-{1\over 2}}\,\,\td x_{1,3}\,+\,h^{{1\over 2}}\,\,\td s^2_6\,\,,
\eeq
where $h$ is a warp factor and $\td s^2_6$ is the internal six-dimensional metric of the flavored deformed conifold \cite{Benini:2007gx}. In order to write explicitly this metric, let us introduce the five one-forms $g^i$  $(i=i\ldots 5)$ as:
\bear
g^1 &=& \frac{-\sin\theta\,\td\phi -\cos\psi\sin\tilde\theta\,\td\tilde\phi +\sin\psi\,\td\tilde\theta}{\sqrt{2}}\,,\quad
g^2 = \frac{\td\theta-\sin\psi\sin\tilde\theta\,\td\tilde\phi -\cos\psi\,\td\tilde\theta}{\sqrt{2}}\,,
\rc\rc
g^3 &=& \frac{-\sin\theta\,\td\phi +\cos\psi\sin\tilde\theta\,\td\tilde\phi-\sin\psi\,\td\tilde\theta}{\sqrt{2}}\,,\quad
g^4 = \frac{\td\theta+\sin\psi\sin\tilde\theta\,\td\tilde\phi +\cos\psi\,\td\tilde\theta}{\sqrt{2}}\,,\rc\rc
g^5&=& \td\psi +\cos\theta\,\td\phi + \cos\tilde\theta\,\td\tilde\phi\,.
\label{gis}
\eear
Then,  $\td s^2_6$ can be written as:
\beq
\td s^2_6\,=\,\Gamma(r)\,e^{2G_1(r)}\,\Big[\,
(g^1)^2\,+\,(g^2)^2\,\Big]\,+\,{e^{2G_2(r)}\over \Gamma(r)}\,
\Big[\,(g^3)^2\,+\,(g^4)^2\,\Big]\,+\,
{e^{2G_3(r)}\over 9}\,\Big[\,(dr)^2\,+\,(g^5)^2\,\Big]\,\,,
\eeq
where the functions $G_i(r)$ satisfy a system of first-order BPS equations and
$\Gamma(r)$ is the function:
\beq
\Gamma(r)\,=\,{\cosh(r)\over \cosh(r)+1}\,\,.
\eeq
This background has a running dilaton $\Phi(r)$, a NSNS  field $H_{(3)}=\td B_{(2)}$ and is endowed with RR one-, three- and five-forms $F_{(1)}$, $F_{(3)}$ and $F_{(5)}$.  We refer to the article \cite{Bigazzi:2008qq} for a complete account of this massive flavored solution. Here we will only need the expression of $B_{(2)}$ and $F_{(1)}$, which are given by:
\beq
B_{(2 )}\,=\, \frac{M}{2} \Bigl[ f\, g^1 \wedge g^2\,+\,k\, g^3 \wedge 
g^4 \Bigr]\,,\qquad\qquad
F_{(1)}={N_f\,S(r)\over 4\pi}\,\,g^5\,,
\eeq
where $M$ is the  fractional D3-brane Page charge, $S(r)$ is the D7-brane charge density function (similar to the one used in the main text for the model dual to ${\cal N}=1$ SYM) and $f$ and $g$ are determined in terms of the dilaton $\Phi$ as follows:
\beq
f\,=\, e^{\Phi}\,\,\frac{r \coth r -1}{2 \sinh r} (\cosh r -1)\,\,,\qquad\qquad
k= e^{\Phi}\,\,\frac{r \coth r-1}{2 \sinh r} (\cosh r +1)\,\,.
\label{fkexplicit}
\eeq
Notice that $\td F_{(1)}=- \Omega$, where $\Omega$ is the 
symmetry preserving D7-brane density distribution two-form, which is given by:
\begin{equation}
\Omega =\frac{N_f}{4\pi}\Big[\,S(r)\,\big(\,g^1\wedge g^4\,-\,g^2\wedge g^3\,\big)\,
- \,S'(r) \,\td r\wedge g^5\,\Big]\,.
\label{massive_2form}
\end{equation}
The full background can be determined if the profile function $S(r)$ is known (see section 3.3 in \cite{Bigazzi:2008qq}). In what follows we will show how one can find $S(r)$ by applying the technique developed in sections \ref{charge-distributions}-\ref{Smoothing} of the main text. In particular, we will be able to reproduce and simplify the results written in eqs. (2.14) and (2.15) of ref. \cite{Bigazzi:2008qq} by using a holomorphic formulation similar to the one employed in the main text. 

The K\"ahler form of the transverse 6d manifold can be written as \cite{Bigazzi:2008qq}:
\beq
h^{-{1\over 2}}J\,=\,e^{G_1+G_2}\,\Big(\,g^1\wedge g^4\,-\,g^2\wedge g^3\,\Big)\,-\,
{e^{G_3}\over 3}\,\td r\wedge g^5\,\,.
\label{J-flavoredconifold}
\eeq
 The holomorphic variables $z_1, z_2, z_3, z_4$ that  we will use in this case are just the same as those defined in (\ref{eqn:MNz}) with the substitution $r\to r/2$.  Similarly, 
the $SO(4)$-invariant (1,1)-forms  $\eta_1$, $\eta_2$ and $\eta_3$ are defined in (\ref{eqn:etas}). Their expressions in terms of $r$ and the angles are just the ones written in (\ref{etas-angles}) with the changes $r\to r/2$ and $\td r\to \td r/2$.  One can easily show  that it is possible to express $J$ and $B_2$ in terms of these forms as:
\bear
J&=&-2i\frac{\sqrt{h}}{\sinh r}\left(\eta_1\,e^{G_1+G_2}+\eta_2\left(\frac{\cosh r}{\sinh^2 r}e^{G_1+G_2}-\frac{1}{9}\frac{e^{2G_3}}{\sinh r}\right)\right)\,,\rc\rc
B_{(2)}&=&i\,M\,e^{\Phi}\,\frac{r\,\coth r\,-1}{\sinh^2 r}\,\eta_3\,.
\label{J-B2-KS}
\eear
The smearing form $\Omega$ in this case, written in  (\ref{massive_2form}), can also be expressed in terms of the $\eta_i$ forms, namely:
\beq
\Omega\,=\,-{i N_f\over 2\pi}\,\,{S(r)\over \sinh r}\,\,
\Big(\,\eta_1\,+\,{\cosh r\over \sinh^2 r}\,\eta_2\,\Big)\,+\,
{i N_f\over 2\pi}\,{S'(r)\over \sinh^2 r}\,\,\eta_2\,\,.
\eeq
The D7-branes wrap a non-compact four-cycle on the transverse space. For a supersymmetric configuration the corresponding eight-dimensional worldvolume is calibrated by the eight-form:
\beq
{\cal K}\,=\,{\td^4 x \over 2h}\,\,\wedge\big[\,
J\wedge J\,-\,e^{-\Phi}\,\,B_{(2)}\wedge B_{(2)}\,\big]\,. 
\eeq
It follows from (\ref{J-B2-KS}) that ${\cal K}$  will also be expressed in terms of the $SO(4)$-invariant (1,1)-forms $\eta_i$ , namely: 
\beq
{\cal K}=\sum C_{ij}(r)\,\td^4x\,\wedge\,\h_i\wedge\h_j\,\,,
\eeq
with $C_{ij}$ being the functions:
\beq
\begin{aligned}
C_{11}&=-\frac{2\,e^{2G_1+2G_2}}{\sinh^2 r}\,,\\[0.1in]
C_{12}&=-\frac{4\,e^{G_1+G_2}}{\sinh^3 r}\left(\coth r\,e^{G_1+G_2}-\frac{1}{9}e^{2G_3}\right)\,,\\[0.1in]
C_{22}&=-\frac{2\,e^{G_1+G_2}}{\sinh^4 r}\left(\coth r\,e^{G_1+G_2}-\frac{1}{9}e^{2G_3}\right)^2\,,\\[0.1in]
C_{33}&=M^2\frac{e^{\Phi}}{2h}\frac{(r\,\coth r\,-1)^2}{\sinh^4r}\,.
\end{aligned}
\label{Cij}
\eeq
The integration of the calibration form over the D7-brane worldvolume will basically give the WZ  term of the action. Our method  to compute $S(r)$ amounts to compare this term of the action for a smeared set of equivalent flavor branes with the same quantity evaluated for a single brane of the set. The former is given in terms of ${\cal K}$ and $\Omega$ as:
\beq
S^{\textrm{smeared}}_{WZ}=T_7\,\int\,e^{\Phi}\,{\cal K}\wedge\Omega\,\,,
\label{WZ-smeared-KS}
\eeq
while the latter is just:
\beq
S^{\textrm{single}}_{WZ}=T_7\,\int_{{\cal M}_8}\,
\pbi{e^{\Phi}\,{\cal K}}\,\,.
\eeq
Clearly, if the total number of flavor branes is $N_f$, one should have:
\beq
S^{\textrm{smeared}}_{WZ}\,=\,N_f\,S^{\textrm{single}}_{WZ}\,\,.
\label{micro-macro}
\eeq
Let us show how (\ref{micro-macro}) can be used to determine $S(r)$. 
In order to compute the WZ term of the smeared action (\ref{WZ-smeared-KS}), 
let us define the functions $F_1$ and $F_2$ as follows:
\beq
F_1(r)={8\over 9}\,e^{G_1+G_2+2G_3}\,,\qquad F_2(r)={M^2e^{\Phi}\over h}\,\,
\left(r\,\coth r\,-1\right)^2\,+\,4\,e^{2G_1+2G_2}\,.
\label{F12}
\eeq
Then, one can check by direct calculation that:
\beq
e^{\Phi}\,{\cal K}\wedge\Omega\,=\,-{N_f\over 16\pi}\,e^{\Phi}\,
\big[\,F_1\,S\,+\,F_2\,S'\,\big]\,\td^4x\,\wedge\,\td r\,\wedge \td^5\Theta\,\,,
\eeq
where $\td^5\Theta$ is the angular five-form:
\beq
\td^5\Theta\,=\,\sin\theta\sin\tilde\theta\,\td\theta\wedge \td\phi\wedge 
\td\tilde\theta\wedge \td\tilde\phi\wedge \td\psi\,\,.
\eeq
Using these results in (\ref{WZ-smeared-KS}), we obtain after integrating over the angular coordinates:
\beq
S^{\textrm{smeared}}_{WZ}\,=\,-4\pi^2\,N_f\,T_7\,\int\,
\td^4x\,\td r\,e^{\Phi}\,\big[\,F_1\,S\,+\,F_2\,S'\,\big]\,\,.
\label{S-smeared}
\eeq

The next step is then to compute the pullback of the calibration form for one of the embeddings within the family used for the smearing in \cite{Bigazzi:2008qq}. Let us consider the following particular representative embedding \cite{kuper}:
\beq
z_1-z_2=2\hat{\m}=2\m\,e^{i\a}\,,
\label{eqn:Kup.emb}
\eeq
with $\m$ being the modulus of $\hat\m$, and $\a$ its phase. This embedding defines a four-dimensional submanifold inside the deformed conifold. Accordingly, we choose the complex coordinates $\tz_2$ and $z_3$ to parameterize it, where $\tz_2=z_2+\hat{\m}$:
\beq
z_1=\tz_2+\hat\m\,,\qquad\qquad z_4=\frac{\tz_2^2-1-\hat\m^2}{z_3}\,.
\eeq
The relation between $r$ and the complex coordinates follows from:
\beq \label{eq:rInTermsOfz}
2\cosh r=\sum |z_i|^2=2|\tz_2|^2+2\m^2+|z_3|^2+\frac{|\tz_2^2-1-\hat\m^2|^2}{|z_3|^2}\,.
\eeq
It is easy to compute the minimum of this expression. Let us denote by $r_q$ the minimal value of $r$. Then, one has:
\begin{multline}
2\cosh r_q=\min\left\{2|\tz_2|^2+2\m^2+\min\left\{|z_3|^2+\frac{|\tz_2^2-1-\hat\m^2|^2}{|z_3|^2}\right\}_{z_3}\right\}_{\tz_2}=\\\\
=\min\left\{2|\tz_2|^2+2\m^2+2|\tz_2^2-1-\hat\m^2|\right\}_{\tz_2}=2\left(\m^2+|1+\hat\m^2|\right)\,.
\label{eqn:minimum}
\end{multline}
Varying the phase of $\hat{\m}$, the minimum of this expression is always attained for a purely imaginary $\hat\m$. Notice that when $|\hat\m|<1$, it is possible to achieve $r_{q}=0$. This matches the discussion in \cite{Bigazzi:2008qq}. Moreover, for given $\hat\m$, solving for the phase of $\hat\m$ in \eqref{eqn:minimum}, we get:
\beq
\tan\a=\sqrt{\frac{\cosh r_{q}\,+1}{\cosh r_{q}\,-1}}\sqrt{\frac{1+2\m^2-\cosh r_{q}}{1-2\m^2+\cosh r_{q}}}\,,
\label{eqn:tan.alpha}
\eeq
which is exactly the expression (A.10) of the paper \cite{Bigazzi:2008qq} and tells us that actually the embeddings \eqref{eqn:Kup.emb} are inequivalent for different values of the phase of $\hat\m$.

In order to compute the pullback of ${\cal K}$ needed to calculate $S^{\textrm{single}}_{WZ}$, we have  to calculate the pullbacks of the 4-forms $\h_i\wedge\h_j$. We find:
\beq
\pbi{\h_i\wedge\h_j}=A_{ij}(r)\frac{1}{|z_3|^2}\td\tz_2\wedge\td \bar{\tz}_2\wedge\td z_3\wedge \td\bar{z}_3\,,
\eeq
where the only non-zero $A_{ij}^{\phantom{ij}'}s$ are:
\beq\begin{aligned}
A_{11}&=2\left(\cosh r\,-\m^2\right)\,,\\[0.1in]
A_{12}&=2\m^2\left(\cosh r\,+\cos2\a\right)-\sinh r\,,\\[0.1in]
A_{33}&=-2\m^2\left(1+\cosh^2r+2\cosh r\,\cos2\a\right)\,.
\label{Aij}
\end{aligned}
\eeq
Now we have all the information to carry out  our procedure. 
Let us write:
\beq
\pbi{e^{\Phi}\,{\cal K}}=\left(e^{\Phi}\,\sum C_{ij}(r)A_{ij}(r)\right)\,\,
\frac{1}{|z_3|^2}\,\,\td^4x\wedge
\td\tz_2\wedge\td \bar{\tz}_2\wedge\td z_3\wedge \td\bar{z}_3\,\,.
\label{pullback-K-KS}
\eeq
Let us first compute the terms  in parenthesis on the right-hand side of  (\ref{pullback-K-KS}). From the values of the $C_{ij}$'s written in (\ref{Cij}) and those of the $A_{ij}$'s displayed in  (\ref{Aij}) one readily gets:
\bear
&&e^{\Phi}\,\sum C_{ij}(r)A_{ij}(r)\,=-\,e^{\Phi}\left[\frac{1}{2\sinh^3 r}\left(\sinh^2 r\,-2\m^2\cosh r\,-2\m^2\cos2\a\right)
\,F_1\,
+\,\right.\rc\rc
&&\qquad\qquad\qquad\qquad\qquad\qquad
\left.\,+\,\m^2\frac{1+\cosh^2 r\,+2\cosh r\,\cos 2\a}{\sinh^4r}\,\,F_2
\,\,\right]\,\,,
\eear
where $F_1$ and $F_2$ are the functions defined in (\ref{F12}). This has a very suggestive form if we compare it with the result we obtained for the smeared action in (\ref{S-smeared}). In order to continue with our calculation of $S^{\textrm{single}}_{WZ}$ we use the fact that  
for any function $F(r)$ we have:
\beq
\int 
\frac{\td\tz_2\wedge\td \bar{\tz}_2\wedge\td z_3\wedge \td\bar{z}_3}{|z_3|^2}\,
F(r)\,=\,
\int_{r_q}^{\infty}\,\td r\,
{\cal J}(r)\,F(r)\,\,,
\label{z-integral}
\eeq
where the jacobian (which is computed below in subsection \ref{Jacobian-KS}) is given by:
\beq
{\cal J}(r)\,=\,8\pi^2\,\sinh r\,\,.
\label{jacobian}
\eeq
With these results we can write the action for a single embedding:
\bear
&&S^{\textrm{single}}_{WZ}\,=\,-4\pi^2\,T_f\,\int \td^4x \,\td r\,e^{\Phi}\,
\Big[\,{\sinh^2 r\,-2\m^2\cosh r\,-2\m^2\cos 2\alpha\over \sinh^2 r}
\,\,F_1\,+\,\qquad\qquad\qquad\rc\rc
&&\qquad\qquad\qquad\qquad\qquad\qquad\qquad\qquad
+\,2\m^2\,\frac{1+\cosh^2 r\,+2\cosh r\,\cos 2\a}{\sinh^3r}\,\,F_2\,\,\Big]\,\,.
\label{S-single}
\eear
By plugging  (\ref{S-smeared}) and  (\ref{S-single})  into  (\ref{micro-macro}), one immediately extracts the value of the profile function $S(r)$:
\beq
S(r)\,=\,{\sinh^2 r\,-2\m^2\cosh r\,-2\m^2\cos 2\alpha\over \sinh^2 r}\,\,\Theta(r-r_q)\,\,.
\label{S-KS-alpha}
\eeq
An important consistency check of this identification is  provided by the relation:
\beq
{\td\over \td r}\,\,\Bigg[\,
{\sinh^2 r\,-2\m^2\cosh r\,-2\m^2\cos2\alpha\over \sinh^2 r}\,\Bigg]\,=\,
2\m^2\,\frac{1+\cosh^2 r\,+2\cosh r\,\cos 2\a}{\sinh^3r}\,\,.
\eeq
Let us eliminate in (\ref{S-single}) the phase $\alpha$ in favor of the minimal distance $r_q$. From (\ref{eqn:tan.alpha}) one can show that $\alpha$ is given by:
\beq
2\mu^2\cos 2\alpha\,=\,\sinh^2 r_q\,-\, 2\mu^2\,\cosh r_q\,\,.
\eeq
Then, one can rewrite the profile function in (\ref{S-KS-alpha}) as:
\beq
S(r)\,=\,{(\cosh r-\cosh r_q)\,(\cosh r\,+\,\cosh r_q\,-\,2\mu^2)\over 
\sinh^2 r}\,\,\Theta(r-r_q)\,\,. 
\label{elementary-S-KS}
\eeq
Clearly $S(r_q)\,=\,0$ as it should. Actually, near $r=r_q$ the profile function behaves as:
\beq
S(r)\,\sim\,{2(\cosh r_q\,-\,2\mu^2)\over \sinh r_q}\,\,(r-r_q)\,\,,
\eeq
which in particular means that $S'(r_q)\not=0$ and, thus, $S'(r)$ has a finite jump at $r=r_q$ and the Ricci tensor will have the same type of singularity at the position  tip of the branes. 

\subsection{Taming the threshold singularity}

As in section \ref{Smoothing} we are going to find regular solutions by considering a brane distribution in which the position $r_q$ of the tip of the branes is not fixed but distributed with a measure. First of all, let us define $x$ and $x_q$ as:
\beq
x\,=\,\cosh r\,\,,\qquad\qquad x_q\,=\,\cosh r_q\,\,.
\eeq
The density distribution function $S(x)$ will be obtained by superposing the ``elementary" distributions of eq. (\ref{elementary-S-KS}). One gets:
\beq
S(x)\,=\,\int_{x_0}^{x}\,\td x_q\,\rho(x_q)\,\,
{(x-x_q)(x+x_q-2\mu^2)\over x^2-1}\,\,,
\label{average-S-KS}
\eeq
where $x_0$ is the minimal value of $x$ and $\rho(x_q)$ is the measure function, normalized as in (\ref{rho-normalization}). 
In the next two subsections we will consider two possible elections for the function $\rho(x_q)$.

\subsubsection{Phase average}

As mentioned above, for the embeddings (\ref{eqn:Kup.emb}) the position $x_q$ of the tip of the branes is related to the phase $\alpha$ of the parameter $\hat \mu$ by means of (\ref{eqn:minimum}), which now we rewrite as:
\beq
x_q\,=\,\mu^2\,+\, \sqrt{1+\mu^4\,+\,2\mu^2\cos (2\alpha)}\,\,.
\label{xq-alpha-KS}
\eeq
The prescription proposed in ref. \cite{Bigazzi:2008qq} consists in averaging over all possible values of $\alpha$ with a flat measure $\rho(x_q) \td x_q\sim \td\alpha$. Actually, taking into account that $x_q$  in (\ref{xq-alpha-KS}) depends on $\cos 2\alpha$, it is enough to consider $\alpha$ in the range $0\le \alpha\le \pi/2$. Let us denote by $x_0$ and $x_1$ the minimal and maximal values of $x_q$ respectively. For $\mu\ge 1$ (which we will assume in what follows) $x_0$ and $x_1$ are given by:
\beq
x_0\,=\,2\mu^2-1\,\,,\qquad\qquad x_1\,=\,2\mu^2\,+\,1\,\,.
\label{x0-x1-mu}
\eeq
Then, the measure for this prescription is given by:
\beq
\rho(x_q)\,\td x_q\,=\,{2\over \pi}\,\Big[\,
\Theta(x_q-x_0)\,-\,\Theta(x_q-x_1)\,\Big]\,\td\alpha\,\,,
\eeq
where the $x_q$'s on the right-hand side should be regarded as the functions of $\alpha$  written in (\ref{xq-alpha-KS}). 

Let us compute the profile function $S$ as a function of the radial variable $x$. From equation \eqref{eqn:tan.alpha}, we can see that the branes are lying in the range $2\m^2-1\leq x_q\leq2\m^2+1$. We can distinguish three regions.

\paragraph{Region I: $x\geq2\m^2+1$.}

In this region, the effect of all the branes ($0\leq\a<\pi/2$) can be felt. So, the range of integration in $\a$ will be from $\p/2$ to $0$. Therefore, we can write:
\beq
S(x)\,=\,{2\over \pi}\,\,
\int_{0}^{{\pi\over 2}}\,\td\alpha\,
{x^2-1-2\mu^2x\,-\,2\mu^2\,\cos 2\alpha\over x^2-1}\,\,.
\eeq
The integration over $\alpha$  kills the term with $\cos 2\a$  and one gets:
\beq
S(x)\,=\,1-2\m^2\frac{x}{x^2-1}\,\,,\qquad\qquad
x\ge2\mu^2+1
\eeq
which is exactly the expression (2.14) found in \cite{Bigazzi:2008qq} for region I.

\paragraph{Region II: $2\m^2-1\leq x\leq2\m^2+1$.}

In this region, not all of the branes are contributing, only those which reach a minimum distance smaller than $x$. Since the minimum value of $x_q$ is attained for $\a=\pi/2$  it suffices with integrating in the region $\a(x)\leq\a\leq\frac{\p}{2}$, where $\a(x)$ is obtained solving in \eqref{eqn:tan.alpha}, namely:
\beq
\a(x)=\arctan\left(\sqrt{\frac{(x+1)(1+2\m^2-x)}{(x-1)(1-2\m^2+x)}}\right)\,.
\eeq
Therefore, one has:
\beq
S(x)\,=\,{2\over \pi}\,\,
\int_{\alpha(x)}^{\pi/2}\,\td\alpha\,
{x^2-1-2\mu^2x\,-\,2\mu^2\,\cos 2\alpha\over x^2-1}\,\,.
\eeq
Performing the integral one obtains:
\beq
S(x)\,=\,{2\over \pi}\,\,\Big(\,{\pi\over 2}\,-\,\alpha(x)\,\Big)\,\Big(\,
1-2\m^2\frac{x}{x^2-1}\,\Big)\,-\,{2\mu^2\over \pi}\,\,{\sin 2\alpha(x)\over x^2-1}\,\,,
\eeq
which can be rewritten as:
\begin{align}
S(x)=&\frac{2}{\p}\left[\,\left(\,\frac{\p}{2}\,-\,
\arctan\left(\sqrt{\frac{(x+1)(1+2\m^2-x)}{(x-1)(1-2\m^2+x)}}\right)\right)\left(1-2\m^2\frac{x}{x^2-1}\right)+\right.\nonumber\\
&\left.+{1\over 2}\,\sqrt{{(1+2\m^2-x) (1-2\m^2+x)\over x^2-1}}
\right]\,.
\end{align}
This looks much simpler than the expression written in \cite{Bigazzi:2008qq}. However, one can check that both are equivalent. That is a consequence of the following non-trivial identity involving elliptic functions:
\beq
2\m^2\left((A_2(x,\m^2)-A_2(2\m^2+1,\m^2)\right)=\arctan\left(\sqrt{\frac{(x+1)(1+2\m^2-x)}{(x-1)(1-2\m^2+x)}}\right)\,,
\eeq
where the function $A_2(x,\m^2)$ has been defined in equation (2.16) of  \cite{Bigazzi:2008qq}.  Remarkably, the charge density function $S$ can be further simplified in this region, namely:
\beq
S(x)={1\over \pi}
\Bigg[\Big(1\,-\,2\mu^2\,{x\over x^2\,-\,1}\,\Big)\arccos
\big(x+{1-x^2\over 2\mu^2}\big)+
\sqrt{{(1+2\mu^2-x)\,(1-2\mu^2+x)\over x^2-1}}\,\Bigg]\,\,.
\label{KS-simpler}
\eeq

\paragraph{Region III: $x\leq2\m^2-1$.} Here no branes are contributing, so $S(x)=0$.

\subsubsection{Rectangular measure}

We will now consider the possibility of distributing the branes homogeneously in $x_q$ in some interval $x_0\le x_q\le x_1$, where $x_0=2\mu^2-1$  and $x_1-x_0\le 2$. The corresponding measure is just the same as in (\ref{step-measure})  with $\delta=x_1-x_0$ and the resulting profile function $S(x)$ can be straightforwardly obtained by performing the integral (\ref{average-S-KS}). One gets:
\beq
\begin{aligned}
S(x)\,&=\,{1\over 3}\,\,{2x+x_0-3\mu^2\over (x_1-x_0)\,(x^2-1)}\,(x-x_0)^2\,\,,
\qquad\qquad x_0\le x\le x_1\,\,,\\[0.1in]
S(x)\,&=\,{3x(x-2\mu^2)\,+\,x_0\,x_1\,+\,(x_0+x_1)\,(3\mu^2-x_0-x_1)
\over3( x^2-1)}\,\,,
\qquad x\ge x_1\,\,.
\end{aligned}
\eeq
We can expand this result near $x\approx x_0$, with the result:
\beq
S(x)\,\sim\,{x_0-\mu^2\over (x_1-x_0)\,(x_0^2-1)}\,\,(x-x_0)^2\,+\,\cdots\,\,,
\eeq
which leads to a regular background at $x=x_0$.

In the particular case in which $x_0$ and $x_1$ are taken as in the case of the phase average (\ie\ given by (\ref{x0-x1-mu})), the above expression for $S(x)$ reduces to:
\beq
\begin{aligned}
S(x)\,&=\,{2x-1-\mu^2\over 6(x^2-1)}\,\,(x+1-2\mu^2)^2\,\,,
\qquad\qquad 2\mu^2-1\le x\le 2\mu^2+1\,,\\[0.1in]
S(x)\,&=\,1\,-\,{2\over 3}\,\,{3\mu^2 x-1\over x^2-1}\,\,,\qquad\qquad
x\ge 2\mu^2+1\,\,.
\end{aligned}
\label{KS-flat}
\eeq
Now, near the endpoint $x=x_0=2\mu^2-1$, the behavior of $S(x)$ is simply:
\beq
S(x)\sim {1\over 8\mu^2}\,\,(x-2\mu^2+1)^2\,+\,\cdots\,\,.
\eeq
By plotting (\ref{KS-flat}) and (\ref{KS-simpler}) together one concludes that the profile given by the simpler expression (\ref{KS-flat}) is indeed very close to the one given by the more involved expression (\ref{KS-simpler}).

\subsection{Details of the calculation of the Jacobian}
\label{Jacobian-KS}

In this section  we will detail the calculation leading to  equations (\ref{z-integral}) and (\ref{jacobian}). 
Let us start by reparameterizing the complex variables $\tz_2$ and $z_3$, which take values in the entire complex plane, in terms of their modulus and phase:
\beq
	\tz_2 = \sqrt{u_2} \, e^{i \b_2} \,, \qquad \qquad z_3 = \sqrt{u_3} \, e^{i \b_3}\,\,.
\eeq
In terms of those new variables, one can rewrite $x = \cosh r$, from \eqref{eq:rInTermsOfz}, as:
\beq
	x = \frac{1+\m^4 +2 \m^2 u_3 +(u_2+u_3)^2 + 2 \m^2 \cos (2\a) -2u_2 \left( \m^2 \cos (2\a-2\b_2)+\cos (2\b_2)\right)}{2u_3}\,\,,
\eeq
and the integral is now:
\beq
{\cal I}\,\equiv\,
	\int \frac{1}{|z_3|^2} \td \tz_2 \wedge \td \bar{\tz}_2 \wedge \td z_3 \wedge \td \bar{z}_3\,\,F(r)\,=\, \int \frac{1}{u_3} \td u_2 \wedge \td u_3 \wedge \td \b_2 \wedge \td \b_3
	 \,\,F(r)\,\,.
	 \label{cal-I}
\eeq
Now we replace $u_3$ by $x$ in the integral (\ref{cal-I}). The relation  between these two variables is:
\beq
	u_3 = x - \m^2 - u_2 \pm M\,\,, 
\eeq
where $M$ is defined as:
\beq
M\,=\, \sqrt{x^2 - 1 + 2 u_2 (\cos(2\b_2) - x) + 2 \m^2 (u_2-x-\cos(2\a)+u_2 \cos(2\a-2\b_2))}\,\,.
\eeq

There are two different zones in the integral over $u_3$ in (\ref{cal-I}), one for $0<u_3<u_{3,0}$ (where one uses the relation with the minus sign) and the other one for $u_{3,0}<u_3$ (where one chooses the plus sign). $u_{3,0}$ is the value of $u_3$ for which $x$ is minimum, which is given by:
\beq
	u_{3,0} = \sqrt{1+\m^4 +u_2^2 + 2 \m^2 \cos (2\a) -2u_2 \left( \m^2 \cos (2\a-2\b_2)+\cos (2\b_2)\right)}\,\,.
\eeq
The value of $x$ at that point is just $x_{3,0} = \m^2 + u_2 +u_{3,0}$. Therefore, the integral over $u_3$ can be written as:
\beq
\int_0^\infty \frac{\td u_3}{u_3} = \int_0^{u_{3,0}} \frac{\td u_3}{u_3} + \int_{u_{3,0}}^\infty \frac{\td u_3}{u_3}\,=\,2\,\int_{x_{3,0}}^\infty {\td x\over M}\,\,.
\eeq
Since the integrand  in (\ref{cal-I}) does not depend on the phase $\beta_3$, we can write:
\beq
{\cal I}\,=\,4\pi\,\int_0^{2\pi}\,\td\beta_2\,
\int_0^{\infty}\,\td u_2\,\int_{x_{3,0}}^\infty \,\,{\td x\over M}\,F(x)\,\,.
\eeq
In order to perform next the integral over $u_2$ we must take into account that 
$x_{3,0}$ depends on $u_2$. By looking at \eqref{eqn:tan.alpha} one has that  $\min \{x_{3,0} \}_{u_2} = x_q$. Moreover  $x = x_{3,0}$ for $u=u_{2,0}$, with $u_{2,0}$ being:
\beq
	u_2 = u_{2,0} =\frac{1}{2}\frac{1+2\m^2 x - x^2 +2 \m^2 \cos (2\a)}{ \m^2 - x +\m^2 \cos(2\a-2\b_2) + \cos (2\b_2)}\,\,.
\eeq
Therefore, one can exchange the order of the integrations over $u_2$ and $x$ as follows:
\beq
	\int_0^\infty \td u_2 \int_{x_{3,0}}^\infty \td x = \int_{x_q}^\infty \td x \int_0^{u_{2,0}} \td u_2\,\,,
\eeq
and the integral over $u_2$ can be performed with the result:
\beq
\int_0^{u_{2,0}} \,\frac{\td u_2}{M}\,=\, \frac{\sqrt{x^2-1-2\m^2 x -2\m^2 \cos(2\a)}}{x - \m^2 - \m^2 \sin(2\a) \sin(2\b_2) - (\m^2 \cos(2\a) +1) \cos(2\b_2)}\,\,.
\eeq
The only remaining integral to perform is the one on $\b_2$. By employing the standard integration techniques in the complex plane, this integral can be easily computed with the result:
\beq
	\int_0^{2\p}  \td \b_2\frac{4\p\sqrt{x^2-1-2\m^2 x -2\m^2 \cos(2\a)}}{x - \m^2 - \m^2 \sin(2\a) \sin(2\b_2) - (\m^2 \cos(2\a) +1) \cos(2\b_2)}\,=\,8\pi^2\,\,.
\label{beta2-integral}
\eeq
Remarkably, the right-hand side of (\ref{beta2-integral})  does not depend on $x$, $\alpha$ or $\mu$. It follows that:
\beq
{\cal I}\,=\,8\pi^2\,\int_{x_q}^{\infty}\,\td x\,F(x)\,=\,8\pi^2\,\int_{r_q}^{\infty}\,
\td r\,\sinh r \,F(r)\,\,,
\eeq
in agreement with (\ref{z-integral}) and (\ref{jacobian}).



\begin{thebibliography}{99}
 
 
\bibitem{Maldacena:1997re}
  J.~M.~Maldacena,
  ``The large N limit of superconformal field theories and 
supergravity'',
  Adv.\ Theor.\ Math.\ Phys.\  {\bf 2}, 231 (1998)
  [Int.\ J.\ Theor.\ Phys.\  {\bf 38}, 1113 (1999)];
  hep-th/9711200.

 
 
\bibitem{Aharony:1999ti}
  O.~Aharony, S.~S.~Gubser, J.~M.~Maldacena, H.~Ooguri and Y.~Oz,
  ``Large N field theories, string theory and gravity,''
  Phys.\ Rept.\  {\bf 323}, 183 (2000)
  [arXiv:hep-th/9905111].


\bibitem{'tHooft:1973jz}
  G.~'t Hooft,
  ``A Planar Diagram Theory for Strong Interactions,''
  Nucl.\ Phys.\  {\bf B72}, 461 (1974).
  
 \bibitem{Malda-Nunez:N=1}
J.~M.~Maldacena and C.~N\'u\~nez
``Towards the large N limit of pure $\cN=1$ super Yang Mills,''
[arXiv:hep-th/0008001].

\bibitem{Chamseddine:1997nm}
  A.~H.~Chamseddine, M.~S.~Volkov,
  ``NonAbelian BPS monopoles in N=4 gauged supergravity,''
  Phys.\ Rev.\ Lett.\  {\bf 79}, 3343-3346 (1997).
  [hep-th/9707176].



\bibitem{MNreviews}
  M.~Bertolini,
  ``Four Lectures On The Gauge/Gravity Correspondence,''
  Int.\ J.\ Mod.\ Phys.\  A {\bf 18}, 5647 (2003)
  [arXiv:hep-th/0303160];
  F.~Bigazzi, A.~L.~Cotrone, M.~Petrini and A.~Zaffaroni,
  ``Supergravity duals of supersymmetric four dimensional gauge 
theories,''
  Riv.\ Nuovo Cim.\  {\bf 25N12}, 1 (2002)
  [arXiv:hep-th/0303191];
  A.~Paredes,
  ``Supersymmetric solutions of supergravity from wrapped branes,''
  arXiv:hep-th/0407013;
  J.~D.~Edelstein, R.~Portugues,
  ``Gauge/string duality in confining theories,''
  Fortsch.\ Phys.\  {\bf 54}, 525-579 (2006).
  [hep-th/0602021].



\bibitem{Karch:2002sh}
  A.~Karch and E.~Katz,
  ``Adding flavor to AdS/CFT'',
  JHEP {\bf 0206}, 043 (2002)
  [arXiv:hep-th/0205236]. 

 \bibitem{Nunez:2003cf}
C.~N\'u\~nez, A.~Paredes and A.~V.~Ramallo,
``Flavoring the gravity dual of $\cN = 1$ Yang-Mills with probes'',
[arXiv:hep-th/0311201].

 
\bibitem{Erdmenger:2007cm}
  J.~Erdmenger, N.~Evans, I.~Kirsch, E.~Threlfall,
  ``Mesons in Gauge/Gravity Duals - A Review,''
  Eur.\ Phys.\ J.\  {\bf A35}, 81-133 (2008).
  [arXiv:0711.4467 [hep-th]].
 
 
 
 
 
 
\bibitem{Casero:2006pt}
  R.~Casero, C.~Nunez, A.~Paredes,
  ``Towards the string dual of N=1 SQCD-like theories,''
  Phys.\ Rev.\  {\bf D73 } (2006)  086005.
  [hep-th/0602027].
 
 
\bibitem{Bigazzi:2005md}
  F.~Bigazzi, R.~Casero, A.~L.~Cotrone, E.~Kiritsis, A.~Paredes,
 ``Non-critical holography and four-dimensional CFT's with fundamentals,''
  JHEP {\bf 0510}, 012 (2005).
  [hep-th/0505140].


 \bibitem{Casero:2007pt}
 R.~Casero, C.~N\'u\~nez and A.~Paredes,
 ``Elaborations on the string dual of $\cN = 1$ SQCD'',
  Phys.\ Rev.\  {\bf D77 } (2008)  046003.
  [arXiv:0709.3421 [hep-th]].






\bibitem{HoyosBadajoz:2008fw}
 C.~Hoyos, C.~N\'u\~nez and I.~Papadimitriou,
 ``Comments on the String dual to $\cN = 1$ SQCD'',
[arXiv: hep-th/0807.3039].


\bibitem{Caceres}
  E.~Caceres, R.~Flauger, M.~Ihl, T.~Wrase,
  ``New supergravity backgrounds dual to N=1 SQCD-like theories with N(f) = 2N(c),''
  JHEP {\bf 0803}, 020 (2008).
  [arXiv:0711.4878 [hep-th]];
  E.~Caceres, R.~Flauger, T.~Wrase,
  ``Hagedorn Systems from Backreacted Finite Temperature $N_f=2N_c$ Backgrounds,''  [arXiv:0908.4483 [hep-th]].






\bibitem{Nunez:2010sf}
  C.~N\'u\~nez, A.~Paredes, A.~V.~Ramallo,
  ``Unquenched flavor in the gauge/gravity correspondence,''
  Adv.\ High Energy Phys.\  {\bf 2010}, 196714 (2010).
  [arXiv:1002.1088 [hep-th]].
  
  
  
\bibitem{Benini:2006hh}
  F.~Benini, F.~Canoura, S.~Cremonesi, C.~Nunez and A.~V.~Ramallo,
  ``Unquenched flavors in the Klebanov-Witten model,''
  JHEP {\bf 0702} (2007) 090
  [arXiv:hep-th/0612118].


 \bibitem{Benini:2007gx}
  F.~Benini, F.~Canoura, S.~Cremonesi, C.~N\'u\~nez and A.~V.~Ramallo,
  ``Backreacting Flavors in the Klebanov-Strassler Background,''
  JHEP {\bf 0709}, 109 (2007)
  [arXiv:0706.1238 [hep-th]].

\bibitem{Benini:2007kg}
  F.~Benini,
  ``A Chiral cascade via backreacting D7-branes with flux,''
  JHEP {\bf 0810} (2008) 051
  [arXiv:0710.0374 [hep-th]].

 


\bibitem{Conde:2011sw}
  E.~Conde and A.~V.~Ramallo,
  ``On the gravity dual of Chern-Simons-matter theories with unquenched
  flavor,''
  arXiv:1105.6045 [hep-th].



\bibitem{Bigazzi:2009bk}
  F.~Bigazzi, A.~L.~Cotrone, J.~Mas, A.~Paredes, A.~V.~Ramallo, J.~Tarrio,
  ``D3-D7 Quark-Gluon Plasmas,''
  JHEP {\bf 0911}, 117 (2009).
  [arXiv:0909.2865 [hep-th]].







\bibitem{smearingmethod}
  F.~Bigazzi, A.~L.~Cotrone and A.~Paredes,
  ``Klebanov-Witten theory with massive dynamical flavors'',
  JHEP {\bf 0809} (2008) 048
  [arXiv:hep-th/0807.0298].
  
  
  
\bibitem{Bigazzi:2008ie}
  F.~Bigazzi, A.~L.~Cotrone, A.~Paredes, A.~Ramallo,
  ``Non chiral dynamical flavors and screening on the conifold,''
  Fortsch.\ Phys.\  {\bf 57}, 514-520 (2009).
  [arXiv:0810.5220 [hep-th]].
  
  
\bibitem{Bigazzi:2008qq}
 F.~Bigazzi, A.~L.~Cotrone, A.~Paredes and A.~V.~Ramallo,
 ``The Klebanov-Strassler model with massive dynamical flavors,''
 JHEP {\bf 0903}, 153 (2009) [arXiv:0812.3399 [hep-th]].



\bibitem{Malda-Nunez:nogo}
J.~M.~Maldacena and C.~N\'u\~nez
``Supergravity description of field theories on curved manifolds and a no  go
theorem,''
[arXiv:hep-th/0007018].





\bibitem{AndrewsDorey}
  R.~P.~Andrews, N.~Dorey,
  ``Deconstruction of the Maldacena-Nunez compactification,''
  Nucl.\ Phys.\  {\bf B751}, 304-341 (2006).
  [hep-th/0601098];
  R.~P.~Andrews, N.~Dorey,
  ``Spherical deconstruction,''
  Phys.\ Lett.\  {\bf B631}, 74-82 (2005).
  [hep-th/0505107].






 





\bibitem{Herzog:2001xk}
  C.~P.~Herzog, I.~R.~Klebanov, P.~Ouyang,
  ``Remarks on the warped deformed conifold,''
  [hep-th/0108101].


\bibitem{MetodoAngel}
Angel Paredes, unpublished notes.



\bibitem{Seiberg:1994pq}
  N.~Seiberg,
  ``Electric - magnetic duality in supersymmetric nonAbelian gauge theories,''
  Nucl.\ Phys.\  {\bf B435}, 129-146 (1995).
  [hep-th/9411149].





\bibitem{Strassler:2005qs}
  M.~J.~Strassler,  ``The Duality cascade,'' [hep-th/0505153].



 \bibitem{Bigazzi:2008gd}
  F.~Bigazzi, A.~L.~Cotrone, C.~N\'u\~nez and A.~Paredes,
  ``Heavy quark potential with dynamical flavors: a first order transition,''
  Phys.\ Rev.\  D {\bf 78} (2008) 114012
  [arXiv:0806.1741 [hep-th]].  
 


\bibitem{Nunez:2009da}
  C.~N\'u\~nez, M.~Piai and A.~Rago,
  ``Wilson Loops in string duals of Walking and Flavored Systems,''
  arXiv:0909.0748 [hep-th].

  


  
   
  
   

 
\bibitem{Herzog:2001fq}
  C.~P.~Herzog, I.~R.~Klebanov,
  ``On string tensions in supersymmetric SU(M) gauge theory,''
  Phys.\ Lett.\  {\bf B526}, 388-392 (2002).
  [hep-th/0111078].
 
 
\bibitem{Bachas:2000ik}
  C.~Bachas, M.~R.~Douglas, C.~Schweigert,
  ``Flux stabilization of D-branes,''
  JHEP {\bf 0005}, 048 (2000).
  [hep-th/0003037]. 
 
 
 
\bibitem{Bertolini&Merlatti}
M.~Bertolini and P.~Merlatti,
``A note on the dual of $\cN=1$ super Yang-Mills theory'',
[arXiv:hep-th/0211142].
 
 
 
\bibitem{Ridgway:2007vh}
  J.~M.~Ridgway,
  ``Confining k-string tensions with D-Branes in Super Yang-Mills theories,''
  Phys.\ Lett.\  {\bf B648}, 76-83 (2007).
  [hep-th/0701079].

 
 
 
\bibitem{Camino:2001at}
  J.~M.~Camino, A.~Paredes, A.~V.~Ramallo,
  ``Stable wrapped branes,''
  JHEP {\bf 0105}, 011 (2001).
  [hep-th/0104082].
 



\bibitem{Barranco}
  A.~Barranco, E.~Pallante and J.~G.~Russo,
  ``N=1 SQCD-like theories with $N_f$ massive flavors from AdS/CFT and beta
  functions,'' arXiv:1107.4002 [hep-th].

 
 
 
\bibitem{Gaillard:2010qg}
  J.~Gaillard, D.~Martelli, C.~Nunez and I.~Papadimitriou,
  ``The warped, resolved, deformed conifold gets flavoured,''
  Nucl.\ Phys.\  B {\bf 843} (2011) 1
  [arXiv:1004.4638 [hep-th]].
 
 
 
\bibitem{Nunez:2008wi}
  C.~Nunez, I.~Papadimitriou and M.~Piai,
  ``Walking Dynamics from String Duals,''
  Int.\ J.\ Mod.\ Phys.\  A {\bf 25} (2010) 2837
  [arXiv:0812.3655 [hep-th]].





\bibitem{Elander:2009pk}
  D.~Elander, C.~Nunez and M.~Piai,
``A Light scalar from walking solutions in gauge-string duality,''
  Phys.\ Lett.\  B {\bf 686}, 64 (2010)
  [arXiv:0908.2808 [hep-th]].

  
\bibitem{Piai:2010ma}
  M.~Piai,
  ``Lectures on walking technicolor, holography and gauge/gravity dualities,''
  Adv.\ High Energy Phys.\  {\bf 2010}, 464302 (2010).
  [arXiv:1004.0176 [hep-ph]].
  
  
  

\bibitem{ks}I.~R.~Klebanov and M.~J.~Strassler,
  ``Supergravity and a confining gauge theory: Duality cascades and
  chiSB-resolution of naked singularities,''
  JHEP {\bf 0008}, 052 (2000)
  [arXiv:hep-th/0007191].

  





\bibitem{kuper}
  S.~Kuperstein,
  ``Meson spectroscopy from holomorphic probes on the warped deformed conifold,''
  JHEP {\bf 0503}, 014 (2005).
  [hep-th/0411097].




    
    
    
    
  
\end{thebibliography}
\end{document}